\documentclass[12pt,titlepage]{article}
\usepackage{cite}
\def\href#1{}
\def\IZ{\mathrm{Z}}
\newcommand{\sectiono}[1]{\section{#1}\setcounter{equation}{0}}

\def\eq{\begin{equation}}
\def\en{\end{equation}}
\def\eqa{\begin{eqnarray}}
\def\ena{\end{eqnarray}}
\newcommand{\me}{\mathrm{e}}
\newcommand{\dif}{\mathrm{d}}
\def\Dlambda{\mathrm{D}\lambda}
\def\D{\mathrm{D}}
\def\gst{g_s}
\def\abs#1{\left | #1 \right |}
\def\expval#1{\left\langle #1 \right\rangle}
\def\lambdar{\lambda_{r}}
\def\partialby#1{\frac{\partial\hfill}{\partial#1}}
\def\Dlambdar{\mathrm{D}\lambdar}
\def\drho{\dif \rho }
\def\dvol{\mathrm{dvol}}
\def\del{\nabla}
\def\drhoequil{\dif \rho_{\mathit eq}}
\def\Lambdazero{\Lambda_{0}}
\def\lPlanck{l_{\mathrm P}}
\def\gev{{\mathrm GeV}}
\def\apm{{\it a priori} measure}
\def\citeF{\cite{Friedan-1,Friedan-2,Friedan-3}}
\def\M#1{M(#1) }
\def\barMinfinity{\overline{\M{\infty}}}
\def\eff{e}
\def\lambdaeff{\lambda_{\eff}}
\def\lambdazero{\lambda_{0}}
\def\phieff{\phi^{\Lambda,\eff}}
\def\betaeff{\beta_{\eff}}
\def\geff{g^{\eff}}
\def\Dlambdaeff{\mathrm{D}\lambdaeff}
\def\Seff{S_{\eff}}
\def\deleff{\del^{\eff}}
\def\gsubeff{g_{\eff}}
\def\Teff{T_{\eff}}
\def\aeff{a_{\eff}}
\def\dvoleff{\dvol_{\eff}}
\def\ad{\gamma}
\def\alphaprime{\alpha^{\prime}}
\def\substar{_{\ast}}
\def\betastar{\beta\substar}
\def\Z{Z}
\def\L{{\mathcal L}}
\def\Deff{D_{\eff}}
\def\Leff{{\mathcal L}_{\eff}}
\def\adeff{\ad_{\eff}}
\def\rhor{\rho_{r}}
\def\drhor{\dif \rhor}
\def\rhoeff{\rho_{\eff}}
\def\drhoeff{\dif \rhoeff}
\def\expvaleff#1{\expval{#1}_{\eff}}
\def\NS{N-S}
\def\t{t}
\def\that{\hat{t}}
\def\K{K}
\def\Khat{\hat{K}}
\def\Sgso{S_{+}}
\def\ket#1{ \left . | #1 \right \rangle}
\def\bra#1{\left \langle#1 | \right .}
\def\inprod#1#2{\left \langle #1 | #2 \right \rangle}
\def\F{F}
\def\G{G}
\def\pslash{\not \! p}
\def\Imag{\mathrm{Im}}
\begin{document}
%
%
%
\begin{titlepage}
\hfill{\vbox{\hbox{hep-th/0204131}
 				\hrule width0pt height1.5\smallskipamount
 				\hbox{RUNHETC-2002-12}
 				\hrule width0pt height5\medskipamount}}
\vspace{12ex}
\begin{center}
{\Large A tentative theory of large distance physics}\\[8ex]
{\large Daniel Friedan}\\[4ex]
         Department of Physics and Astronomy\\
		 Rutgers, The State University of New Jersey\\
		 Piscataway, New Jersey, USA\\[1ex]
		 and\\[1ex]
		 Raunv\'{\i}sindastofnum H\'ask\'olans \'Islands\\
		 Reykjav\'{\i}k, \'{I}sland\\
		 The Natural Science Institute of the University of Iceland\\
		 Reykjavik, Iceland\\[2ex]
		 email: friedan@physics.rutgers.edu\\[12ex]
April 17, 2002
\end{center}
\end{titlepage}
\begin{abstract}
A theoretical mechanism is devised to determine the large distance physics of 
spacetime.  It is a two dimensional nonlinear model, the lambda model, set to 
govern the string worldsurface to remedy the failure of string theory.  The 
lambda model is formulated to cancel the infrared divergent effects of handles 
at short distance on the worldsurface.  The target manifold is the manifold of 
background spacetimes.  The coupling strength is the spacetime coupling 
constant.  The lambda model operates at 2d distance $\Lambda^{-1}$, very much 
shorter than the 2d distance $\mu^{-1}$ where the worldsurface is seen.  A large 
characteristic spacetime distance $L$ is given by $L^2=\ln(\Lambda/\mu)$.  
Spacetime fields of wave number up to $1/L$ are the local coordinates for the 
manifold of spacetimes.  The distribution of fluctuations at 2d distances 
shorter than $\Lambda^{-1}$ gives the {\it a priori} measure on the target 
manifold, the manifold of spacetimes.  If this measure concentrates at a 
macroscopic spacetime, then, nearby, it is a measure on the spacetime fields.  
The lambda model thereby constructs a spacetime quantum field theory, cutoff at 
ultraviolet distance $L$, describing physics at distances larger than $L$.  The 
lambda model also constructs an effective string theory with infrared cutoff 
$L$, describing physics at distances smaller than $L$.  The lambda model evolves 
outward from zero 2d distance, $\Lambda^{-1} = 0$, building spacetime physics 
starting from $L=\infty$ and proceeding downward in $L$.  $L$ can be taken 
smaller than any distance practical for experiments, so the lambda model, if 
right, gives all actually observable physics.  The harmonic surfaces in the 
manifold of spacetimes are expected to have novel nonperturbative effects at 
large distances.%

\end{abstract}
\newpage
\thispagestyle{empty}
\tableofcontents
\newpage
\setcounter{page}{1}
%
%
%
\sectiono{Introduction}
\label{sect:intro}

I propose here a systematic, mechanical theory of large distance 
physics.  The mechanism is a two dimensional nonlinear model, the 
{\em lambda model}, whose target manifold is a manifold of 
spacetimes.  Each spacetime is characterized by its riemannian 
metric and certain other spacetime fields.  In the lambda model, 
spacetime as a whole fluctuates locally in two dimensions.  The 
distribution of the fluctuations at short two dimensional distance 
is a measure on the manifold of spacetimes.  If this measure 
concentrates at a macroscopic spacetime, then, nearby, it is a 
measure on the spacetime fields in that macroscopic spacetime.  
Spacetime quantum field theory is thereby constructed as the 
effective description of large distance physics.  But the dynamics 
that governs the large distance physics is the local dynamics of the 
two dimensional nonlinear model, the lambda model.

I only formulate the theory here.  I describe its structure and 
speculate about its prospects.  I do no calculations in the theory.  
The arguments are based on abstract general principles.  Most of the 
technical details are left to be filled in.  I concentrate on the 
task of formulating a well-defined theoretical structure that is 
capable of providing a comprehensive and useful theory of the large 
distance physics of the real world.  The theory that I am proposing 
does appear capable of selecting a specific discrete set of 
macroscopic spacetimes, producing a specific spacetime quantum field 
theory in each.  In particular, the theory appears capable of 
producing specific, calculable, nonperturbatively small mass 
parameters in the effective spacetime quantum field theories.  But 
calculations are needed to check whether the theory actually does 
accomplish this.  If the theory does work as envisioned, it will be 
a comprehensive, definitive, predictive theory of large distance 
physics, whose reliability can be checked by detailed comparison 
with existing experimental knowledge of the real world.

\subsection{Renormalization of the general nonlinear model}

This work began with the renormalization of the two dimensional 
general nonlinear model~\citeF. The general nonlinear model is a two 
dimensional quantum field theory.  It is defined as a functional 
integral
\eq
\int \D x \; \me^{-A(x)}
\en
over all maps $x(z,\bar z)$ from the plane to a fixed compact 
riemannian manifold, called the target manifold.  The couplings of 
the general nonlinear model are comprised in a riemannian metric 
$h_{\mu\nu}(x)$ on the target manifold, called the target metric or 
the metric coupling.  The classical action is
\eq
\int \dif^{2}z \, \frac1{2\pi} \,
h_{\mu\nu}(x) \, \partial x^{\mu} \, \bar \partial x^{\nu}
\:.
\en
Each wave mode $\delta h_{\mu\nu}(x)$ of the reimannian metric on 
the target manifold is a coupling constant $\lambda^{i}$ in the two 
dimensional quantum field theory, parametrizing a perturbation
\eq
\delta A(x) = \int \dif^{2}z \, \frac1{2\pi} \,
\delta h_{\mu\nu}(x) \, \partial x^{\mu} \, \bar \partial x^{\nu}
\en
of the action.  The general nonlinear model is `general' in the 
sense that the riemannian metric on the target manifold is not 
assumed to have any special symmetries.

The general nonlinear model was shown to be renormalizable~\citeF. 
The renormalized couplings of the model were shown to comprise an 
effective riemannian metric on the target manifold, at every two 
dimensional distance.  The renormalization group was shown to act as 
a flow on the manifold of riemannian metrics.  The infinitesimal 
renormalization group generator $\beta^{i}(\lambda)$ became a vector 
field on the manifold of riemannian metrics.  The renormalization 
group fixed point equation $\beta = 0$, expressing two dimensional 
scale invariance, became, at large distance in the target manifold, 
the equation $R_{\mu\nu}=0$.  This was recognized as Einstein's 
equation for the spacetime metric in general relativity, without 
matter.

Renormalization is based on an extremely large ratio between two 
distances.  The quantum field theory is constructed at a short 
distance $\Lambda^{-1}$.  The theory is used, its properties 
calculated, at a long distance $\mu^{-1}$.  Inverse powers of the 
extremely large ratio $\Lambda/\mu$ act to suppress the effects of 
all coupling constants having negative scaling dimensions.  In the 
general nonlinear model, the distances $\Lambda^{-1}$ and $\mu^{-1}$ 
are two dimensional distances.

The metric coupling of the general nonlinear model is naively 
dimensionless.  The fluctuations in the model give each coupling 
constant $\lambda^{i}$ an anomalous scaling dimension $-\ad(i)$.  It 
was shown that the anomalous scaling dimensions $-\ad(i)$ are the 
eigenvalues of a covariant second order differential operator on the 
target manifold, acting on the wave modes of the riemannian target 
metric.  Each coupling constant $\lambda^{i}$ is an eigenmode with 
eigenvalue $-\ad(i)$.  The numbers $\ad(i)$ take the form $\ad(i)= 
p(i)^{2}$ up to corrections for the curvature of the target 
manifold, where $p(i)$ is the spacetime wave number of the wave mode 
$\lambda^{i}$.
	
In a renormalized quantum field theory, the coupling constants 
$\lambda^{i}$ having $\ad(i)>0$ are irrelevant.  Their effects are 
suppressed by factors $(\Lambda/\mu)^{-\ad(i)}$.  The quantum field 
theory depends only on the $\lambda^{i}$ having $\ad(i)=0$, which 
are the marginal coupling constants, and the $\lambda^{i}$ having 
$\ad(i)<0$, which are the relevant coupling constants.  Thus the 
small distance modes of the target riemannian metric, the wave modes 
of high wave number, became irrelevant coupling constants 
$\lambda^{i}$ in the renormalized general nonlinear model.  The 
large distance wave modes of the riemannian metric became the 
marginal and relevant coupling constants in the general nonlinear 
model.

The target manifold of the general nonlinear model was taken to be 
compact and riemannian so that the model would be well-defined as a 
two dimensional quantum field theory.  Assuming a riemannian target 
manifold ensured that the action $A(x)$ would be bounded below.  
Assuming a compact target manifold ensured a discrete spectrum of 
anomalous scaling dimensions $-\ad(i)$.  It followed from these 
assumptions that only a finite number of marginal and relevant 
coupling constants $\lambda^{i}$ could occur in the general 
nonlinear model.  The marginal and relevant coupling constants 
$\lambda^{i}$ are the parameters for variations of the quantum field 
theory.  So the space of general nonlinear models was shown to be a 
finite dimensional manifold.

When spacetime geometry was translated into the language of the 
renormalization of the general nonlinear model, it became possible 
to imagine that the physics of real spacetime might be found encoded 
within that abstract machinery.  It became possible to imagine that 
real spacetime might in fact be the target manifold of the general 
nonlinear model, and that Einstein's equation on the physical metric 
of spacetime might in fact be the fixed point equation $\beta=0$ of 
the renormalization group acting on the metric coupling of the 
general nonlinear model.

Renormalized two dimensional quantum field theory offers a small set 
of abstract basic principles which are distinguished, definitive and 
tractable in comparison with the possible principles of spacetime 
physics that they would replace.  The wave modes of the spacetime 
metric become the coupling constants $\lambda^{i}$ which parametrize 
the two dimensional quantum field theory.  The equation of motion 
$R_{\mu\nu}=0$ on the spacetime metric becomes the renormalization 
group fixed point equation $\beta^{i}(\lambda) =0$, expressing scale 
invariance of the two dimensional quantum field theory.  Positivity 
of the spacetime metric becomes unitarity of the two dimensional 
quantum field theory.  The compactness of spacetime becomes the 
discreteness of the spectrum of two dimensional scaling fields.  
Small distance in spacetime becomes irrelevance in the two 
dimensional quantum field theory.  Geometric conditions on spacetime 
became natural regularity conditions on two dimensional quantum 
field theories.  The renormalization of the coupling constants 
$\lambda^{i}$ of the general nonlinear model is a systematic and 
reliable calculus.  A construction can be formulated in the language 
of the renormalized general nonlinear model with confidence in its 
coherence, although explicit calculations might remain technically 
difficult.

When the general nonlinear model was shown to be renormalizable, it 
was pointed out~\citeF\ that a manifold of nontrivial compact 
riemannian solutions to the one loop fixed point equation 
$R_{\mu\nu}=0$ were already known to exist, namely the Calabi-Yau 
spaces~\cite{Calabi,Yau}.  But the two dimensional scale invariance 
of a general nonlinear model with a Calabi-Yau target manifold was 
violated by the two loop contribution to the beta function.  
Nontrivial two dimensional scale invariance was discovered in the 
supersymmetric version of the general nonlinear model with 
Calabi-Yau target manifold, when the remarkable cancellation among 
the two loop contributions to the beta function was 
discovered~\cite{AGF}.

\subsection{Application in string theory}

The nontrivial scale invariant general nonlinear models found a role 
in perturbative string theory~\cite{Lovelace-1}.  The general 
nonlinear model constructs the string worldsurface in a curved 
background spacetime.  The target manifold of the nonlinear model is 
the background spacetime in which strings scatter.  The two 
dimensional plane gives the local two dimensional patches out of 
which the string worldsurface is made.  Consistency of the string 
theory requires the string worldsurface to be scale invariant, so 
the coupling constants $\lambda^{i}$ in the general nonlinear model 
of the worldsurface must satisfy the fixed point equation 
$\beta^{i}(\lambda) = 0$.  The manifold of scale invariant general 
nonlinear models forms the manifold of possible background 
spacetimes.

\subsection{The failure of string theory}

String theory failed as a theory of physics because of the existence 
of a manifold of possible background spacetimes.  All potentially 
observable properties of string theory depend on the geometry and 
topology of the background spacetime in which the strings scatter.  
In string theory, a specific background spacetime has to be selected 
by hand, or by ``initial conditions,'' from among the manifold of 
possibilities.  Many continuously adjustable parameters must be 
dialed arbitrarily to specify the background spacetime.  The 
existence of a manifold of possible background spacetimes renders 
string theory powerless to say anything definite that can be 
checked.
 
\subsection{Physics is reliable knowledge}

Physics is reliable knowledge of the real world, based on 
experiment.  A new theory of physics must establish its reliability 
first by explaining existing knowledge of the real world.  New 
theories of physics build on existing reliable theories.  A 
candidate theory of physics obtains credibility first by giving 
definite explanations of established theories.  A new theory 
inherits the reliability of the theories it explains.  For example, 
special relativity explained newtonian mechanics.  General 
relativity explained special relativity and newtonian gravity.  
Quantum mechanics explained classical mechanics.  Bohr's 
correspondence principle, which guided the formation of quantum 
mechanics, was an explicit statement that a candidate theory of 
physics must explain the existing reliable theory.

Present knowledge of the laws of physics is summarized in the 
combination of classical general relativity and the standard model 
of elementary particles, to the extent that the standard model has 
been confirmed by experiment.  A candidate theory of physics must 
establish its reliability by explaining this currently successful 
theory.  It must explain classical general relativity and the 
standard model in detail.  At the very least, a candidate theory of 
physics must contain fewer adjustable parameters than does the 
standard model, and must give reliable methods to calculate precise 
numerical values for the masses and coupling constants in the 
standard model.  A candidate theory of physics that is not capable 
of explaining the standard model and classical general relativity 
cannot obtain reliability, because it cannot be checked against the 
existing knowledge of the real world.

The standard model of elementary particles is a quantum field 
theory.  General relativity is a classical field theory, but can be 
regarded equally well as a quantum field theory which is accurately 
approximated by its classical field theoretic limit at the spacetime 
distances where gravity is observed.  A theory of physics should 
explain quantum field theory.  It should explain why quantum field 
theory in spacetime has been so successful at the spacetime 
distances accessible to observation.

A candidate theory of physics must be capable of {\em producing} 
spacetime quantum field theory.  More, it must be capable of 
producing one specific quantum field theory, containing specific, 
nontrivial, calculable mass parameters and coupling constants.  One 
specific quantum field theory, the standard model, has been 
successful in physics, not quantum field theory in general.  Quantum 
field theory in general has too many free parameters to be a useful 
search space in which to find a definite explanation of the standard 
model.  A mechanism is needed that is capable of producing a 
specific spacetime quantum field theory, the one that is actually 
seen in the real world.  The unnaturally small value of the 
cosmological constant suggests that, if such a mechanism is at work 
in the real world, it does not work generically, but rather in a 
very specific fashion, to produce a very specific spacetime quantum 
field theory.

I stress {\em capability}.  The first step in forming a theory of 
physics is to find a well-defined theoretical structure {\em 
capable} of producing a specific spacetime quantum field theory.  
Only then is there a chance of explaining the standard model and 
general relativity, and of making definite predictions.  When such a 
theoretical structure is found, it becomes worthwhile to perform 
calculations to determine whether the capabilities are realized.  Of 
course, success in physics requires actually giving definite 
explanations of existing knowledge and actually making definite 
predictions that are verified.  But a first prerequisite in a 
candidate theory of physics is a structure {\em capable} of 
providing definite, unequivocal explanations and predictions.  A 
theory of physics must be capable of making definite statements that 
can be checked.  It must be capable of doing useful work in the real 
physical world.

\subsection{Only large distance physics is observable}

In units of the Planck length, $\lPlanck=(1\times 10^{19} \,
\gev)^{-1}$, the smallest distance probed by feasible experiments is 
a very large dimensionless number, on the order of $1 \times 10^{16} 
= (1\times 10^{3} \,\gev \, \lPlanck)^{-1}$, or perhaps $1 \times 
10^{14} = (1\times 10^{5} \,\gev \, \lPlanck)^{-1}$.  In any theory 
of physics in which spacetime distances are dimensionless numbers 
and in which the unit of distance lies within a few orders of 
magnitude of the Planck length, the only theoretical explanations 
and predictions that can be checked against experiment are those 
made in the large distance limit of the theory.

\subsection{The long-standing crisis of string theory}

The long-standing crisis of string theory is its complete failure to 
explain or predict any large distance physics.  String theory cannot 
say anything definite about large distance physics.  String theory 
is incapable of determining the dimension, geometry, particle 
spectrum and coupling constants of macroscopic spacetime.  String 
theory cannot give any definite explanations of existing knowledge 
of the real world and cannot make any definite predictions.  The 
reliability of string theory cannot be evaluated, much less 
established.  String theory has no credibility as a candidate theory 
of physics.

Recognizing failure is a useful part of the scientific strategy.  
Only when failure is recognized can dead ends be abandoned and 
useable pieces of failed programs be recycled.  Aside from possible 
utility, there is a responsibility to recognize failure.  
Recognizing failure is an essential part of the scientific ethos.  
Complete scientific failure must be recognized eventually.

String theory fails to explain even the existence of a macroscopic 
spacetime, much less its dimension, geometry and particle physics.  
The size of the generic possible background spacetime is of order 
$1$ in dimensionless units.  Large distances occur only in 
macroscopic spacetimes, which are found near the boundary of the 
manifold of background spacetimes.  String theory, being incapable 
of selecting from among the manifold of possible background 
spacetimes, cannot explain the existence of a macroscopic spacetime.

Even if some particular macroscopic background spacetime is chosen 
arbitrarily, by hand or by ``initial conditions,'' string theory 
still fails to be realistic at large distance.  The large distance 
limit of string theory consists of the perturbative scattering 
amplitudes of the low energy string modes, which are particle-like.  
But the particle masses are exactly zero, and the low energy 
scattering amplitudes are exactly supersymmetric.  String theory 
fails to provide any mechanism to generate the very small nonzero 
masses that are observed in nature, or to remove the exact spacetime 
supersymmetry, which is not observed in nature.  More broadly, 
string theory is incapable of generating the variety of large 
characteristic spacetime distances seen in the real world.  At best, 
for each macroscopic background spacetime in the manifold of 
possibilities, string theory gives large distance scattering 
amplitudes that form a caricature of the scattering amplitudes of 
the standard model of particle physics.

The massless string modes are the manifestations, locally in the 
macroscopic spacetime, of the continuous degeneracy of the manifold 
of background spacetimes.  The failure of string theory to generate 
nonzero small particle masses is a consequence of its failure to 
resolve the continuous degeneracy of the manifold of spacetimes.  
The continuous degeneracy of the manifold of background spacetimes 
makes string theory unacceptable as a candidate theory of physics.  
If the continuous degeneracy were accepted, then, by assumption, it 
would be impossible to determine the dimension and geometry of 
macroscopic spacetime or the masses and coupling constants of the 
elementary particles.

String theory fails to produce spacetime quantum field theory at 
large distance.  String theory gives only scattering amplitudes.  
String theory cannot explain the standard model, or general 
relativity, because it cannot {\em produce} a spacetime quantum 
field theory as an effective description of large distance physics.  
The practice in string theory is to {\em assume} that spacetime 
quantum field theory describes the large distance physics.  First, a 
macroscopic background spacetime is chosen by hand, arbitrarily, 
from among the manifold of possibilities.  Then string theory 
scattering amplitudes are calculated perturbatively in the chosen 
background spacetime.  The perturbative string theory is invariant 
under some spacetime supersymmetries.  The massless particle-like 
states and their perturbative large distance scattering amplitudes 
are identical to the perturbative large distance scattering 
amplitudes derived from a supersymmetric field theory lagrangian in 
the arbitrarily chosen macroscopic spacetime.  It is then {\em 
assumed} that the large distance physics in the chosen macroscopic 
spacetime is given by some quantized version of the supersymmetric 
spacetime field theory.  The continuous degeneracy of the manifold 
of background spacetimes appears as a continuous degeneracy of the 
manifold of ground states of the spacetime field theory, as a 
continuous degeneracy of the manifold of possible vacuum expectation 
values of the spacetime fields.  The supersymmetric spacetime field 
theory is then examined for possible nonperturbative effects that 
might break the degeneracy of the manifold of ground states.

The assumption that spacetime quantum field theory governs the large 
distance physics is not justified.  There is no derivation of 
spacetime quantum field theory from string theory.  There is no 
construction from string theory of any effective spacetime quantum 
field theory governing the large distance physics, even given an 
arbitrary choice of background spacetime.  String theory is 
incapable of explaining any spacetime field theory, classical or 
quantum mechanical.  String theory provides nothing at large 
distance but perturbative scattering amplitudes for gravitons and 
other massless particles.  It is true that the same perturbative 
scattering amplitudes for massless particles can be derived from 
massless supersymmetric quantum field theories, but this formal 
coincidence does not justify the claim that string theory explains 
quantum field theory, or the claim that string theory implies 
quantum field theory at large distances.

In particular, there is no justification for the claim that string 
theory explains or predicts gravity.  String theory gives 
perturbative scattering amplitudes of gravitons.  Gravitons have 
never been observed.  Gravity in the real world is accurately 
described by general relativity, which is a classical field theory.  
There is no derivation of general relativity from string theory.  
General relativity can be regarded as the large distance classical 
limit of quantum general relativity, if an ultraviolet cutoff is 
imposed to make sense of quantum general relativity.  A cutoff 
quantum general relativity would give the same formal perturbative 
low energy scattering amplitudes for massless gravitons as does 
string theory.  But it is illogical to claim, from this formal 
coincidence between two technical methods of calculating unobserved 
graviton scattering amplitudes, that string theory explains 
classical general relativity, or that string theory explains 
gravity, or that string theory is a quantum theory of gravity.  
String theory does not produce any mechanical theory of gravity, 
much less a quantum mechanical theory.

In any case, a quantum theory of gravity is unnecessary.  No 
physical effects of quantum gravity have been observed, and there is 
no credible possibility of observing any.  What {\em is} needed is a 
theory which produces general relativity as an effective classical 
field theory at large distance.  It might produce classical general 
relativity by producing at large distance an effective quantized 
general relativity that is deep in its classical regime.  But what 
is essential to produce is the classical, mechanical spacetime field 
theory of gravity.

String theory is only a perturbative theory.  The widespread 
practice is to assume that there exists a nonperturbative 
formulation of string theory, and that this hypothetical 
nonperturbative formulation would be a quantum mechanical theory, 
microscopic in spacetime, invariant under some exact, fundamental 
spacetime supersymmetries.  If such a nonperturbative formulation of 
string theory did exist, then it might well follow that the large 
distance physics in that hypothetical theory would be governed by 
supersymmetric spacetime quantum field theory, and that the fate of 
the degeneracy of the manifold of background spacetimes would be 
determined by nonperturbative field theoretic effects at large 
distance in spacetime in that supersymmetric quantum field theory.  
But it is only an assumption that there exists such a 
nonperturbative, microscopic, quantum mechanical formulation of 
string theory.  Any reasoning about a hypothetical nonperturbative 
version of string theory is unreliable if it rests on the assumption 
of spacetime quantum field theory at large distance, without any way 
to derive spacetime quantum field theory from string theory.

The assumption of fundamental, exact, quantum mechanical spacetime 
supersymmetry is a very strong extrapolation from perturbative 
string theory, where spacetime supersymmetry is only a perturbative 
symmetry of the scattering amplitudes in individual background 
spacetimes.  Adopting this assumption requires accepting as 
inevitable the continuous degeneracy of the manifold of background 
spacetimes.  An assumption as strong as fundamental spacetime 
supersymmetry loses credibility as a guide in searching for a theory 
of physics if it cannot lead to definite explanations of existing 
knowledge and definite predictions.  There is certainly no physical 
evidence to support the assumption that spacetime supersymmetry is a 
fundamental property of nature.  At most, it is possible that 
indications of approximate spacetime supersymmetry might be found 
experimentally in the not so distant future.  Contrast the radical 
assumptions of the old quantum theory, which obtained credibility by 
giving definite explanations of the black body spectrum, the 
photoelectric effect, the Balmer series, the Rydberg constant, and 
much more of atomic physics, before eventually leading to quantum 
mechanics.

Any reasoning about a hypothetical nonperturbative version of string 
theory is unreliable if it assumes fundamental spacetime 
supersymmetry but is unable to make definite, unequivocal 
explanations or predictions that could be used to check that 
assumption.  The search for a theory of physics should not be based 
on dogma.  Certain symmetries are observed in the real world, to a 
certain accuracy, in a certain range of spacetime distances.  This 
does not justify a dogma of fundamental symmetry in theoretical 
physics, much less a dogma of fundamental spacetime supersymmetry.  
It has rarely proved fruitful in physics to cling indefinitely to 
assumptions that are incapable of producing definite explanations of 
existing knowledge or definite predictions.

Spacetime supersymmetry does give beneficial formal effects in 
spacetime quantum field theories of particle physics, but these 
benefits could as well be provided by accidental, approximate 
spacetime supersymmetry.  In a weakly coupled theory, perturbative 
spacetime supersymmetry would be enough to protect mass parameters 
that are perturbatively zero, so that very small masses could be 
produced by nonperturbative, supersymmetry violating effects.  
Spacetime supersymmetry provides benefits for formal calculation, 
giving powerful analytic control over quantum mechanical theories 
and especially over spacetime quantum field theories.  But the price 
of control is the supersymmetry itself.  Supersymmetry is not 
observed in nature, and the theoretical control is lost with the 
loss of supersymmetry.  Useful theoretical control must come from 
some other source.

The assumption that physics has a microscopic quantum mechanical 
formulation is of course supported by an enormous body of physical 
evidence.  Microscopic quantum mechanics has had triumphant success, 
culminating in the local quantum field theory that is the standard 
model of elementary particles.  But the evidence for microscopic 
quantum mechanics is entirely at very large distance in spacetime.  
However strong is the evidence for quantum mechanics at large 
distance, that evidence does not require that microscopic quantum 
mechanics in spacetime must be the fundamental language of physics.  
The evidence only requires that microscopic quantum mechanics be 
produced at large distance in any theory of physics.  Quantum 
mechanics in spacetime is the language in which large distance 
physics is to be read out, but it is not necessarily the language in 
which physics is to be written.

Likewise, the fact that certain beautiful mathematical forms were 
used in the period 1905-1974 to make the presently successful theory 
of physics does not imply that any particular standard of 
mathematical beauty is fundamental to nature.  The evidence is for 
certain specific mathematical forms, of group theory, differential 
geometry and operator theory.  The evidence comes from a limited 
range of spacetime distances.  That range of distances grew so large 
by historical standards, and the successes of certain specific 
mathematical forms were so impressive, that there has been an 
understandable psychological impulse in physicists responsible for 
the triumph, and in their successors, to believe in a certain 
standard of mathematical beauty.  But history suggests that it is 
unwise to extrapolate to fundamental principles of nature from the 
mathematical forms used by theoretical physics in any particular 
epoch of its history, no matter how impressive their success.  
Mathematical beauty in physics cannot be separated from usefulness 
in the real world.  The historical exemplars of mathematical beauty 
in physics, the theory of general relativity and the Dirac equation, 
obtained their credibility first by explaining prior knowledge.  
General relativity explained newtonian gravity and special 
relativity.  The Dirac equation explained the non-relativistic, 
quantum mechanical spinning electron.  Both theories then made 
definite predictions that could be checked.  Mathematical beauty in 
physics cannot be appreciated until after it has proved useful.  
Past programs in theoretical physics that have attempted to follow a 
particular standard of mathematical beauty, detached from the 
requirement of correspondence with existing knowledge, have failed.  
The evidence for beautiful mathematical forms in nature requires 
only that a candidate theory of physics explain those specific 
mathematical forms that have actually been found, within the range 
of distances where they have been seen, to an approximation 
consistent with the accuracy of their observation.

\subsection{Formal clues}

The search for an explanation of the standard model and of general 
relativity has become a speculative enterprise, because the guiding 
theoretical principles of local spacetime field theory, quantum 
mechanics and symmetry have proved inadequate.  A strategy must be 
chosen.  It is necessary to decide what formal clues might be 
useful.  No particular choice of strategy is inherently valid.  Only 
the outcome of the search can give validity.

My primary formal clue to a possible theory of large distance 
physics has been the expression of spacetime geometry in the 
renormalization of the general nonlinear model.  The appearance of 
the field equation $R_{\mu\nu}=0$ of general relativity as the fixed 
point equation $\beta = 0$ of the general nonlinear model suggested 
that spacetime field theory might somehow be derived from the 
general nonlinear model.  The renormalization of the general 
nonlinear model isolates the large distance wave modes of its target 
manifold, decoupling the irrelevant small distance wave modes.  The 
renormalization uses the extreme shortness of the two dimensional 
distance $\Lambda^{-1}$ where the model is constructed, compared to 
the two dimensional distance $\mu^{-1}$ where the properties of the 
model are seen.  The large distance physics of spacetime is to be 
found in the short distance structure of the renormalized general 
nonlinear model.

String theory was the second clue.  String theory gave a specific 
technical context, the string worldsurface, in which to place the 
general nonlinear model.  The fixed point equation $\beta = 0$ for 
the general nonlinear model of the string worldsurface is the 
condition of two dimensional scale invariance, which is needed for 
string theory to be consistent.  The possible background spacetimes 
for string theory are determined by imposing the equation $\beta = 
0$ on the general nonlinear model of the worldsurface.  Thus the 
Einstein equation $R_{\mu\nu}=0$ arises as a consistency condition 
in string theory.

But a consistency condition is not an equation of motion.  A 
consistency condition is not the mechanical dynamics of a field 
theory in spacetime.  String theory does not have a dynamical 
mechanism that constructs an effective quantum field theory at large 
distance in spacetime, whose equation of motion is $\beta = 0$.

Nor does string theory have a dynamical mechanism that selects the 
background spacetimes to be those in which the worldsurface is scale 
invariant.  Without such a mechanism, there cannot be a reliable 
characterization of the possible background spacetimes for string 
theory.

The failure of string theory at large distance was the third clue.  
The failure of string theory at large distance provides a formal 
task for a theory of large distance physics to accomplish, the task 
of determining dynamically the possible background spacetimes for 
string theory.  String theory, while useless at large distance, is 
formally successful in the ultraviolet as a technical perturbative 
algorithm for calculating ultraviolet scattering amplitudes in a 
given background spacetime.  But scattering amplitudes are not 
sufficient for physics.  A theory of physics must produce an 
effective mechanical theory at large distance, if it is to explain 
existing knowledge.  The evidence for the reliability of theoretical 
physics includes all the evidence for newtonian mechanics, newtonian 
gravity, classical electromagnetism, special relativity, general 
relativity, non-relativistic quantum mechanics and the standard 
model.  Any candidate theory of physics must be capable of producing 
each of those mechanical theories as an approximation in the 
appropriate regime.  A theory that gives only scattering amplitudes 
is not capable of this.  Scattering amplitudes can be derived from a 
mechanical theory, but mechanics cannot be derived from a theory of 
scattering amplitudes.  Scattering amplitudes intrinsically 
represent small distance physics as observed by a relatively large 
experimentalist.  String theory might well serve adequately as a 
technical perturbative algorithm for calculating ultraviolet 
scattering amplitudes.  It might serve as a formal representation of 
unobservable small distance physics.  But a reliable and effective 
mechanism outside string theory is needed to determine the large 
distance physics of spacetime.

\subsection{The lambda model}

My strategy has been to analyze the failure of string theory at 
large distance in an arbitrarily fixed background spacetime.  The 
technical symptom of failure is a short distance pathology in the 
string worldsurface, a logarithmic divergence at short two 
dimensional distance, bi-local in form, produced by degenerating 
handles attached locally to the worldsurface.  The divergence is due 
to the existence of marginal coupling constants in the general 
nonlinear model of the worldsurface.  Marginal coupling constants 
express the continuous degeneracy of the manifold of possible 
spacetimes.  The divergence is an infrared problem in spacetime, 
because the marginal coupling constants are the large distance wave 
modes of spacetime.

A theoretical mechanism is then devised to cancel the bi-local 
divergence.  The mechanism is a two dimensional nonlinear model, the 
{\it lambda model}.  The target space of the lambda model is the 
manifold of spacetimes, which is the manifold of renormalized 
general nonlinear models of the string worldsurface.  In the lambda 
model, spacetime as a whole fluctuates in two dimensions.

The fields $\lambda^{i}(z,\bar z)$ of the lambda model are local 
sources in the general nonlinear model.  They are coupled to the 
marginal and slightly irrelevant two dimensional quantum fields 
$\phi_{i}(z,\bar z)$ of the general nonlinear model.  The lambda 
fields $\lambda^{i}(z,\bar z)$ fluctuate with a propagator designed 
so that, acting as a bi-local source, it cancels the bi-local 
effects of a handle attached locally on the worldsurface.  The 
couplings of the lambda model are completely determined by the 
cancellation requirement.  In particular, the coupling strength of 
the lambda model is equal to the spacetime coupling constant $\gst$ 
of the perturbative string theory.

There are two widely separated two dimensional distances.  The 
coupling constants $\lambda^{i}$ of the renormalized general 
nonlinear model are normalized at $\mu^{-1}$, the long two 
dimensional distance.  The lambda fields $\lambda^{i}(z,\bar z)$ 
fluctuate at short two dimensional distances, up to a sliding 
characteristic two dimensional distance $\Lambda^{-1}$ which stays 
much shorter than $\mu^{-1}$.  The renormalization of the general 
nonlinear model suppresses the effects of the short distance 
fluctuations of the coupling constant $\lambda^{i}$ by a factor 
$(\Lambda/\mu)^{-\ad(i)}$, where $-\ad(i)$ is the anomalous 
dimension of $\lambda^{i}$.  The coupling constants which have 
$\ad(i) \ln(\Lambda/\mu) \gg 1$ are irrelevant at the short two 
dimensional distance $\Lambda^{-1}$.  There are only a finite number 
of non-irrelevant coupling constants $\lambda^{i}$ in the general 
nonlinear model, so the target manifold of the lambda model is 
finite dimensional.

The number $L$ defined by $L^{2}=\ln(\Lambda/\mu)$ is a spacetime 
distance, because the anomalous dimensions $-\ad(i)$ are the 
eigenvalues of differential operators in spacetime that are 
quadratic in spacetime derivatives.  The effects of the spacetime 
wave mode $\lambda^{i}$ are suppressed by factors 
$\me^{-L^{2}\ad(i)}$.  The irrelevant coupling constants in the 
general nonlinear model are the $\lambda^{i}$ with $\ad(i) L^{2} \gg 
1$.  These are the spacetime wave modes at spacetime distances 
$1/p(i)$ smaller than $L$.

The small distance wave modes, being irrelevant coupling constants, 
are decoupled in the renormalization of the general nonlinear model.  
Their fluctuations can be omitted from the lambda model at two 
dimensional distance $\Lambda^{-1}$.  Only the non-irrelevant 
coupling constants fluctuate, the coupling constants $\lambda^{i}$ 
having $\ad(i) L^{2} < 1$.  These are the large distance spacetime 
wave modes, the wave modes at spacetime distances larger than $L$.  
Thus, from the renormalization of the general nonlinear model, the 
lambda model inherits a natural, built-in, sliding ultraviolet 
spacetime cutoff distance $L$.  The spacetime wave modes at 
distances smaller than $L$ are decoupled from the large distance 
wave modes, so this is an ultraviolet cutoff in the strongest sense.

The lambda model is a two dimensional quantum field theory.  As 
such, its construction starts from the short distance limit at 
$\Lambda^{-1}=0$, building outward to nonzero values of the sliding 
characteristic two dimensional distance $\Lambda^{-1}$.  So the 
lambda model builds spacetime physics from the limit at $L=\infty$ 
downward to finite values of the sliding characteristic large 
spacetime distance $L$.

The fluctuations in a nonlinear model at distances shorter than the 
characteristic two dimensional distance $\Lambda^{-1}$ distribute 
themselves to form a measure on the target manifold of the model, 
called the \apm\ of the nonlinear model, following the terminology 
of lattice statistical mechanics, using `{\em a priori}' with its 
literal meaning `from what is before' or `from the earlier 
part'~\citeF. The \apm\ summarizes the short distance fluctuations in 
the nonlinear model.  As the characteristic two dimensional distance 
$\Lambda^{-1}$ increases from zero, the fluctuations in the 
nonlinear model generate the \apm\ by a diffusion process on the 
target manifold.

The target manifold of the lambda model is the manifold of 
spacetimes, so the \apm\ of the lambda model is a measure on the 
manifold of spacetimes.  The manifold of spacetimes is the manifold 
of general nonlinear models.  As $\Lambda^{-1}$ increases, the 
fluctuations cause the \apm\ of the lambda model to diffuse in the 
manifold of general nonlinear models, while simultaneously the 
general nonlinear model is flowing under the renormalization group.  
The \apm\ of the lambda model undergoes a driven diffusion process.  
The generator of the driving flow is the vector field 
$-\beta^{i}(\lambda)$ on the manifold of general nonlinear models.  
The renormalization group flow pushes the \apm\ toward the fixed 
point submanifold where $\beta(\lambda)=0$.  The lambda model 
dynamically imposes the two dimensional scale invariance condition 
$\beta(\lambda)=0$ on the general nonlinear model.

The lambda model is background independent, because, even if a 
particular background spacetime is initially selected by hand, the 
fluctuations in the lambda model diffuse the distribution of 
spacetimes away from the arbitrary initial spacetime.  Whatever the 
arbitrary initial choice of background spacetime, the \apm\ diffuses 
to the unique stable measure of the driven diffusion process.  The 
lambda model eliminates the need in string theory to choose a 
background spacetime by hand.

The lambda model is a nonperturbative two dimensional quantum field 
theory.  Nonperturbative two dimensional effects in the lambda model 
due to harmonic surfaces in the manifold of spacetimes appear 
capable of lifting the continuous degeneracy of the manifold of 
spacetimes, possibly concentrating the \apm\ at some macroscopic 
spacetimes.  If so, then the \apm\ of the lambda model, near such a 
macroscopic spacetime, is a measure on the spacetime wave modes 
$\lambda^{i}$ at spacetime distances larger than $L$.  It is a 
spacetime quantum field theory with ultraviolet cutoff distance $L$.  
The lambda model produces a specific quantum field theory, with 
equation of motion $\beta = 0$, describing the spacetime physics at 
large spacetime distances $L$ in each possible macroscopic 
spacetime.

Although the lambda model works from $L=\infty$ downwards in $L$, it 
constructs spacetime quantum field theory so that it is local in 
spacetime, in the sense that the \apm\ at a spacetime distance 
$L_{1}$ can be obtained from the \apm\ at a smaller distance 
$L_{2}<L_{1}$ by integrating out the spacetime wave modes at the 
distances between $L_{2}$ and $L_{1}$.  The lambda model 
accomplishes this in reverse fashion.  The lambda model constructs 
the \apm\ with {\em increasing} two dimensional distance 
$\Lambda^{-1}$, so with {\em decreasing} spacetime distance $L$.  
The lambda model makes the \apm\ at the smaller distance $L_{2}$ 
from the \apm\ at the larger distance $L_{1}$ by diffusion of the 
wave modes at the intermediate spacetime distances, between $L_{2}$ 
and $L_{1}$.  Locality is ensured because integrating out the 
intermediate wave modes merely undoes the diffusion.  The spacetime 
quantum field theory is constructed so as to be local as a measure 
on the spacetime wave modes, but there is no guarantee that the 
effective lagrangian of the spacetime field theory at the smaller 
distance can be used to determine the effective spacetime quantum 
field theory at larger distances, except perturbatively.  
Nonperturbative two dimensional effects in the lambda model might 
intervene in the construction of the effective spacetime action.

The proposed theory, if successful, will call into question the 
atomistic assumption that the effective laws of physics at large 
distance can be deduced from the laws of physics at small distance.  
It will call into question the atomistic assumption there is a 
fundamental microscopic formulation of physics.  If the proposed 
theory works, the observed quantum mechanical hamiltonian will be 
explained, but there will not be a fundamental quantum mechanical 
hamiltonian.

The lambda model also produces an effective worldsurface at two 
dimensional distances longer than $\Lambda^{-1}$.  The effective 
worldsurface can be used to calculate effective string scattering 
amplitudes, cut off in the infrared at spacetime distance $L$.  For 
each large spacetime distance $L$, the lambda model gives two 
complementary descriptions of spacetime physics.  The spacetime 
physics at distances larger than $L$ is described by an effective 
spacetime quantum field theory.  The spacetime physics at distances 
smaller than $L$ is described by effective string scattering 
amplitudes.  The two descriptions of spacetime physics are 
consistent.

The \apm\ on the manifold of spacetimes is the effective spacetime 
background in which the effective string scattering takes place, at 
every large spacetime distance $L$.  The relation between the 
effective string scattering amplitudes and the effective spacetime 
quantum field theory is not the relation commonly assumed in string 
theory.  The large distance spacetime physics is not derived from 
any microscopic, small distance physics.  In particular, it is not 
derived from string theory.  There is no microscopic quantum 
mechanical system underlying string theory, that has spacetime 
quantum field theory as its effective description at large distance.  
The lambda model provides the spacetime background for the effective 
string theory, constructing it starting from the limit at spacetime 
distance $L=\infty$.  The fluctuations of the lambda model make 
quantum corrections to the effective worldsurface in tandem with 
corrections to the effective metric coupling and \apm\ of the lambda 
model itself, thus ensuring that the effective string scattering 
amplitudes match, at every large spacetime distance $L$, the 
particle scattering amplitudes calculated from the effective quantum 
field theory.

The lambda model, if right, determines all actually observable 
physics.  The sliding characteristic spacetime distance $L$ can be 
taken smaller than any distance actually accessible to experiment, 
while still remaining a large number.  As a practical matter, only 
the large distance physics given by the \apm\ can be checked.  All 
calculations of large distance physics can be done in the lambda 
model.  No string calculations are necessary.  The fluctuations of 
the lambda model completely replace the effects of handles at short 
distance in the worldsurface, so string calculations, and especially 
string loop calculations, are entirely unnecessary, as far as large 
distance physics is concerned.

String loop calculations are needed only for perturbative 
calculations of the effective string scattering amplitudes at 
unobservably small spacetime distances.  There is no practical use 
for these effective string scattering amplitudes.  The only 
information that the effective string scattering amplitudes give, 
beyond what is given by the \apm\ of the lambda model, is 
information about physics at unobservably small distances.  There is 
no practical way to make any independent test of the small distance 
effective string scattering amplitudes.

The lambda model is a nonperturbative theory, while the small 
distance effective string theory is only perturbative.  String 
theory calculations of scattering amplitudes at spacetime distances 
smaller than $L$ will not be reliably accurate, because 
nonperturbative effects in the lambda model at spacetime distances 
smaller than $L$ will not yet have been taken into account.  The 
only reliable calculations will be the nonperturbative calculations 
of large distance physics that are made in the lambda model.

String theory is used in three ways in the lambda model.  First, the 
string worldsurface gives a specific technical context in which to 
place the general nonlinear model.  The manifold of spacetimes is 
the manifold of general nonlinear models of the worldsurface.  The 
detailed specification of the manifold of spacetimes depends on the 
detailed technical form of the worldsurface in which the general 
nonlinear model is placed.

Second, string theory gives an algorithm for calculating 
perturbative corrections to scattering amplitudes in terms of 
handles in the worldsurface, an algorithm that is formally 
consistent at small spacetime distance.  The lambda model is 
constructed to cancel the perturbative corrections due to handles 
attached locally on the worldsurface, so the details of the 
technical form of the worldsurface determine the detailed definition 
of the target manifold and the metric coupling of the lambda model.

Once the lambda model is defined, string theory becomes an auxiliary 
technical apparatus.  String theory is used only as a formal 
representation of unobservable small distance physics in the 
spacetime constructed by the lambda model.  The correspondence 
between the effective general nonlinear model of the worldsurface 
and the effective lambda model is used for technical purposes.  The 
correspondence constrains the renormalization of the lambda model.  
The two dimensional scaling properties of the effective general 
nonlinear model implies the two dimensional scale invariance of the 
lambda model.

It appears that, for entirely technical reasons, the heterotic 
string worldsurface~\cite{heterotic} is the only form of the 
worldsurface that is suitable for the lambda model.  The manifold of 
spacetimes is a graded manifold.  Its bosonic and fermionic 
coordinates are the bosonic and fermionic coupling constants 
$\lambda^{i}$, which are the wave modes of the bosonic and fermionic 
spacetime fields.  A two dimensional nonlinear model such as the 
lambda model is well-defined only if its metric coupling is positive 
definite on the bosonic part of its target manifold.  Only for the 
heterotic string worldsurface is the metric on the manifold of 
spacetimes positive definite in the bosonic directions.  For this 
technical reason, it seems that only the heterotic string 
worldsurface is suitable for the lambda model.

Fortunately, the heterotic worldsurface is the only one that is 
suitable for a formal representation of weakly coupled small 
distance spacetime physics.  The heterotic string theory gives small 
distance scattering amplitudes of massless chiral fermions, vector 
bosons, scalar bosons, and gravitons.  The perturbative spacetime 
superymmetry of the heterotic string theory ensures that the general 
nonlinear model of the heterotic worldsurface contains only marginal 
and irrelevant coupling constants.  There are no relevant coupling 
constants, which would be the wave modes of spacetime tachyon 
fields.  The perturbative spacetime supersymmetry of the heterotic 
theory ensures that the perturbative string theory is consistent at 
small distance.  The perturbative spacetime supersymmetry is 
inherited by the lambda model, where it protects the zero mass 
spacetime fields against perturbative mass corrections, allowing the 
possibility of small spacetime masses generated by nonperturbative 
effects in the lambda model violating the perturbative spacetime 
supersymmetry.

My most optimistic hope, amounting only to wishful thinking at 
present, is that nonperturbative weak coupling effects in the lambda 
model will produce a calculable spectrum of large distances in the 
spacetime physics generated by the lambda model.  Each harmonic 
surface $\lambda_{H}(z,\bar z)$ in the manifold of spacetimes is an 
instanton in the lambda model, a {\em lambda instanton}.  It is easy 
to point to the existence of harmonic surfaces in the manifold of 
spacetimes, but their effects remain speculative until calculated.

Harmonic surfaces in the manifold of spacetimes appear capable of 
making small contributions to the effective action of the spacetime 
quantum field theory, giving small masses to the elementary 
particles.  More broadly, they appear capable of eliminating the 
continuous degeneracies, including spacetime supersymmetries and 
ordinary gauge symmetries.  Spacetime symmetries would then be seen 
as only accidental and approximate attributes of individual 
spacetimes.

A lambda instanton $\lambda_{H}(z,\bar z)$ that is localized in a 
macroscopic spacetime can be expected to generate spacetime masses 
of the form $m^{2}=\me^{-S(\lambda_{H})}$, where $S(\lambda_{H})$ is 
the classical action of the harmonic surface $\lambda_{H}$.  The 
coupling strength of the lambda model is the spacetime coupling 
constant $\gst$ so $S(\lambda_{H})$ is proportional to $\gst^{-2}$.  
The actual numerical values of the elementary particle masses might 
be produced in this way, if $\gst^{2}$ is on the order of 1/100 and 
if the unit of length is logarithmically close to the Planck length.  
For example, the mass-squared of the W vector boson is approximately 
$m_{W}^{2} = \me^{-78}$ in Planck units, and that of the electron is 
$m_{e}^{2} = \me^{-101}$.  If the lambda model does in fact produce 
such calculable small particle masses, it will become of interest to 
check whether mass generation by the lambda model can be 
distinguished experimentally from the quantum field theoretic Higgs 
mechanism.

Even more fanciful hopes are evoked by writing the inverse square of 
the Hubble length in Planck units, approximately $\me^{-281}$.  It 
is hard to imagine where such a number might come from, if not a 
semiclassical, nonperturbative, weak coupling effect.  It would be 
wonderful, though rather much to expect, if semiclassical 
nonperturbative effects in the lambda model could explain 
systematically the essential features of the rich spectrum of large 
characteristic spacetime distances observed in the real 
world.

If the lambda model does succeed in reducing the continuous 
degeneracy of the manifold of spacetimes at least to a discrete 
degeneracy, then the remaining uncertainty would be acceptable, as 
long as a finite number of experiments could serve to decide which, 
if any, of the remaining discrete collection of possible spacetimes 
matches the real world.  Definite explanations and predictions could 
then be made, and tested definitively against existing knowledge and 
future experiments.

Even if all goes well, even if effects can be found in the 
lambda model that concentrate the \apm\ on a discrete set of 
macroscopic spacetimes, produce small particle masses, and that 
fix the spacetime coupling constant $\gst^{2}$ in the 
macroscopic spacetimes at a small value, it will of course still 
not be guaranteed that one of the resulting spacetime quantum 
field theories matches the real world.  It will be necessary to 
check that one of the macroscopic quantum field theories 
produced by the lambda model matches in detail the standard 
model and the observed cosmology.  Formal capabilities do not 
guarantee that a theory will be successful as physics.

The proposed theory of large distance physics, if it succeeds, will 
still be only approximate.  It will be a weak coupling, 
semiclassical approximation, unless effective methods can be found 
to do strongly coupled two dimensional quantum field theory 
calculations in the lambda model.  Moreover, the theory is 
intrinsically only approximate for $L<\infty$, because the 
renormalized general nonlinear model is taken away from the strict 
two dimensional continuum limit at $\Lambda^{-1}=0$.  The 
renormalization of the general nonlinear model is justified by the 
divergence of $L^{2} = \ln(\Lambda/\mu)$.  Renormalization is exact 
only in the limit of an infinitely wide gulf between the short two 
dimensional distance $\Lambda^{-1}$ and the long two dimensional 
distance $\mu^{-1}$.  Renormalization of the general nonlinear model 
is only approximate when $\ln(\Lambda/\mu)<\infty$.

Feasible experiments in the real world are at values of $L^{2}$ 
larger than some extremely large number, at least $10^{28}$, if the 
unit of distance is within a few orders of magnitude of the Planck 
length.  I suspect that $\ln(\Lambda/\mu) \approx 10^{28}$ is close 
enough to the two dimensional continuum limit that the theory will 
be quite precise, intrinsically, at all spacetime distances 
accessible to experiment.

If successful, this theory of large distance physics will not be a 
fundamental theory of physics, but it will describe with sufficient 
accuracy everything that can be checked, unless and until 
experiments are able to probe physics at spacetime distances 
approaching the Planck length.  If spacetime quantum field theory is 
explained as merely an epiphenomenon of the lambda model, then 
quantum field theory, and quantum mechanics, will be seen to be 
effective descriptions of spacetime physics only at large distance, 
and will be seen to be inherently approximate except at infinite 
distance in spacetime.  The possibility will then arise that 
theoretical physics in general might be inherently approximate at 
finite distance in spacetime.  Alternatively, if the lambda model 
does succeed in giving a very accurate approximate description of 
large distance physics, it will become the touchstone for candidate 
exact theories of physics.  The challenge will become to find more 
exact theories that have the lambda model as approximation, and to 
find experiments capable of distinguishing between the lambda model 
and any such candidate exact theories, in order to establish the 
greater reliability of a more exact candidate theory.

\subsection{Many questions remain}

Many questions about the theory remain.  The most immediate 
questions concern calculations of the local properties of the 
spacetime quantum field theories produced by the lambda model.  Do 
the lambda instantons that are local in a macroscopic spacetime 
succeed in removing spacetime supersymmetry, removing local gauge 
symmetries and generating nonperturbatively small particle masses?  
Are there effects in the lambda model that fix the spacetime 
coupling constant at a small numerical value?  If the lambda model 
succeeds in doing these things, then the question becomes, is the 
verified part of the standard model to be found among the spacetime 
quantum field theories constructed by the lambda model?

A reasonable strategy is to assume a macroscopic spacetime, 
temporarily, in order to do the urgent local calculations in 
spacetime.  Eventually, the existence and dimension of macroscopic 
spacetime must be settled by calculation in the lambda model.  The 
lambda model has to explain the observation of macroscopic 
spacetime.  The technical question is, do harmonic surfaces in the 
manifold of spacetimes succeed in concentrating the \apm\ at 
macroscopic spacetimes, of which some are four dimensional?

A theory of large distance physics must give a definite explanation 
of cosmology.  It must explain the observed cosmological data, 
especially the essential features of the rich spectrum of 
characteristic distance scales that are found in the observed 
universe.  But cosmology is still a diffuse, data rich subject 
compared to high energy particle physics.  The standard model of 
particle physics is a sharp theoretical target.  The standard model 
provides a definite, succinct {\em theoretical} structure to be 
explained, and a small set of precisely measured parameters to be 
calculated.  A candidate theory of physics needs credibility before 
it can usefully take on the relatively nebulous theoretical problems 
of cosmology.  The only reliable way I can see for a theory of 
physics to establish credibility is to explain the verified part of 
the standard model in detail.  If that can be done in the lambda 
model, then the project of extracting cosmology from the lambda 
model will become promising.

An explanation of cosmology will require a construction of 
cosmological time.  The general nonlinear model, to be well-defined 
as a two dimensional quantum field theory, needed its target 
manifold, spacetime, to be riemannian and compact.  The general 
nonlinear model has to be well-defined in order that the manifold of 
general nonlinear models be usable as the target manifold of the 
lambda model.  The problem is to construct real time.

A real time quantum field theory can be obtained from the \apm\ of 
the lambda model by making an ad hoc Wick rotation locally in 
spacetime, at distances where the spacetime curvature is 
insignificant, if a macroscopic spacetime is singled out by the 
lambda model.  But cosmological time presumably needs a global 
construction that is everywhere consistent with local Wick rotation.  
I have no clear idea how this might be done.  Perhaps there is a 
global analytic continuation of the manifold of spacetimes, which 
looks like Wick rotation locally in any macroscopic spacetime.  
Perhaps cosmological time can be related to the sliding large 
spacetime distance $L$, through the relation between $L^{2}$ and the 
logarithm of the characteristic two dimensional distance.

In the lambda model, where spacetime is taken to be riemannian for 
technical reasons, an {\em explanation} of real time is needed.  
There should be principle that explains {\em why} Wick rotation 
should be done locally in a macroscopic riemannian spacetime.

The basic question is the existence of the short distance limit in 
two dimensions, the limit at $\Lambda^{-1}=0$.  The lambda model is 
built as a two dimensional quantum field theory by integrating out 
the fluctuations at short two dimensional distances, starting from 
$\Lambda^{-1}=0$.  So the lambda model is well-defined as a two 
dimensional quantum field theory only if the limit exists.  The 
short distance limit $\Lambda^{-1}=0$ in two dimensions is the limit 
$L=\infty$ in spacetime, the limit in which only the spacetime zero 
modes fluctuate.  The lambda model generates and controls the large 
distance spacetime physics, acting downward in spacetime distance 
from the limit at $L=\infty$.  The theory is well-founded only if 
the $L=\infty$ limit exists.

My argument that the lambda model has a scale invariant limit at 
asymptotically short two dimensional distance is only formal.  In 
the limit $\Lambda^{-1}=0$, the fluctuating two dimensional fields 
of the lambda model are dimensionless.  Their fluctuations can be 
expected to explore the entire manifold of spacetimes.  Control over 
the global structure of the manifold of spacetimes will be needed 
before a rigorous argument can be made for the scale invariant short 
distance limit of the lambda model.

The question is probably not a practical one.  For practical 
purposes, it is enough to construct the lambda model for 
sufficiently large finite values of $L$.  The limit at $L=\infty$ is 
only needed to make the theory secure.

%
%
%
%
%
\sectiono{The structure of the theory}

\subsection{The lambda model}
The lambda model is a two dimensional nonlinear model.  Its target 
manifold is the manifold of renormalized general nonlinear models of 
the string worldsurface, which is the manifold of compact riemannian 
background spacetimes.  The lambda model acts on the general 
nonlinear model at short two dimensional distances, determining the 
large distance physics of spacetime.

Consider a renormalized general nonlinear model of the worldsurface.  
The perturbations of this reference general nonlinear model are 
parametrized by the nearly dimensionless coupling constants 
$\lambda^{i}$.  The $\lambda^{i}$ are coupled to the approximately 
marginal spin 0 quantum fields $\phi_{i}(z,\bar z)$, the fields that 
have scaling dimension near $2$.  The nearby general nonlinear 
models are made by inserting
\eq
\me^{-\int \dif^{2}z \, \mu^{2} \frac1{2\pi} \,
\lambda^{i} \phi_i(z,\bar z)}
\en
into the reference general nonlinear model.  The two dimensional 
distance $\mu^{-1}$ is the distance at which the general nonlinear 
model is normalized.  The coupling constants $\lambda^{i}$ are local 
coordinates for the manifold of spacetimes.  The $\lambda^{i}$ are 
the large distance wave modes of the spacetime metric and other 
spacetime fields.

The lambda model is a nonlinear model whose field is a fluctuating 
map $\lambda(z,\bar z)$ from the worldsurface to the manifold of 
spacetimes.  In coordinates, the lambda field $\lambda(z,\bar z)$ is 
expressed by component lambda fields $\lambda^{i}(z,\bar z)$ which 
act as local sources coupled to the quantum fields of the reference 
general nonlinear model.  The worldsurface is defined locally as a 
function of the lambda field by inserting
\eq
\me^{-\int \dif^{2}z \, \mu^{2} \frac1{2\pi} \,
\lambda^{i}(z,\bar z) \phi_i(z,\bar z)}
\en
into the reference general nonlinear model.

The map $\lambda(z,\bar z)$ fluctuates at short two dimensional 
distances, from a two dimensional cutoff distance $\Lambdazero^{-1}$ 
up to a sliding characteristic two dimensional distance 
$\Lambda^{-1}$ which is still very much shorter than $\mu^{-1}$.  
The fluctuations are described by a functional integral
\eq
\label{eq:lambda-model-a}
\int \Dlambda
\; \me^{- S(\lambda)} \;
\me^{-\int \dif^{2}z \, \mu^{2} \frac1{2\pi} \,
\lambda^{i}(z,\bar z) \phi_i(z,\bar z)}
\en
inserted in the reference general nonlinear model.  The action 
$S(\lambda)$ depends on the two dimensional distance.  The action of 
fluctuations at two dimensional distance $\Lambda^{-1}$ is
\eq
S(\Lambda,\lambda) = \int \dif^{2}z \frac1{2\pi}
\,
T^{-1} g_{ij}(\Lambda, \lambda)
\,
\partial \lambda^{i}
\,
\bar \partial \lambda^{j}
\:.
\en
The metric coupling $T^{-1} g_{ij}$ varies with the two dimensional 
distance $\Lambda^{-1}$.

At large spacetime distance, there are only a finite number of 
spacetime wave modes, because of spacetime being assumed compact and 
riemannian.  So there are only a finite number of nearly marginal 
coupling constants $\lambda^{i}$ in the general nonlinear model.  So 
there are only a finite number of fields $\lambda^{i}(z,\bar z)$ in 
the lambda model.  The target manifold of the lambda model is finite 
dimensional.

\subsection{Cancelling handles at short two dimensional distance}

The lambda model is designed to cancel the effects of handles at 
short distance on the worldsurface.  A handle attached to the 
worldsurface at two dimensional distance $\Lambda^{-1}$ has the 
effect of a bi-local insertion
\eq
\label{eq:log-insertion}
\frac12 
\int \dif^{2}z_1\,\mu^{2} \frac1{2\pi}
\;
\int \dif^{2}z_2\,\mu^{2} \frac1{2\pi}
\;
\phi_{i}(z_1,\bar z_1)
\;
T\,g^{ij}(\Lambda,\lambda)
\;
\ln ({\Lambda^2} \abs{z_1-z_2}^2)
\;
\phi_{j}(z_2,\bar z_2)
\en
where $z_{1}$ and $z_{2}$ are the points where the two ends of the 
handle are attached to the worldsurface.  The sum over indices $i,j$ 
is the sum over states flowing through the handle.  The fields 
$\phi_{i}(z_1,\bar z_1)$ and $\phi_{j}(z_2,\bar z_2)$ are produced 
in the worldsurface by the states flowing through the ends of the 
handle.  The handle connects the fields at its two ends by the 
gluing matrix $T g^{ij}(\Lambda,\lambda)$.

The bi-local insertion, equation~\ref{eq:log-insertion}, will 
calculated explicitly in section~\ref{sect:divergence} below.  For 
now, its form follows from general principles of two dimensional 
quantum field theory.  The logarithmic dependence on the 
separation $\abs{z_{1}-z_{2}}$ between the two ends of the handle, 
for separations near $\Lambda^{-1}$, follows from the fact that the 
fields $\phi_{i}(z,\bar z)$ are approximately marginal.

The metric coupling $T^{-1} g_{ij}(\Lambda,\lambda)$ of the lambda 
model is formulated as the inverse of the handle gluing matrix 
$T\,g^{ij}(\Lambda,\lambda)$.  This formulation is designed so that 
the propagator of the lambda fields at two dimensional distance 
$\Lambda^{-1}$ is
\eq
\expval{\,\lambda^i(z_1,\bar z_1) \; \lambda^j(z_2,\bar z_2)\,}
= - T g^{ij}(\Lambda,\lambda)
\,
\ln (\Lambda^{2} \abs{z_{1}-z_{2}}^{2})
\:.
\en
The lambda model, equation~\ref{eq:lambda-model-a}, then produces the bi-local 
insertion
\eq
\frac12 
\int \dif^{2}z_1\,\mu^{2} \frac1{2\pi}
\;
\int \dif^{2}z_2\,\mu^{2} \frac1{2\pi}
\;
\phi_{i}(z_1,\bar z_1)
\;
\expval{\,\lambda^i(z_1,\bar z_1) \; \lambda^j(z_2,\bar z_2)\,}
\;
\phi_{j}(z_2,\bar z_2)
\en
which cancels the effects of the handle.

The properties that define the lambda model -- its form as a two 
dimensional nonlinear model, its field as a map from the 
worldsurface to the manifold of spacetimes, its specific metric 
coupling -- are all naturally determined by the short distance 
properties of the worldsurface, which determine the effects of 
handles at short distance in the worldsurface, which the lambda 
model is designed to cancel.

The number $T^{-1}$ is the partition function of the worldsurface 
without handles, the 2-sphere.  In a macroscopic spacetime of volume 
$V$,
\eq
T^{-1} = \gst^{-2} \, V
\en
where $\gst$ is the spacetime coupling constant.  The metric 
coupling of the lambda model has a form that is local in spacetime,
\eq
T^{-1} g_{ij} = \gst^{-2} \,V g_{ij}
\en
where $V g_{ij}$ is properly normalized so that it is expressible as 
the spacetime integral of the product of the corresponding spacetime 
wave modes.  The coupling strength of the lambda model is therefore 
the spacetime coupling constant $\gst$.

\subsection{Generalized scale invariance}

The general nonlinear model is renormalizable, so it depends on the 
characteristic short distance $\Lambda^{-1}$ only through the 
running coupling constants $\lambdar^{i}(\Lambda/\mu,\lambda)$ 
which satisfy the renormalization group equation
\eq
\Lambda \partialby{\Lambda}_{\left / \mu,\lambda\right.} \lambdar^{i}
= \beta^{i}(\lambdar)
\:.
\en
The running coupling constants couple to the two dimensional quantum 
fields $\phi^{\Lambda}_i(z,\bar z)$ normalized at the short 
two dimensional distance $\Lambda^{-1}$.  The general nonlinear 
model can be described at short distance by the insertion of the 
running couplings,
\eq
\me^{-\int \dif^{2}z \, \mu^{2} \frac1{2\pi} \,
\lambda^{i} \phi_i(z,\bar z)}
=
\me^{-\int \dif^{2}z \, \Lambda^{2} \frac1{2\pi} \,
\lambdar^{i} \phi^{\Lambda}_i(z,\bar z)}
\en
obeying the renormalization group equation
\eq
\left (
\Lambda \partialby{\Lambda}_{\left /\lambdar\right.}
+ \beta^{i}(\lambdar) \partialby{\lambdar^{i}}
\right )
\;
\me^{-\int \dif^{2}z \, \Lambda^{2} \frac1{2\pi} \,
\lambdar^{i} \phi^{\Lambda}_i(z,\bar z)}
=0
\:.
\en
The handle gluing matrix and its inverse matrix, the metric coupling 
$T^{-1}g_{ij}$, are natural structures of the worldsurface at two 
dimensional distance $\Lambda^{-1}$, so they depend on 
$\Lambda^{-1}$ only through the running coupling constants 
$\lambda_{r}$.  The action of the lambda model therefore depends 
only on the running sources $\lambdar^{i}(z,\bar z)$,
\eq
S(\Lambda,\lambda) = S(\lambdar)
\en
\eq
S(\lambdar) = \int \dif^{2}z \, \frac1{2\pi}
\,
T^{-1} g_{ij}(\lambdar)
\,
\partial \lambdar^{i}
\,
\bar \partial \lambdar^{j}
\en
where the metric coupling $T^{-1} g_{ij}(\lambdar)$ is independent 
of the two dimensional distance, as a function of the running 
coupling constants.  The lambda model takes the same form at every 
short two dimensional distance $\Lambda^{-1}$
\eq
\int \Dlambdar
\; \me^{- S(\lambdar)} \;
\me^{-\int \dif^{2}z \, \Lambda^{2} \frac1{2\pi} \,
\lambdar^{i}(z,\bar z) \phi^{\Lambda}_i(z,\bar z)}
\:,
\en
when it is expressed in terms of running fields $\lambdar^{i}(z,\bar 
z)$.  The running fields transform, with an increase of the two 
dimensional distance $\Lambda^{-1}\rightarrow 
(1+\epsilon)\Lambda^{-1}$, by the renormalization group flow 
$\lambdar^{i} \rightarrow \lambdar^{i} - \epsilon 
\beta^{i}(\lambdar)$.

The lambda model is therefore a scale invariant nonlinear model in 
the generalized sense~\citeF. The metric coupling 
$T^{-1}g_{ij}(\Lambda, \lambda)$, written in terms of the original 
renormalized coupling constants, is not literally invariant under a 
change of the characteristic two dimensional distance 
$\Lambda^{-1}$.  Rather, the metric coupling is invariant under the 
combination of changing scale, $\Lambda^{-1}\rightarrow 
(1+\epsilon)\Lambda^{-1}$, and simultaneously flowing in the target 
manifold, $\lambda^{i} \rightarrow \lambda^{i} + \epsilon 
\beta^{i}(\lambda)$.  The transformation of the target manifold is 
only a change of variables in the functional integral that defines 
the nonlinear model, so all observable quantities are scale 
invariant.  The lambda model is novel in that its scale invariance 
is of the generalized kind even at the classical level.

\subsection{The lambda model acts at short two dimensional distance}

The lambda model is constructed, starting at a vanishingly short two 
dimensional cutoff distance $\Lambdazero^{-1}$, by integrating over 
more and more short distance fluctuations, up to the characteristic 
two dimensional distance $\Lambda^{-1}$.  The characteristic two 
dimensional distance $\Lambda^{-1}$ slides {\em outwards}, as in any 
two dimensional quantum field theory.  The lambda model acts 
entirely at short two dimensional distance.  The fluctuations at two 
dimensional distances up to $\Lambda^{-1}$ act on the short distance 
structure of the general nonlinear model to produce an effective 
general nonlinear model at two dimensional distances longer than 
$\Lambda^{-1}$.  The effective general nonlinear model of the 
worldsurface defines an effective string theory.

\subsection{The general nonlinear model at short distance}

Formally, the general nonlinear model is parametrized by infinitely 
many coupling constants $\lambda^{i}$, corresponding to infinitely 
many spacetime wave modes.  But the general nonlinear model is 
renormalizable, so almost all of the coupling constants are 
irrelevant.  The irrelevant coupling constants $\lambda^{i}$ are 
coupled to irrelevant quantum fields $\phi_{i}(z,\bar z)$.  An 
irrelevant quantum field has negligible renormalized effect when 
inserted at short two dimensional distance.  In particular, the 
irrelevant quantum fields that are inserted at short distance at the 
ends of handles have negligible renormalized effect.  There is no 
need to cancel this negligible effect, so there is no need for the 
irrelevant coupling constants $\lambda^{i}$ to fluctuate in the 
lambda model.  It would not matter if the irrelevant $\lambda^{i}$ 
did fluctuate.  Their fluctuations would have negligible effect.

Consider a reference general nonlinear model that is scale 
invariant, satisfying $\beta=0$.  In coordinates around this 
reference point, the beta function takes the form
\eq
\beta^{i}(\lambda) = \ad(i) \lambda^{i} + O(\lambda^{2})
\en
so
\eq
\lambda^{i} = (\mu \Lambda^{-1})^{\ad(i)} \, \lambdar^{i}
\en
up to higher order corrections.  Each coupling constant 
$\lambda^{i}$ has definite scaling dimension $-\ad(i)$.  The 
corresponding two dimensional quantum field $\phi_{i}(z,\bar z)$ is 
a scaling field with scaling dimension $2+\ad(i)$.  The number 
$\ad(i)$ is the anomalous dimension of the field $\phi_{i}$.

All coupling constants with $\ad(i)>0$ are irrelevant in the extreme 
short distance limit $\Lambda^{-1}=0$.  Their effects are driven to 
zero at the long two dimensional distance $\mu^{-1}$.  They have no 
effect in the renormalized two dimensional quantum field theory. 

If a coupling constant $\lambda^{i}$ had $\ad(i)<0$, it would be a 
relevant coupling constant.  But a relevant coupling constant in the 
general nonlinear model of the string worldsurface would correspond 
to a tachyonic spacetime wave mode.  There are no relevant coupling 
constants in a sensible string worldsurface.  That is, all the 
anomalous dimensions satisfy
\eq
\ad(i) \geq 0
\:.
\en
At $\Lambda^{-1}=0$, the general nonlinear model is parametrized by 
the marginal coupling constants, the $\lambda^{i}$ with $\ad(i)=0$.

Because all the anomalous dimensions $\ad(i)$ are nonnegative, the 
renormalization group flow drives the general nonlinear model to the 
submanifold of fixed points, the submanifold where $\beta = 0$.  The 
manifold of scale invariant general nonlinear models is the 
attracting manifold for the renormalization group flow.  
Renormalization {\em forces} the general nonlinear model to be scale 
invariant, so there is no possible choice of spacetime besides the 
manifold of scale invariant general nonlinear models.

The sharp distinction between the irrelevant and the marginal 
coupling constants does not persist when $\Lambda^{-1}$ is greater 
than zero.  Define the number $L$ by
\eq
L^{2} = \ln (\mu^{-1} \Lambda)
\:.
\en
The suppression of irrelevant coupling constants is 
\eq
\lambda^{i} = \me^{-L^{2}\ad(i)} \, \lambdar^{i}
\en
so $\lambda^{i}$ is irrelevant at two dimensional distance 
$\Lambda^{-1}$ if $L^{2}\ad(i)\gg 1$.  The property of 
irrelevance changes with the short two dimensional distance 
$\Lambda^{-1}$, depending on the value of the number $L$.

At a macroscopic spacetime, the coupling constants $\lambda^{i}$ in 
the general nonlinear model are the wave modes of spacetime fields, 
including the spacetime metric.  For wave modes that are localized 
in the macroscopic spacetime, the numbers $\ad(i)$ are the 
eigenvalues of covariant second order differential operators acting 
on the spacetime wave modes.  For wave modes localized at spacetime 
distances where spacetime curvature is insignificant, the anomalous 
dimensions $\ad(i)$, being quadratic in the spacetime derivatives, 
take the form
\eq
\ad(i)= p(i)^{2}+m(i)^{2}
\en
where $p(i)$ is the spacetime wave number and $m(i)$ is the 
spacetime mass of the wave mode $\lambda^{i}$.

The number $L$ is therefore a spacetime distance.  The manifold of 
spacetimes is parametrized by the spacetime wave modes that have 
wave number $p(i)$ and mass $m(i)$ not many times larger than $1/L$.  
These are the coupling constants $\lambda^{i}$ that fluctuate in the 
lambda model, so the number $L$ is the characteristic ultraviolet 
spacetime distance in the lambda model.

Each coupling constant $\lambda^{i}$ is associated to a spacetime 
distance $L(i)$ given by
\eq
\label{eq:Li}
L(i)^{2} = \ad(i)^{-1}
\:.
\en
The coupling constant $\lambda^{i}$ is irrelevant if $L(i)/L \ll 1$.  
To be specific, call $\lambda^{i}$ irrelevant if $L(i)/L< 1/20$.  
With this definition, the irrelevant coupling constants are 
suppressed by drastic scaling factors $\me^{-L^{2}\ad(i)}$, factors 
of $\me^{-400}$ or less.  The effects of the irrelevant coupling 
constants can be omitted without significant loss of accuracy.  The 
ratio $1/20$ is more or less arbitrary.  The details of the 
definition of irrelevance do not matter, as long as enough accuracy 
is maintained.

The coupling constants that are not irrelevant have 
$\delta(i)<400/L^{2}$.  If $L^{2}$ is a large number, the 
non-irrelevant coupling constants are very nearly marginal.  Their 
scaling dimensions $-\ad(i)$ differ only slightly from zero.  They 
might be called the {\it quasi-marginal} or {\it $L$-marginal} 
coupling constants.  The coupling constants that are irrelevant at 
two dimensional distance $\Lambda^{-1}$ might be called the {\it 
$L$-irrelevant} coupling constants.

The renormalization of the general nonlinear model decouples the 
irrelevant coupling constants.  The decoupling is accomplished by 
defining the renormalized quasi-marginal coupling constants so that 
all effects of the $L$-irrelevant coupling constants are absorbed 
into the effects of the quasi-marginal coupling constants.  This is 
the basic principle of renormalization in quantum field theory.

The coupling constants that describe the structure of the general 
nonlinear model at short two dimensional distance $\Lambda^{-1}$ are 
the spacetime wave modes that describe spacetime at spacetime 
distances larger than $L$.  The coupling constants that are 
spacetime wave modes at distances much smaller than $L$ are 
decoupled.  Thus the short distance structure of the renormalized 
general nonlinear model encodes the large distance structure of 
spacetime.  The lambda model acts on the short distance structure of 
the general nonlinear model, constructing an effective short 
distance structure in two dimensions, thereby constructing the 
effective large distance structure of spacetime.

The principles of renormalization can be applied accurately to the 
general nonlinear model at the short distance $\Lambda^{-1}$ as long 
as $L^{2}=\ln(\Lambda/\mu)$ is a very large number, large enough to 
be effectively a divergence in the two dimensional field theory.  
Precisely how large is necessary will have to be settled by detailed 
calculations.  It seems reasonable to assume, tentatively, that 
numbers on the order of $L^{2}= 10^{24}$ or $L^{2}= 10^{20}$ are 
more than large enough.  If that is so, and if the unit of spacetime 
distance is within a few orders of magnitude of the Planck length, 
then $L$ can be taken smaller than any distance practical for 
experiment, while $L^{2}=\ln(\Lambda/\mu)$ still remains large 
enough for the lambda model to work.

Write $\M{L}$ for the manifold of renormalized general nonlinear 
models at two dimensional distance $\Lambda^{-1}$.  The manifold of 
renormalized general nonlinear models depends on $L$ because the 
property of coupling constant irrelevance depends on the ratio 
$\mu\Lambda^{-1} = \me^{-L^{2}}$ between the short two dimensional 
distance $\Lambda^{-1}$ and the long two dimensional distance 
$\mu^{-1}$.  The $L$-irrelevant coupling constants are the coupling 
constants whose scaling dimensions $-\ad(i)$ are far from zero on 
the scale set by $L^{-2}$, say $\ad(i) L^{2} > 400$.  The 
$L$-irrelevant coupling constants are decoupled by the 
renormalization of the general nonlinear model and are ignored, 
without significant loss of accuracy.  The manifold $\M{L}$ is 
parametrized by the coupling constants that are not $L$-irrelevant, 
the quasi-marginal coupling constants $\lambda^{i}$, those whose 
scaling dimensions are not far from zero on the scale set by 
$L^{-2}$, say $\ad(i) L^{2} < 400$.  The structure of the general 
nonlinear model at two dimensional distance $\Lambda^{-1}$ depends 
only on the quasi-marginal coupling constants $\lambda^{i}$.  So 
these $\lambda^{i}$ parametrize the manifold $\M{L}$.

Each coupling constant $\lambda^{i}$ is associated to a spacetime 
distance $L(i)$ by equation~\ref{eq:Li}.  The $L$-irrelevant 
coupling constants are the spacetime wave modes at spacetime 
distances $L(i)$ which are small on the scale set by $L$, say $L(i) 
< L/20$.  The quasi-marginal coupling constants $\lambda^{i}$ are 
the spacetime wave modes at spacetime distances that are not much 
smaller than $L$, say $L(i)>L/20$.  The manifold $\M{L}$ is 
parametrized by the spacetime wave modes at spacetime distances on 
the order of $L$ and larger.  $\M{L}$ is the manifold of spacetimes 
at spacetime distances on the order of $L$ and larger.

$\M{\infty}$ is the manifold of scale invariant general nonlinear 
models, the exact solutions of the fixed point equation $\beta=0$.  
In the strict continuum limit, the limit $\Lambda^{-1}=0$, 
renormalization forces the general nonlinear model to lie in 
$\M{\infty}$, because $\M{\infty}$ is the stable attracting manifold 
of the renormalization group flow.  There are no relevant coupling 
constants in $\M{\infty}$, only marginal and irrelevant coupling 
constants.

$\M{\infty}$ is the foundation on which all the manifolds $\M{L}$ 
are built.  Scaling fields $\phi_{i}(z,\bar z)$ are constructed in a 
scale invariant general nonlinear model, a model in $\M{\infty}$.  
The anomalous dimensions $\ad(i)$ are calculated there.  These 
calculations identify the quasi-marginal coupling constants 
$\lambda^{i}$, out of which the manifolds $\M{L}$ are built.

As $\Lambda^{-1}$ increases from zero, as $L$ decreases from 
infinity, the set of quasi-marginal coupling constants grows.  As 
$L$ becomes smaller, the spacetime wave modes at spacetime distances 
somewhat smaller than $L$ become available as quasi-marginal 
coupling constants.  The dimension of the manifold $\M{L}$ 
increases.

The manifolds $\M{L}$ are built up incrementally as $L$ decreases.  
If $L > L^{\prime}$, the manifold $\M{L^{\prime}}$ is made by 
extending $\M{L}$.

A general nonlinear model $\lambda$ in $\M{L}$ satisfies 
$\beta^{i}(\lambda)=0$ in the directions of the $L$-irrelevant 
coupling constants $\lambda^{i}$.  As a spacetime, $\lambda$ 
satisfies the classical field equation $\beta = 0$ at spacetime 
distances smaller than $L$.  The spacetime $\lambda$ might be 
pictured as composed of spacetime regions or cells, each of linear 
size $L$, satisfying $\beta =0$ inside each cell.  The spacetime 
wave modes localized inside the cells are the $L$-irrelevant 
coupling constants.  They are coupled to the $L$-irrelevant scaling 
fields $\phi_{i}(z,\bar z)$.  These are scaling fields because they 
see only the spacetime at distances smaller than $L$, where 
$\beta=0$.  The properties of the $L$-irrelevant scaling fields 
$\phi_{i}(z,\bar z)$, including the anomalous scaling dimensions 
$\ad(i)$, are calculated locally in the spacetime $\lambda$.  The 
properties of an $L$-irrelevant scaling field localized inside a 
spacetime cell depend on the values of the quasi-marginal coupling 
constants only through the values of the spacetime fields in the 
neighborhood of the spacetime cell where the $L$-irrelevant scaling 
field is localized.

At each $\lambda$ in $\M{L}$, the $L$-irrelevant coupling constants 
are determined by calculations in the spacetime $\lambda$ that are 
local on the spacetime distance scale set by $L$.  In particular, 
the coupling constants that are $L^{\prime}$-marginal but 
$L$-irrelevant are determined locally in spacetime.  These are the 
additional coupling constants that parametrize the extension of 
$\M{L}$ to $\M{L^{\prime}}$ at the spacetime $\lambda$ in $\M{L}$.

In this way, the structure of the manifold of spacetimes is built 
from large spacetime distance towards smaller, as the renormalized 
general nonlinear model is built from short two dimensional distance 
towards longer.  As the characteristic spacetime distance $L$ 
decreases, the spacetime wave modes at the distances below $L$ 
appear as ripples on the larger wave modes of spacetime.

For $L$ larger than $L^{\prime}$, the manifold $\M{L}$ is a 
submanifold in $\M{L^{\prime}}$.  It is the submanifold defined by 
the vanishing of the $L$-irrelevant coupling constants.  It is the 
submanifold of spacetimes in $\M{L^{\prime}}$ that solve the 
classical field equations $\beta=0$ at spacetime distances smaller 
than $L$.

$\M{L}$ is also a quotient manifold of $\M{L^{\prime}}$.  At two 
dimensional distance $\Lambda^{-1}$, the $L$-irrelevant coupling 
constants that extend $\M{L}$ to $\M{L^{\prime}}$ are decoupled from 
the quasi-marginal coupling constants that parametrize $\M{L}$.  
Nothing of the structure of the general nonlinear model at two 
dimensional distance $\Lambda^{-1}$ depends on the $L$-irrelevant 
coupling constants.

A general nonlinear model at particular values of the quasi-marginal 
coupling constants $\lambda^{i}$ describes a string worldsurface at 
two dimensional distances longer than $\Lambda^{-1}$.  The 
properties of the general nonlinear model of the worldsurface depend 
on the values of the quasi-marginal coupling constants 
$\lambda^{i}$.  The quasi-marginal coupling constants parametrize 
the classical background spacetime in which the strings scatter.

The quasi-marginal coupling constants change only very slowly with 
the two dimensional distance, so the worldsurface is approximately 
scale invariant at distances longer than $\Lambda^{-1}$.  String 
scattering amplitudes are computed by integrating scaling fields 
$\phi_{i}(z,\bar z)$ over the worldsurface.  The two dimensional 
distance $\Lambda^{-1}$ acts as short distance cutoff in these 
worldsurface integrals.  As a result, the string propagator has the 
cut off form
\eq
T\,g^{ij} \,\left [ \frac{1-\me^{-L^{2}\ad(i)}}{\ad(i)} \right ]
\:.
\en
The propagation of string modes associated to spacetime distances 
$L(i)$ larger than $L$ is suppressed.  The spacetime distance $L$ 
acts as infrared spacetime cutoff in the string scattering 
amplitudes.  The string scattering is taking place inside a 
spacetime region, or cell, of linear size $L$.

The manifold $\M{L}$ of general nonlinear models at two dimensional 
distance $\Lambda^{-1}$ is the manifold of classical background 
spacetimes where strings scatter within spacetime regions of size 
$L$.  The amplitudes for string scattering within a spacetime cell 
are calculated by integrating over the worldsurface the 
$L$-irrelevant scaling fields that are localized inside the 
spacetime cell.  The worldsurface integrals are cutoff at the short 
distance two dimensional distance $\Lambda^{-1}$.  The spacetime 
cell of size $L$ might be considered a hypothetical experimental 
region, within which strings are hypothetically scattered at 
spacetime distances smaller than $L$.  The string scattering 
amplitudes are determined by the background spacetime in the 
spacetime neighborhood where the experiment takes place, which is 
parametrized by the quasi-marginal coupling constants.

The quasi-marginal coupling constants are not precisely 
dimensionless.  Some have $\beta(\lambda) \ne 0$, so can act as 
sources and detectors for string modes at spacetime distance $L$.  
It should be possible, in principle, to use the quasi-marginal 
coupling constants in this way to describe the experimental 
apparatus for scattering experiments at spacetime distances smaller 
than $L$.  The theoretical representation of nature is divided, for 
every large spacetime distance $L$, into a background spacetime at 
spacetime distances larger than $L$, containing observers measuring 
hypothetical string scattering amplitudes in the background 
spacetime at spacetime distances smaller than $L$.

From the point of view of an observer situated in spacetime at 
spacetime distance $L$, the condition $\beta(\lambda) = 0$ on the 
$L$-marginal coupling constants is the consistency condition for 
extending tree-level string scattering amplitudes to spacetime 
distances larger than $L$.  It expresses the condition that the 
background spacetime is a classical vacuum at spacetime distances 
$L$ and larger.  When the tree-level string scattering calculations 
are extended to spacetime distances larger than $L$, the general 
nonlinear model of the worldsurface is probed at two dimensional 
distances shorter than $\Lambda^{-1}$.  If the background spacetime 
were not a classical vacuum, if $\beta(\lambda)$ were not zero on 
the $L$-marginal coupling constants, then the running coupling 
constants would blow up at two dimensional distances shorter than 
$\Lambda^{-1}$.  The worldsurface would be pathological at short two 
dimensional distance.  From this point of view, the equation $\beta 
= 0$ is a consistency condition.

But if the classical spacetime is the renormalized general nonlinear 
model, then the condition $\beta(\lambda) = 0$ is inevitable.  It is 
forced by the renormalization.  The renormalized general nonlinear 
model at nonzero two dimensional distance $\Lambda^{-1}$ derives 
from the renormalized model at $\Lambda^{-1}=0$, where necessarily 
$\beta (\lambda) = 0$ because there are no relevant coupling 
constants.  Spacetime is necessarily a classical vacuum.  From this 
point of view, the equation $\beta = 0$ holds necessarily.  In this 
classical picture, there does not seem to be any room for an 
observer, in the absence of fluctuations.

\subsection{What the cancelling does}
\label{subsect:what-the-cancelling-does}

The condition $\beta (\lambda)= 0$ on the quasi-marginal coupling 
constants allows the tree-level string scattering amplitudes to be 
extended to larger spacetime distances.  But string loop corrections 
are divergent in the spacetime infrared.  The divergence is 
logarithmic in the short two dimensional distance $\Lambda^{-1}$.  
Because of the divergence, the two dimensional cutoff cannot be 
removed.  The spacetime infrared cutoff cannot be relaxed.

The divergence, equation~\ref{eq:log-insertion}, is a bi-local 
insertion at short two dimensional distance.  It disturbs the short 
distance structure of the general nonlinear model, disturbing the 
selected background spacetime.  The divergence signals that the 
possible background spacetimes are not properly determined, at any 
large spacetime distance $L$.  A mechanism is missing that will 
determine the background spacetime so as to nullify the effects of 
the string loop divergence at short two dimensional distances.

The lambda model does this.  The fluctuating lambda fields cancel 
the string loop effects at short two dimensional distance, 
eliminating the dependence on $\Lambda^{-1}$.  The lambda model acts 
at two dimensional distances from $\Lambdazero^{-1}$ up to 
$\Lambda^{-1}$, producing an effective general nonlinear model of 
the worldsurface at each two dimensional distance between 
$\Lambdazero^{-1}$ and $\Lambda^{-1}$.  These are effective 
background spacetimes at every spacetime distance from $L_{0}$ down 
to $L$, where $L_{0}$ is defined by
\eq
L_{0}^{2} = \ln (\Lambdazero/\mu)
\:.
\en
The cancelling implies that infrared string loop corrections do not 
ever have to be calculated, because their effects are already built 
into the effective background spacetimes produced by the lambda 
model.  Infrared string loop corrections, starting at spacetime 
distance $L$ in the effective background spacetime, would merely 
undo what the lambda model already did when it built the effective 
background spacetime from larger spacetime distance down to $L$.  
Because of the cancelling, the infrared string loop corrections 
would disturb the effective background spacetime at spacetime 
distance $L$ exactly so as to produce the effective background 
spacetimes at spacetime distances larger than $L$.  The lambda model 
produces the effective background spacetime depending on $L$ exactly 
so it nullifies the infrared divergent string loop corrections.

The lambda model thus acts autonomously at short two dimensional 
distances, from a very short two dimensional cutoff distance 
$\Lambdazero^{-1}$ up to $\Lambda^{-1}$.  It acts autonomously at 
large spacetime distances from a very large infrared spacetime 
cutoff distance $L_{0}$ down to $L$.  The structure of the effective 
general nonlinear model of the worldsurface is determined, at any 
short distance $\Lambda^{-1}$, entirely by two dimensional nonlinear 
model calculations in the lambda model.  Handles are dispensed with 
completely, at two dimensional distances shorter than 
$\Lambda^{-1}$.  Infrared string loop corrections are dispensed 
with, at spacetime distances larger than $L$.  String theory is used 
only at spacetime distances smaller than $L$.  There is no practical 
use for string theory, if $L$ can be pushed smaller than any 
observable spacetime distance.

Moreover, the lambda model makes sense nonperturbatively, as a two 
dimensional nonlinear model, while string theory is formulated only 
perturbatively.  The lambda model constructs the effective 
background spacetime nonperturbatively.  It {\em defines} 
nonperturbative string theory, at large distance in spacetime.  If a 
nonperturbative version of string theory did exist, then its 
infrared quantum corrections, accumulated from small distance to 
large, would undo the work of the lambda model.  Even if a 
nonperturbative version of string theory did exist, there would be 
no need to use it at large spacetime distances.

\subsection{The effective general nonlinear model}

The lambda model constructs the effective general nonlinear model of 
the worldsurface by a local process in two dimensions, acting 
entirely at short distance.  So the effective general nonlinear 
model at two dimensional distance $\Lambda^{-1}$ is independent of 
the two dimensional cutoff distance $\Lambdazero^{-1}$.  It depends 
only on effective coupling constants 
$\lambdaeff^{i}(\Lambda/\Lambdazero,\lambdar)$.  The lambda model 
builds the effective coupling constants starting at the two 
dimensional cutoff distance,
starting from the running coupling constants at that distance,
\eq
\lambdaeff^{i}(\Lambdazero/\Lambdazero,\lambdar) =
\lambdazero^{i} = \lambdar^{i}(\mu/\Lambdazero,\lambda)
\:.
\en
The effective coupling constants $\lambdaeff^{i}$ are coupled to 
effective two dimensional fields $\phieff_{i}(z,\bar z)$ at two 
dimensional distance $\Lambda^{-1}$.  The effective general 
nonlinear model is described by the insertion
\eq
\me^{-\int \dif^{2}z \, \Lambda^{2} \frac1{2\pi} \,
\lambdaeff^{i} \phieff_i(z,\bar z)}
\:.
\en
It satisfies an effective renormalization group equation
\eq
\left (
\Lambda \partialby{\Lambda}_{\left /\lambdaeff\right.}
+ \betaeff^{i}(\lambdaeff) \partialby{\lambdaeff^{i}}
\right )
\;
\me^{-\int \dif^{2}z \, \Lambda^{2} \frac1{2\pi} \,
\lambdaeff^{i} \phieff_i(z,\bar z)}
=0
\:.
\en
The effective beta function $\betaeff$ consists of the beta function 
of the general nonlinear model, $\beta$, plus corrections $\delta 
\beta$ generated by the lambda model,
\eq
\betaeff = \beta + \delta \beta
\:.
\en

\subsection{The effective lambda model}

The lambda fluctuations produce an effective lambda model as well as 
an effective general nonlinear model.  The lambda model is itself a 
two dimensional nonlinear model, so it is renormalizable.  The 
effective lambda model at two dimensional distance $\Lambda^{-1}$ is 
described by the effective metric coupling 
$T^{-1}\geff_{ij}(\Lambda/\Lambdazero,\lambdar)$.

The crucial principle is that the effective lambda model and the 
effective general nonlinear model of the worldsurface evolve in 
tandem as the two dimensional distance $\Lambda^{-1}$ increases.  
The propagator of the effective lambda model cancels handles at 
distance $\Lambda^{-1}$ in the effective worldsurface.  The 
effective metric coupling $T^{-1} 
\geff_{ij}(\Lambda/\Lambdazero,\lambdar)$ is the inverse of the 
effective handle gluing matrix.

This principle of {\it tandem renormalization} follows from the 
cancelling of handles by lambda fluctuations over a finite range of 
two dimensional distances.  The cancelling of handles by lambda 
fluctuations between two dimensional distances $\Lambdazero^{-1}$ 
and $\Lambda_{1}^{-1}$, with $\Lambdazero^{-1}< 
\Lambda^{-1}<\Lambda_{1}^{-1}$, can be broken up into the cancelling 
between $\Lambdazero^{-1}$ and $\Lambda^{-1}$, and the cancelling 
between $\Lambda^{-1}$ and $\Lambda_{1}^{-1}$.  The second 
cancelling can be expressed in terms of handles in the effective 
worldsurface and fluctuations in the effective lambda model, both at 
two dimensional distance $\Lambda^{-1}$.  The second cancelling 
therefore implies that the effective metric coupling $T^{-1} 
\geff_{ij}(\Lambda/\Lambdazero,\lambdar)$ is the inverse of the 
effective handle gluing matrix.

The effective metric coupling of the lambda model is identified with 
a natural structure in the effective general nonlinear model of the 
worldsurface, so it depends only on the effective coupling constants 
$\lambdaeff^{i}$.  It depends on $\Lambda^{-1}$ only through the 
$\lambdaeff^{i}$.  Therefore the effective lambda model is scale 
invariant in the generalized sense, taking the same form
\eq
\int \Dlambdaeff
\; \me^{- \Seff(\lambdaeff)} \;
\me^{-\int \dif^{2}z \, \Lambda^{2} \frac1{2\pi} \,
\lambdaeff^{i}(z,\bar z) \phieff_i(z,\bar z)}
\en
\eq
\Seff(\lambdaeff) = \int \dif^{2}z \, \frac1{2\pi}
\,
T^{-1} \geff_{ij}(\lambdaeff)
\,
\partial \lambdaeff^{i}
\,
\bar \partial \lambdaeff^{j}
\en
at every short two dimensional distance $\Lambda^{-1}$.

\subsection{The \apm}

In a two dimensional nonlinear model, the fluctuations at two 
dimensional distances shorter than the characteristic distance 
$\Lambda^{-1}$ distribute themselves over the target manifold of the 
model to form a measure $\drho(\Lambda,\lambda)$ on the target 
manifold, called the \apm~\citeF. The \apm\ summarizes the 
fluctuations at two dimensional distances shorter than 
$\Lambda^{-1}$.

The \apm\ is calculated as the one point expectation value at two 
dimensional distance $\Lambda^{-1}$,
\eq
\label{eq:one-point}
\int \drho(\Lambda,\lambda) \, f(\lambda)
= \expval{\, f(\lambda(z,\bar z) ) \,}
\en
where the expectation value is calculated by integrating over the 
lambda fluctuations at two dimensional distances up to 
$\Lambda^{-1}$.
Equivalently,
\eq
\drho(\Lambda,\lambda^{\prime})
= \dvol(\lambda^{\prime}) \,
\expval{\, \delta(\lambda^{\prime} - \lambda(z,\bar z) ) \,}
\:.
\en

The \apm\ of the lambda model is a measure on the manifold of 
spacetimes.  Locally on the manifold of spacetimes, it takes the 
form of a measure on the spacetime wave modes $\lambda^{i}$, so it 
has the potential to be a quantum field theory in spacetime.  The 
spacetime quantum field theory correlation functions of the wave 
modes $\lambda^{i}$ are to be the integrals of functions 
$f(\lambda)$ with respect to the \apm\ on the manifold of 
spacetimes.  So the spacetime correlation functions are to be 
calculated as the one point expectation values in the lambda model, 
as in equation~\ref{eq:one-point}.

In a two dimensional nonlinear model, the renormalization group acts 
on the target manifold of the model by a diffusion process~\citeF. 
As the characteristic two dimensional distance $\Lambda^{-1}$ 
increases, additional fluctuations are included in the \apm, causing 
the it to diffuse outwards in the target manifold.  The metric 
governing the diffusion is the effective metric coupling.  When the 
nonlinear model is scale invariant, the diffusion process has 
stationary coefficients.  When the scale invariance is of the 
generalized kind, as in the lambda model, the stationary diffusion 
process is driven by the flow in the target manifold that provides 
the scale invariance.

With each infinitesimal increase in the two dimensional distance, 
$\Lambda^{-1}\rightarrow (1+\epsilon)\Lambda^{-1}$, the \apm\ 
diffuses outwards because of the additional fluctuations.  At the 
same time, the effective coupling constants flow, 
$\lambdaeff^{i}\rightarrow \lambdaeff^{i} - \epsilon 
\betaeff^{i}(\lambdaeff)$, along the driving vector field 
$-\betaeff^{i}$.

Writing the \apm\ in the variables $\lambdaeff^{i}$ as 
$\drhoeff(\Lambda,\lambdaeff)$, the driven diffusion process is
\eq
- \Lambda \partialby\Lambda_{\left / \lambdaeff \right .}
\,
\drhoeff(\Lambda,\lambdaeff) 
=
\deleff_{i} \left ( T\,\gsubeff^{ij}(\lambdaeff) \deleff_{j}
+\betaeff^{i}(\lambdaeff)
\right )
\,
\drhoeff(\Lambda,\lambdaeff) 
\en
where $\deleff_{i}$ is the covariant derivative with respect to the 
effective metric $T^{-1}\geff_{ij}$.  The coefficients, 
$T\,\gsubeff^{ij}$ and $\betaeff^{i}$, of the diffusion process are 
stationary, independent of $\Lambda^{-1}$, because of the 
generalized scale invariance of the effective lambda model.

In the very long diffusion time $\ln(\Lambdazero/\Lambda)$, the 
\apm\ converges to the equilibrium measure $\drhoequil(\lambdaeff)$ 
of the stationary diffusion process, no matter what its initial 
value at $\Lambdazero^{-1}$.  The equilibrium \apm\ is peaked near 
the attracting submanifold where $\betaeff = 0$.  If the equilibrium 
\apm\ concentrates near a macroscopic spacetime, then the \apm, as a 
measure on the wave modes of spacetime fields, is potentially a 
quantum field theory in spacetime, uniquely produced by the lambda 
model.

Assuming that the lambda model makes only small corrections to 
$\beta$, the \apm\ is driven by $-\betaeff$ first to the submanifold 
where $\beta=0$.  Then the corrections to $\beta$ determine the 
subset of the $\beta=0$ submanifold at which the \apm\ actually 
concentrates.

The \apm\ of the lambda model is generally covariant in spacetime, 
the target manifold of the general nonlinear model, because the 
renormalization of the general nonlinear model is carried out in a 
way that is invariant under reparametrization of its target 
manifold~\citeF.

If a particular background spacetime is chosen by hand, arbitrarily, 
then it serves as the initial condition for the diffusion of the 
\apm.  By the time a finite value of $\Lambda^{-1}$ is reached, the 
\apm\ has diffused to the equilibrium measure, erasing all 
dependence on the initial choice of spacetime.  The lambda model 
dynamically produces independence from the arbitrary choice of 
background spacetime.

\subsection{The spacetime action principle}

The beta function of the general nonlinear model, 
$\beta^{i}(\lambda)$, is a gradient vector field on the manifold of 
general nonlinear models~\cite{Friedan-1,Friedan-2,Friedan-3, 
Fradkin-Tseytlin,CFMP,Zamolodchikov-1,Zamolodchikov-2}.  In the 
proof of the gradient 
property~\cite{Zamolodchikov-1,Zamolodchikov-2}, the beta function 
is expressed as a gradient with respect to an intrinsic metric on 
the manifold of two dimensional quantum field theories, defined by 
the two point correlation function of the fields $\phi_{i}(z,\bar 
z)$ on the plane.  The metric coupling of the lambda model, 
$T^{-1}g_{ij}(\lambda)$, defined as the inverse of the handle gluing 
matrix, is that same intrinsic metric, multiplied by $T^{-1}$.  So 
the beta function of the general nonlinear model is the gradient 
with respect to the metric coupling $T^{-1}g_{ij}(\lambda)$
\eq
T^{-1} g_{ij} (\lambda) \, \beta^{j} (\lambda) = 
\partialby{\lambda^{i}} \, T^{-1} a(\lambda)
\en
of the {\it potential function} $T^{-1} a(\lambda)$ on the manifold 
of spacetimes.

In a macroscopic spacetime of volume $V$,
\eq
T^{-1} a(\lambda) = \gst^{-2} \, V \, a(\lambda)
\en
where $V \, a(\lambda)$ is properly normalized so as to be the 
spacetime integral of a local functional of the spacetime wave modes 
$\lambda^{i}$.  The potential function $T^{-1} a(\lambda)$ is the 
spacetime action of the massless spacetime field theory whose 
scattering amplitudes are the same as the tree-level string 
scattering amplitudes at large distance~\cite{Lovelace-2,CFMP}.

The gradient property will be derived as well for the effective beta 
function of the effective general nonlinear model,
\eq
T^{-1} \geff_{ij} (\lambdaeff) \, \betaeff^{j} (\lambdaeff) = 
\partialby{\lambdaeff^{i}} \, T^{-1} \aeff(\lambdaeff)
\:.
\en
The stationary diffusion process for the \apm\ is therefore driven 
by a gradient flow.  The equilibrium measure then satisfies the 
first order differential equation
\eq
\label{eq:first-order-equil}
0 = \left ( T\,\gsubeff^{ij}(\lambdaeff) \deleff_{j}
+\betaeff^{i}(\lambdaeff)
\right )
\,
\drhoequil(\lambdaeff) 
\en
whose solution is
\eq
\drhoequil(\lambdaeff) =
\dvoleff(\lambdaeff) \; e^{-T^{-1}\aeff(\lambdaeff)}
\en
where $\dvoleff(\lambdaeff)$ is the metric volume element associated 
to the effective metric coupling.

The first order equation~\ref{eq:first-order-equil} for the \apm\ is 
the equation of motion $\betaeff^{i}(\lambdaeff) = 0$ in the sense 
of spacetime quantum field theory.  If the lambda model is 
successful, spacetime quantum field theory will be explained as the 
\apm\ of the lambda model.  The quantum action principle of 
spacetime physics will then derive from the gradient property of the 
beta function of the general nonlinear model.

The \apm\ of the lambda model is nontrivial even at the classical 
level, because the lambda model is scale invariant in the 
generalized sense even at the classical level.  The effective 
potential function is the classical potential function plus 
corrections generated by the lambda fluctuations
\eq
T^{-1}\aeff = T^{-1}a
+ \delta (T^{-1}a)
\:.
\en

Before quantum corrections are taken into account, the \apm\ and the 
diffusion process are written in terms of the uncorrected running 
coupling constants $\lambdar^{i}$, around a spacetime satisfying 
$\beta=0$, to leading order in the $\lambdar^{i}$,
\eq
\label{eq:leading-diffusion}
- \Lambda \partialby\Lambda_{\left / \lambdar \right .}
\,
\drhor(\Lambda,\lambdar) 
=
\partial_{i} \left ( T g^{ij} \partial_{j}
+ \ad(i) \lambdar^{i}
\right )
\,
\drhor(\Lambda,\lambdar)
\:.
\en
The uncorrected potential function is
\eq 
T^{-1}a(\lambdar) =
\frac12 \, T^{-1} g_{ij} \ad(i) \lambdar^{i}
\lambdar^{j}
+O(\lambdar^{3})
\:.
\en
The uncorrected equilibrium \apm\ is
\eq
\dvol(\lambdar) \; e^{-\frac12 \, T^{-1} g_{ij} \ad(i) \lambdar^{i}
\lambdar^{j}}
\en
in the gaussian approximation.  The equilibration time for the wave 
mode $\lambda^{i}$ is $1/\ad(i)$.

\subsection{Complementarity between spacetime quantum field theory 
and string theory}

For each value of the characteristic spacetime distance $L$, the 
lambda model produces two complementary descriptions of spacetime 
physics.  The \apm\ describes the spacetime physics at distances 
larger than $L$ as spacetime quantum field theory.  The effective 
general nonlinear model of the worldsurface describes spacetime 
physics at distances smaller than $L$ by effective string scattering 
amplitudes.  Both descriptions apply in local regions of spacetime, 
on the scale of spacetime distance set by $L$.  $L$ is the 
characteristic ultraviolet distance in the effective spacetime 
quantum field theory, and the characteristic infrared distance in 
the effective string scattering amplitudes.

The lambda model constructs the \apm\ and the effective general 
nonlinear model so that the two descriptions agree at spacetime 
distance $L$, where both apply.  By the tandem renormalization 
principle, the data of the effective lambda model matches the data 
of the effective general nonlinear model.  The effective potential 
function on the wave modes $\lambdaeff^{i}$ at spacetime distance 
$L$ will be the generating functional for the effective string 
scattering amplitudes at distance $L$, by a version of the 
argument~\cite{Lovelace-2} that connected the beta function 
$\beta^{i}(\lambda)$ of the general nonlinear model to the 
tree-level string scattering amplitudes at large distance.

Given that the effective string scattering amplitudes match the 
scattering amplitudes of the effective spacetime quantum field 
theory, and that the effective spacetime quantum field theory has to 
be produced by the lambda model before the effective string 
scattering amplitudes become nonperturbatively reliable, there does 
not seem to be any practical use for the string scattering 
amplitudes at any spacetime distance $L$ large enough to be reached 
by the lambda model.  On the other hand, there could be 
circumstances where the effective string theory would be useful as 
an alternate technical algorithm for calculating scattering 
amplitudes.

\subsection{The fermionic spacetime modes}

The target manifold of the lambda model is a graded manifold.  The 
manifold of spacetimes has both bosonic and fermionic dimensions.  
The bosonic coupling constants $\lambda^{i}$ are the wave modes of 
bosonic spacetime fields, the fermionic $\lambda^{i}$ are the wave 
modes of fermionic spacetime fields.  The \apm\ is a measure on the 
graded manifold of spacetimes, a quantum field theory of bosonic and 
fermionic fields in spacetime.

The lambda model needs a technical construction of the general 
nonlinear model of the worldsurface in which the bosonic and 
fermionic coupling constants $\lambda^{i}$ occur on an equal 
footing.  The bosonic $\lambda^{i}$ must couple to bosonic fields 
$\phi_{i}(z,\bar z)$, the fermionic $\lambda^{i}$ to fermionic 
fields $\phi_{i}(z,\bar z)$.  The metric coupling $T^{-1}g_{ij}$ 
must be symmetric in the bosonic directions and antisymmetric in the 
fermionic directions.  And the construction must accomodate a 
general compact riemannian spacetime.

A construction of the fermionic coupling constants $\lambda^{i}$ and 
their antisymmetric metric coupling $T^{-1}g_{ij}$ is given below, 
in section~\ref{sect:fermionmodes}.  The construction is based on 
the locally supersymmetric string worldsurface, in which 
worldsurface reparametrization invariance is implemented by means of 
supersymmetric worldsurface ghost fields~\cite{Polyakov}.  The odd 
parameters of the supersymmetric worldsurface are eliminated, and 
the bosonic worldsurface ghost fields are used to construct an 
ordinary scale invariant worldsurface that is spacetime 
covariant~\cite{FMS}.  Bosonic and fermionic scaling fields 
$\phi_{i}(z,\bar z)$ couple to the spacetime wave modes 
$\lambda^{i}$, which are correspondingly bosonic and fermionic.

The technical drawback of the covariant worldsurface is the picture 
ambiguity.  The worldsurface scaling fields fall into infinitely 
many formally equivalent subspaces called pictures, distinguished by 
a discrete picture charge.  The bosonic scaling fields have integer 
picture charge.  The fermionic scaling fields have picture charge 
integer plus half.  The canonical pictures are distinguished by the 
condition that the dimensions of the scaling fields are bounded 
below.  These are the natural pictures to use in analyzing the 
effects of handles at short distance.  For the bosonic scaling 
fields, there is exactly one canonical picture, which has picture 
charge $-1$.  The metric is symmetric on the canonical bosonic 
picture.  For the fermionic scaling fields, there are two canonical 
pictures, the pictures of charge $-1/2$ and $-3/2$.  The metric 
pairs the two canonical fermionic pictures.  In the sum over states 
flowing through a handle, one of the canonical fermionic pictures is 
inserted at one end of the handle, the other canonical fermionic 
picture at the other end of the handle.  There is no single space of 
fermionic coupling constants $\lambda^{i}$ with an antisymmetric 
metric coupling $T^{-1}g_{ij}$.

A small technical innovation is devised to put the canonical 
fermionic scaling fields effectively in a single picture that 
effectively has picture charge $-1$, and on which there is an 
antisymmetric metric $T^{-1}g_{ij}$.  The bosonic and fermionic 
coupling constants $\lambda^{i}$ then combine to form a single 
graded space, with a graded metric coupling $T^{-1}g_{ij}$.

When the bosonic and fermionic coupling constants $\lambda^{i}$ are 
put on the same footing as coordinates for the graded manifold of 
spacetimes, the spacetime dynamics of the fermionic wave modes takes 
a nonstandard form.  The wave operators acting on the fermionic wave 
modes are quadratic in the spacetime derivatives, like the wave 
operators acting on the bosonic wave modes.  They are not the 
standard first order Dirac wave operators.  The unphysical degrees 
of freedom of the fermionic spacetime fields are eliminated by gauge 
symmetries, leaving the standard physical content of the Dirac 
operators.

The lambda model needs the metric coupling $T^{-1}g_{ij}$ to be 
positive definite in the bosonic directions.  Otherwise, there would 
be instability under short distance fluctuations of the bosonic 
lambda fields $\lambda^{i}(z,\bar z)$.  This positivity condition is 
not satisfied if the worldsurface contains a Ramond-Ramond sector, 
because the metric on that sector is the tensor product of two 
antisymmetric matrices, which is not positive definite.  Only the 
heterotic string worldsurface~\cite{heterotic} is without a 
Ramond-Ramond sector.  For this purely technical reason, it appears 
that the lambda model can only work in the heterotic string 
worldsurface.  The metric coupling $T^{-1}g_{ij}$ on the manifold of 
general nonlinear models of the heterotic worldsurface is positive 
definite in the bosonic directions, because there is no 
Ramond-Ramond sector and because spacetime is assumed riemannian.

\subsection{Physics is built from large distance towards small}

The degrees of freedom of the lambda model are the coupling 
constants $\lambda^{i}$ of the renormalized general nonlinear model, 
varying locally in two dimensions as fields $\lambda^{i}(z,\bar z)$.  
The target manifold of the lambda model at two dimensional distance 
$\Lambda^{-1}$ is the manifold $\M{L}$.  The renormalized general 
nonlinear model at two dimensional distance $\Lambda^{-1}$ provides 
the data on the target manifold $\M{L}$, the metric 
$T^{-1}g_{ij}(\lambdar)$ and the vector field $\beta^{i}(\lambdar)$, 
which specify the couplings of the lambda model.  The renormalized 
general nonlinear model provides the lambda model with its degrees 
of freedom and its interactions.  The renormalized general nonlinear 
model is the raw material on which the lambda model works.

The coupling constants $\lambda^{i}$ are the wave modes of 
spacetime.  The renormalization of the general nonlinear model 
arranges the degrees of freedom $\lambda^{i}$ over the range of 
short two dimensional distances $\Lambda^{-1}$, in a hierarchy 
organized according to spacetime distance $L$, following the formula 
$L^{2} = \ln (\Lambda/\mu)$.  At the shortest two dimensional 
distances, the degrees of freedom $\lambda^{i}$ are at the largest 
spacetime distances.  The renormalization of the general nonlinear 
model decouples the small distance spacetime wave modes at short two 
dimensional distance $\Lambda^{-1}$.  As the two dimensional 
distance $\Lambda^{-1}$ increases, as the spacetime distance $L$ 
decreases, the renormalization introduces, as additional degrees of 
freedom, the spacetime wave modes at smaller spacetime distances on 
the distance scale set by $L$.

The hierarchy of degrees of freedom is put in place by the 
renormalization of the general nonlinear model, before the lambda 
model is set to work.  The lambda model acts autonomously on the 
renormalized general nonlinear model, from short two dimensional 
distance towards long.  The lambda model operates on the degrees of 
freedom $\lambda^{i}$ as it finds them, distributed by the 
renormalization of the general nonlinear model across the range of 
short two dimensional distances.  The lambda model begins to operate 
at the extremely short two dimensional cutoff distance 
$\Lambdazero^{-1}\approx 0$, seeing only the extreme infrared wave 
modes in spacetime.  As the lambda model operates at longer two 
dimensional distance $\Lambda^{-1}$, it sees spacetime wave modes at 
decreasing spacetime distance $L$.  At each stage, at each 
characteristic spacetime distance $L$, the lambda model can ignore 
the decoupled spacetime wave modes at distances smaller than $L$, 
without significant loss of accuracy.  The lambda model never 
notices the small distance structure of spacetime.

Locality in spacetime is expressed by the hierarchy of degrees of 
freedom $\lambda^{i}$, distributed in the two dimensional distance 
$\Lambda^{-1}$ according to the spacetime distance $L$.  This form 
of locality in spacetime is left undisturbed by the lambda model.  
The lambda model produces corrections to the interactions at each 
two dimensional distance $\Lambda^{-1}$, at each characteristic 
spacetime distance $L$.  Nonperturbative effects in the lambda model 
might possibly change the form in which the degrees of freedom 
effectively appear.  But the lambda model does not change the 
distribution of the basic degrees of freedom $\lambda^{i}$ in the 
two dimensional distance $\Lambda^{-1}$, and in the spacetime 
distance $L$.  Nor does it change the decoupling of small distance 
spacetime wave modes, for every large value of $L$.

The \apm\ of the lambda model is the distribution of the 
fluctuations of the fields $\lambda^{i}(z,\bar z)$ at two 
dimensional distances shorter than $\Lambda^{-1}$.  It is a measure 
on the target manifold of the lambda model, so it is a measure, for 
each $L$, on the graded manifold of spacetimes, $\M{L}$.  The \apm\ 
is thus a measure on the bosonic and fermionic spacetime wave modes 
$\lambda^{i}$ at spacetime distances larger than $L$.  If this 
measure has appropriate technical properties, then it is a quantum 
field theory in spacetime, cut off in the ultraviolet at spacetime 
distance $L$, describing physics at spacetime distances larger than 
$L$.

The lambda model is a local two dimensional quantum field theory, so 
it is built starting from its short distance limit at 
$\Lambda^{-1}=0$, and proceeding outwards to longer two dimensional 
distances $\Lambda^{-1}$ by integrating over the short distance 
modes of the fields $\lambda^{i}(z,\bar z)$.

As the two dimensional distance $\Lambda^{-1}$ {\em increases}, the 
spacetime distance $L$ {\em decreases}.  Additional spacetime modes 
$\lambda^{i}$ begin to fluctuate, at smaller and smaller spacetime 
distances.  Each wave mode $\lambda^{i}$ is at a characteristic 
spacetime distance $L(i)$.  The mode $\lambda^{i}$ starts 
fluctuating in the lambda model when $L$ has dropped below a value 
$L_{0}(i)$ not much larger than $L(i)$, say $L_{0}(i)= 20 L(i)$.  
The fluctuations of the two dimensional field $\lambda^{i}(z,\bar 
z)$ are cut off at a short two dimensional distance 
$\Lambda_{0}(i)^{-1}$ given by 
$L_{0}(i)^{2}=\ln(\Lambda_{0}(i)/\mu)$.  The field 
$\lambda^{i}(z,\bar z)$ fluctuates only at two dimensional distances 
$\Lambda^{-1}$ longer than its individualized cutoff distance 
$\Lambda_{0}(i)^{-1}$.

The \apm\ of the lambda model evolves with the increasing two 
dimensional distance $\Lambda^{-1}$, starting from the cutoff 
distance $\Lambdazero^{-1}$.  As $\Lambda^{-1}$ increases, as $L$ 
decreases, the spacetime wave modes $\lambda^{i}$ at distance $L$ 
begin to fluctuate, and are {\em integrated in} to the \apm.  The 
\apm\ is effectively a delta function in the variable $\lambda^{i}$ 
concentrated at $\lambda^{i}=0$, until $L$ drops below $L_{0}(i)$, 
until $\Lambda^{-1}$ becomes longer than $\Lambda_{0}(i)^{-1}$.  
When $L$ drops below $L_{0}(i)$, the \apm\ begins to diffuse outward 
in the spacetime wave mode $\lambda^{i}$.

The characteristic equilibration time of the variable $\lambda^{i}$ 
is $1/\ad(i)=L(i)^{2}$, according to 
equation~\ref{eq:leading-diffusion}.  The available diffusion time, 
$L_{0}(i)^{2}-L(i)^{2}$, is more than enough to allow $\lambda^{i}$ 
to reach equilibrium well before $L$ decreases from $L_{0}(i)$ to 
$L(i)$.  The \apm\ at two dimensional distance $\Lambda^{-1}$ 
therefore gives an accurate description of the physics at all 
spacetime distances greater than $L$, the physics of the spacetime 
wave modes $\lambda^{i}$ at all spacetime distances $L(i) > L$.  
The assumption here is that the equilibration times estimated from 
the uncorrected diffusion process are not significantly different 
from the actual diffusion times in the effective diffusion process.

The lambda model thus builds its \apm, which is to be spacetime 
quantum field theory, starting from the large distance limit at 
$L=\infty$, by {\em integrating in} the spacetime wave modes at 
smaller and smaller spacetime distances $L$.  Despite this top down 
method of construction, from large spacetime distance to small, the 
resulting quantum field theory appears local in spacetime.  
Spacetime locality is expressed in the functional integral 
formulation of quantum field theory by the possibility of 
integrating out the small distance wave modes of the spacetime 
fields without losing information about the functional measure on 
the wave modes at larger distances.  The \apm\ of the lambda model 
is local in spacetime, in this sense, because integrating out the 
spacetime wave modes at small spacetime distance merely reverses the 
process of {\em integrating in} that was performed by the lambda 
model.  Integrating out a small distance wave mode $\lambda^{i}$ 
replaces the equilibrium \apm\ in that variable with the value of 
the integral over $\lambda^{i}$, multiplied by the delta function 
concentrated at $\lambda^{i}=0$.  This simply reverses the diffusion 
accomplished by the lambda model, which starts from the delta 
function concentrated at $\beta^{i}(\lambda)=0$ and tends to the 
equilibrium measure, since diffusion conserves the total weight of 
the measure.  The \apm\ at a larger spacetime distance 
$L>L^{\prime}$ is recovered from the \apm\ at the smaller spacetime 
distance $L^{\prime}$, by integrating out the wave modes 
$\lambda^{i}$ at spacetime distances $L(i)$ between $L$ and 
$L^{\prime}$.

This appearance of locality in the spacetime quantum field theory 
depends crucially on the fact that, as $\Lambda^{-1}$ increases, the 
newly fluctuating fields $\lambda^{i}(z,\bar z)$ at spacetime 
distances smaller than $L$ do not disturb the equilibrium \apm\ on 
the wave modes at spacetime distances larger than $L$.  As $L$ 
decreases, the \apm\ stabilizes on the spacetime wave modes at 
distances larger than $L$.  Stability at large distance is ensured 
by the decoupling of irrelevant coupling constants in the 
renormalization of the general nonlinear model.  The spacetime wave 
modes $\lambda^{i}$ at small distance on the scale set by $L$ are 
relatively irrelevant coupling constants in the renormalized general 
nonlinear model.  The small distance wave modes are decoupled from 
the large distance wave modes, which are less irrelevant.  The 
fluctuations of the small distance wave modes have no effect on the 
\apm\ at distances larger than $L$, as a measure on the large 
distance spacetime wave modes.

The building of the \apm\, from short two dimensional distance 
$\Lambda^{-1}$ towards longer, is based on the hierarchy of 
submanifolds $\M{L}\rightarrow \M{L^{\prime}}$, $L>L^{\prime}$.  As 
$\Lambda^{-1}$ increases, the \apm\ diffuses outwards from the 
submanifold $\M{L}$ in $\M{L^{\prime}}$, along the $L$-irrelevant 
directions $\lambda^{i}$ that parametrize the extension of $\M{L}$ 
to $\M{L^{\prime}}$.

The stability of the \apm\ on the large distance wave modes, its 
independence from the additional degrees of freedom that enter at 
relatively small spacetime distance, is based on the hierarchy of 
quotient manifolds $\M{L^{\prime}}\rightarrow \M{L}$.  The 
decoupling by renormalization, expressed in the quotient structure, 
ensures that integration over the fibers of the quotient reverses 
the diffusion of the \apm, which gives the property of spacetime 
locality to the \apm\ as spacetime quantum field theory.

Because of the stability of the \apm\ on the large distance wave 
modes, it should not be necessary to calculate quantum corrections 
in the lambda model at all spacetime distances from $L=\infty$ down 
to $L$, in order to find the effective spacetime physics at distance 
$L$.  That much calculation would not be practical.  It would 
accumulate enormous amounts of information about very large distance 
physics that would not be relevant to the physics at spacetime 
distance $L$.  The only new calculations that are needed in the 
lambda model at two dimensional distance $\Lambda^{-1}$ are to be 
done at spacetime distances on the order of $L$.  The effective 
interactions in the lambda model of the degrees of freedom at larger 
spacetime distances have already been calculated at two dimensional 
distances shorter than $\Lambda^{-1}$.  The effective lambda model 
has already stabilized in the large distance degrees of freedom, so 
the large distance calculations do not need to be redone.  Only the 
properties of the coupling constants $\lambda^{i}$, on the border 
between $L$-marginal and $L$-irrelevant, are in flux at two 
dimensional distance $\Lambda^{-1}$.  What new calculations are 
needed can be made in local spacetime regions, on the spacetime 
distance scale set by $L$, which is where the borderline coupling 
constants $\lambda^{i}$ are found.

Spacetime quantum field theory, as constructed by the lambda model, 
is local in spacetime, but it is not constructed from a microscopic 
dynamics.  There is no guarantee, outside the perturbation 
expansion, that the dynamical laws of spacetime physics at large 
distance can be derived from the dynamical laws at small distance.  
The effective potential function $\gst^{-2}\,V\,\aeff(\lambdaeff)$ 
at spacetime distance $L$ cannot be derived from the effective 
potential function at a smaller spacetime distance $L^{\prime}$, 
except perturbatively.  The actual form of the effective degrees of 
freedom $\lambdaeff^{i}$ at spacetime distance $L$ are not 
determined by spacetime quantum field theory effects acting on the 
degrees of freedom at spacetime distances smaller than $L$, but 
rather by nonperturbative two dimensional effects in the lambda 
model acting at two dimensional distances shorter than 
$\Lambda^{-1}$, so at spacetime distances larger than $L$.  
Perturbatively, the effective degrees of freedom and the effective 
potential function of the lambda model will be consistent with 
perturbative spacetime quantum field theory.  But nonperturbative 
effects (such as quark confinement) will have to be found in the 
lambda model.
 
For the lambda model to be well-defined, its target manifold should 
be finite dimensional.  Calculations requiring an infinite number of 
fields $\lambda^{i}(z,\bar z)$ would be difficult to control.  The 
target spacetime of the general nonlinear model is assumed to be 
compact and riemannian.  It follows that the general nonlinear model 
has a discrete spectrum of scaling dimensions $\ad(i)$.  The 
renormalization of the general nonlinear model then suppresses the 
ultraviolet wave modes in spacetime by factors $e^{-L^{2}\ad(i)}$, 
leaving only a finite number of quasi-marginal coupling constants 
$\lambda^{i}$ to parametrize the target manifold $\M{L}$.  The 
target manifold $\M{L}$ is therefore finite dimensional, at each 
point $\lambda$ in $\M{L}$.  The lambda model is well-defined, at 
least for small fluctuations around any point $\lambda$ in its 
target manifold, $\M{L}$, the manifold of compact riemannian 
spacetimes.

For any finite $L$, no matter how large, it is possible that 
fluctuations in $\M{L}$ will lead to spacetimes of linear size much 
larger than $L$, spacetimes which are macroscopic on the distance 
scale set by $L$.  Fluctuations in the lambda model can lead to the 
places in $\M{L}$ where spacetime grows arbitrarily large.  The 
dimension of $\M{L}$ would then increase without bound.  In 
particular, there are harmonic surfaces $\lambda_{H}(z,\bar z)$ in 
$\M{L}$ that pass through such places, as described in 
section~\ref{sect:lambda-instantons} below.  Weak coupling 
nonperturbative effects in the lambda model will be dominated by 
harmonic surfaces in the target manifold, including such {\it 
decompactifying} harmonic surfaces.  So it will be necessary to 
control the effects of the unbounded growth in the dimension of 
$\M{L}$ on a decompactifying harmonic surface.

The potential difficulty should only arise when the limit 
$L\rightarrow \infty$ is studied.  At any finite value of $L$, 
calculations in the lambda model see only spacetime regions that are 
bounded on the distance scale set by $L$.  Only a finite number of 
quasi-marginal spacetime wave modes $\lambda^{i}$ can fit into any 
such local spacetime region.  The difficulty posed by the unbounded 
dimensionality of the target manifold is a problem of the extreme 
infrared in spacetime, relevant to the problem of sending the two 
dimensional cutoff distance $\Lambdazero^{-1}$ to zero.

To finding the spacetime physics at finite distance $L$, the lambda 
model must be built from the two dimensional cutoff distance all the 
way out to $\Lambda^{-1}$.  This would seem to require starting with 
control over the extreme infrared in spacetime, the limit 
$L\rightarrow \infty$ at zero two dimensional distance.  But, in the 
most favorable circumstance, it might be possible to postpone the 
problem of controlling the limit $L\rightarrow \infty$.  The limit 
$L\rightarrow \infty$ might become purely a question of principle.  
At any finite value of $L$, it might be possible to obtain the 
spacetime physics of a local region of spacetime, on the distance 
scale set by $L$, in terms of the spacetime fields over a bounded 
neighborhood of that region.  The spacetime wave modes $\lambda^{i}$ 
in the extreme infrared, however many there are, will only make 
their influence known through the values of the local spacetime 
fields in the spacetime neighborhood under study.  The two 
dimensional cutoff distance can be removed, so far as the local 
spacetime physics is concerned, without needing to worry about the 
possibly infinite number of spacetime wave modes in the extreme 
infrared.

Formally, the scale invariance of the effective lambda model allows 
the two dimensional cutoff distance $\Lambdazero^{-1}$ to be taken 
to zero.  According to the principle of tandem renormalization, the 
effective lambda model and the effective general nonlinear model 
evolve together in the two dimensional distance $\Lambda^{-1}$.  It 
follows that the effective lambda model is automatically scale 
invariant, in the generalized sense.  This {\em tautological scale 
invariance} means that any divergent dependence on the two 
dimensional cutoff distance $\Lambdazero^{-1}$ can be absorbed into 
a change of integration variable in the functional integral defining 
the lambda model.  The lambda model is thus finite in the limit 
$\Lambdazero^{-1}\rightarrow 0$.  This formal argument should be 
effective in calculations of spacetime physics at finite spacetime 
distance $L$, allowing the two dimensional cutoff to be removed in 
such calculations.  But it will be necessary to control the extreme 
infrared spacetime wave modes in arbitrarily large spacetimes, 
before it will be possible to establish the tautological generalized 
scale invariance of the lambda model at asymptotically short two 
dimensional distance.  The existence of the limit $L\rightarrow 
\infty$ will need to be established in order to make the foundation 
of the theory secure.

%
%
%
%
%
\sectiono{The infrared divergence in string theory}
\label{sect:divergence}

The infrared failure of string theory, which is due to the existence 
of a manifold of possible background spacetimes, expresses itself as 
a technical disease of the string worldsurface, a divergence at 
short two dimensional distance.  Each degenerating handle in the 
string worldsurface depends logarithmically on the two dimensional 
cutoff distance.  The logarithmic divergence is due to marginal 
scaling fields flowing as string states through the degenerating 
handle.  The divergence is in the infrared in spacetime, because the 
marginal scaling fields of the general nonlinear model correspond to 
the zero modes of the spacetime fields that characterize spacetime.  
The marginal scaling fields of the general nonlinear model generate 
the infinitesimal variations in the manifold of spacetimes.  The 
infrared divergence occurs precisely when there is infinitesimal 
continuous degeneracy of the possible background spacetimes.  The 
divergence occurs because of the existence of a continuous manifold 
of spacetimes.  A background spacetime cannot be chosen from the 
manifold of possible background spacetimes, because perturbative 
string theory is divergent in any one of the manifold of possible 
background spacetimes.

\subsection{A degenerating handle}

Suppose that a background spacetime is chosen, satisfying $\beta = 0$ 
at all spacetime distances.  The string worldsurface is described by 
an exactly scale invariant general nonlinear model.  Consider a 
degenerating handle in the scale invariant worldsurface.

A handle is a tube that connects the worldsurface to itself.  Making 
a transverse cut through the tube displays the handle to be formed 
by gluing together the boundaries of two holes in the worldsurface.  
The two holes can be anywhere on the worldsurface.  The handle 
degenerates when the two holes shrink, each to a single point.  The 
limit is a node, consisting of two distinct points on the 
worldsurface identified together as a single point.

A degenerating handle is parametrized by its two endpoints on the 
worldsurface, $z_{1}$ and $z_{2}$, and by a complex number $q$.  The 
absolute value of $q$ measures the thickness of the handle.  Each 
hole has radius $\abs{q}^{1/2}$.  The phase of $q$ measures the 
twist imparted when the two holes in the worldsurface are glued to 
form the handle.  The endpoints $z_{1}$ and $z_{2}$ are the centers 
of the two holes.  Let $w_{1}$ and $w_{2}$ be complex coordinates 
for the two regions of the worldsurface where the holes are located.  
The first hole is formed by removing the disk 
$\abs{w_{1}-z_{1}}<\abs{q}^{1/2}$ from the first region, the second 
hole by removing the disk $\abs{w_{2}-z_{2}}<\abs{q}^{1/2}$ from the 
second region.  The boundaries of the two resulting holes are 
identified by the equation $(w_1-z_1)(w_2-z_2)=q$.  The result is an 
almost degenerate handle whose complex structure is parametrized by 
the two points, $z_{1}$ and $z_{2}$, and by the complex number $q$ 
which lies near $0$.  At $q=0$, the handle degenerates to a node.  
The two points $z_{1}$ and $z_{2}$ become identified together as a 
single point.

A degenerating handle can also be pictured as a long tube of length 
$- \ln \abs{q}$ connecting the two regions of the worldsurface.  The 
long tube is parametrized by the complex coordinate $u = \ln 
(w_{1}-z_{1})-\frac12 \ln q = - \ln (w_{1}-z_{1}) + \frac12 \ln q $.  
In this view, the degenerating handle represents string states 
propagating between two regions of the worldsurface in the very long 
world time $- \ln \abs{q}$, during which the string can explore the 
largest distances in spacetime.

A string scattering amplitude is calculated in perturbative string 
theory by integrating the partition function of the worldsurface 
with respect to all the parameters of its complex structure.  The 
worldsurface partition function is non-singular in the integration 
parameters, except where a handle in the worldsurface degenerates to 
a node.  Only there can the integral diverge.

\subsection{The contribution of a degenerating handle}

The contribution of a degenerating handle to the worldsurface 
partition function is made explicit by summing over a complete set 
of string states flowing through the handle.  Each end of the handle 
is the boundary circle of a hole in the worldsurface.  A string 
state flowing through an end of the handle shows itself on the 
worldsurface as a boundary condition on the boundary of the hole.  A 
hole in the worldsurface with a boundary condition on the boundary 
of the hole is a local field in the worldsurface.  The local fields 
in a scale invariant worldsurface are linear combinations of the 
scaling fields.  Each sum over string states flowing through an end 
of the handle is a sum over scaling fields inserted at the point in 
the worldsurface where the end of the handle is attached.  The 
integral over the phase of $q$ eliminates the scaling fields of 
nonzero spin, leaving only spin 0 scaling fields at the ends of the 
handle.

Summing over string states flowing through a degenerating handle 
replaces the handle with a double insertion of scaling fields in the 
worldsurface,
\eqa
\label{eq:degen-handle}
\frac12 
\int \dif^{2}z_1\,\mu^{2} \frac1{2\pi}
\;
\int \dif^{2}z_2\,\mu^{2} \frac1{2\pi}
\;
\int \dif^{2}q\,\mu^{4} \frac1{2\pi}
\; 
(\mu \abs{q}^{1/2})^{-8}
\nonumber \\[2ex]
\phi_{i}(z_1,\bar z_1) \: (\mu \abs{q}^{1/2})^{2+\ad(i)} \; T g^{ij} 
\; (\mu \abs{q}^{1/2})^{2+\ad(j)} \, \phi_{j}(z_2,\bar z_2) \:.  
\ena
The $\phi_{i}(z,\bar z)$ form a complete set of linearly independent 
spin 0 scaling fields, normalized at two dimensional distance 
$\mu^{-1}$.  The scaling dimension of the field $\phi_{i}$ is 
$2+\ad(i)$, the anomalous dimension is $\ad(i)$.  The sums over 
indices $i,j$ are sums over the string states flowing through the 
two ends of the handle.  The summation convention is used for 
indices $i,j$ in place of explicit sums.  The scaling field 
$\phi_{i}(z_{1},\bar z_{1})$ represents the string state flowing 
through the handle at endpoint $z_{1}$.  The scaling field 
$\phi_{j}(z_2,\bar z_2)$ represents the string state flowing through 
the handle at endpoint $z_{2}$.  The endpoints $z_{1}$ and $z_{2}$ 
range over the entire worldsurface.

The factor $(\mu \abs{q}^{1/2})^{2+\ad(i)}$ scales the field 
$\phi_{i}$ from the circle of radius $\mu^{-1}$ to the circle of 
radius $\abs{q^{1/2}}$.  Similarly, $\phi_{j}$ is scaled by $(\mu 
\abs{q}^{1/2})^{2+\ad(j)}$.

The factor $(\mu \abs{q}^{1/2})^{-8}$ is for two dimensional scale 
invariance.  The gluing equation $(w_1-z_1)(w_2-z_2)=q$ is left 
invariant when $w_{1}$, $w_{2}$, $z_{1}$, $z_{2}$, and $q^{1/2}$ are 
simultaneously scaled by the same scaling factor, so the integral 
over the handle parameters must also be invariant.

There is an overall factor $1/2$ because the two ends of the handle 
are indistinguishable.  The factor $1/2\pi$ in each two dimensional 
integral is conventional.

The matrix $T\,g^{ij}$ implements the gluing of the two boundary 
circles to form the handle, tying together the string states passing 
through the two ends of the handle.  A more specific identification 
of the gluing matrix $T\,g^{ij}$ is given below, in 
section~\ref{sect:metric-coupling}.  One property is needed now, the 
fact that $T\,g^{ij}=0$ if $\ad(i) \ne \ad(j)$.  This follows from 
scale invariance of the gluing process when the two holes are scaled 
inversely.

The handle degenerates at $q=0$.  The integral over the parameter 
$\abs{q}$ diverges near $q=0$ if and only if there are spin 0 
scaling fields $\phi_{i}(z,\bar z)$ with anomalous scaling dimension 
$\ad(i)=0$.  These are the {\em marginal} scaling fields, the 
scaling fields with scaling dimension exactly equal to $2$.  The 
spin 0 scaling fields are the possible infinitesimal variations
\eq
\epsilon \int \dif^{2}z \frac1{2\pi} \, \phi_{i}(z,\bar z)
\en
of the action of the general nonlinear model.  The marginal scaling 
fields are the infinitesimal variations that preserve two 
dimensional scale invariance.  The marginal scaling fields are the 
infinitesimal variations of the background spacetime.  So a 
degenerating handle produces a divergence if and only if there is an 
infinitesimal continuous degeneracy in the set of possible 
background spacetimes.

The divergence is in the infrared in spacetime, because the marginal 
scaling fields correspond to the zero modes of the spacetime field 
equations $\beta =0$.  Also, a handle with thickness parameter $q$ 
near $0$ describes a string propagating for very long world time, 
exploring the largest distances in spacetime.

In order to regulate the perturbative string theory, integrals over 
worldsurface parameters are cut off at a short two dimensional 
distance $\Lambdazero^{-1}$.  In particular, the radius of the hole 
at each end of a degenerating handle is bounded below by 
$\Lambdazero^{-1}$.  The integral over $q$ in 
equation~\ref{eq:degen-handle} is regulated by the cutoff 
$\abs{q}^{1/2} > \Lambdazero^{-1}$.  The cutoff dependence is 
extracted by integrating up to some limit $\abs{q}^{1/2}= 
\Lambda_{1}^{-1}$, where $\Lambda_{1}^{-1}$ is a fixed worldsurface 
distance, independent of the cutoff.  The cutoff dependent part of 
the handle insertion, integral~\ref{eq:degen-handle}, becomes
\eqa
\lefteqn{\int \dif^{2}z_1\,\mu^{2} \frac1{2\pi}
\int \dif^{2}z_2\,\mu^{2} \frac1{2\pi}
}
\nonumber \\[2ex]
&&
\phi_{i}(z_1,\bar z_1)
\;
T\,g^{ij}\; 
\left [
\frac{(\mu^{2}\Lambda_{1}^{-2})^{\ad(i)}- 
(\mu^{2}\Lambdazero^{-2})^{\ad(i)}} {\ad(i)}
\right ]
\;
\phi_{j}(z_2,\bar z_2)
\:.
\ena
The expression
\eq
T\,g^{ij}\; 
\left [
\frac{(\mu^{2}\Lambda_{1}^{-2})^{\ad(i)}- 
(\mu^{2}\Lambdazero^{-2})^{\ad(i)}} {\ad(i)}
\right ]
\en
is the string propagator of the large distance string modes, with 
the two dimensional short distance cutoff acting as infrared 
regularizer in spacetime.  The infrared spacetime cutoff distance 
$L_{0}$ is given by
\eq
L_{0}^{2}=\ln(\Lambdazero/\mu)
\:.
\en
For small $\ad(i)$, the string propagator behaves as
\eq
T\,g^{ij}\, \frac1{\ad(i)}
\en
until $\ad(i)$ becomes smaller than $L_{0}^{-2}$.  Then the pole is 
regularized, becoming
\eq 
T\,g^{ij}\,\ln(\Lambdazero^{2}\Lambda_{1}^{-2})
\en
a logarithm of the two dimensional cutoff distance 
$\Lambdazero^{-1}$.  The two dimensional cutoff distance, by acting 
as the short distance cutoff in worldsurface integrals, cuts off the 
propagator of the string modes at infrared spacetime distance 
$L_{0}$.

If the spacetime is held fixed, the two dimensional cutoff distance 
$\Lambdazero^{-1}$ can be taken so close to zero that there is 
cutoff dependence only when $\ad(i)=0$.  The divergent part of the 
handle insertion, integral~\ref{eq:degen-handle}, then becomes
\eq
\label{eq:log-divergence}
\frac12 
\int \dif^{2}z_1 \mu^{2} \, \frac1{2\pi}
\;
\int \dif^{2}z_2 \mu^{2} \, \frac1{2\pi}
\;
\phi_{j}(z_2,\bar z_2)
\;
T\,g^{ij}
\;
\ln (\Lambdazero^2 \Lambda_{1}^{-2})
\;
\phi_{i}(z_1,\bar z_1)
\en
where now the indices $i,j$ range only over the marginal scaling 
fields.  They will continue to do so until further notice.

The restriction to marginal scaling fields $\phi_{i}(z,\bar z)$ will 
have to be relaxed, because it will become untenable to assume that 
$\Lambdazero^{-1}$ can be taken to zero with the spacetime held 
fixed.

\subsection{The effects of the divergence at short distance}

The divergence is a symptom of deficiency in the string 
worldsurface.  The divergence signals that string theory is 
incomplete, that the string worldsurface is not adequately 
formulated.  A mechanism is missing to cancel the divergence.  The 
divergence is in the spacetime infrared, so the missing mechanism 
should operate at large distance in spacetime.

The renormalization of the general nonlinear model exhibits the 
large distance spacetime physics to be encoded in the short distance 
structure of the general nonlinear model of the worldsurface.  So 
the missing mechanism should operate at short two dimensional 
distance.  But a degenerating handle is not necessarily local on the 
worldsurface.  A degenerating handle may connect two regions on the 
worldsurface which, in the absence of the handle, are distant from 
each other or even disconnected from each other.  First, it is 
necessary to isolate the divergent effects of degenerating handles 
on the short distance structure of the general nonlinear model of 
the worldsurface.  These are the effects on the large distance 
physics of spacetime.  Then a mechanism can be designed to cancel 
the divergent effects at short two dimensional distance.

The short distance structure of the worldsurface is visible in an 
arbitrary local two dimensional neighborhood.  So the short distance 
effects of degenerating handles are produced by those degenerating 
handles whose two endpoints lie in the same two dimensional 
neighborhood.  These are the {\em local} handles.  The missing 
mechanism can then be designed to cancel the divergence produced by 
the local degenerating handles.

The endpoints $z_{1}$ and $z_{2}$ of a local handle lie in the same 
two dimensional neighborhood.  The two dimensional distance 
$\abs{z_1-z_2}$ is independent of the two dimensional cutoff 
distance, so the divergent effects of a local handle can be 
extracted naturally, by putting $\Lambda_{1}^{-1}=\abs{z_1-z_2}$ as 
upper bound on the handle thickness parameter $\abs{q}^{1/2}$.  The 
availability of the two dimensional distance $\abs{z_{1}-z_{2}}$ 
between the endpoints of a local handle allows the short distance 
effects of the local handle to be isolated naturally.

Substituting $\abs{z_{1}-z_{2}}$ for $\Lambda_{1}^{-1}$ in 
equation~\ref{eq:log-divergence}, the divergent effects of a local 
handle are described by a bi-local insertion in the local two 
dimensional neighborhood
\eq
\frac12 
\int \dif^{2}z_1 \mu^{2} \, \frac1{2\pi}
\;
\int \dif^{2}z_2 \mu^{2} \, \frac1{2\pi}
\;
\phi_{j}(z_2,\bar z_2)
\;
T\,g^{ij}
\;
\ln (\Lambdazero^2 \abs{z_1-z_2}^2)
\;
\phi_{i}(z_1,\bar z_1)
\:.
\en
As mentioned, the indices $i,j$ are now ranging only over the 
marginal scaling fields, but this restriction will have to be 
lifted, because it derives from the assumption that 
$\Lambdazero^{-1}$ can be taken arbitrarily close to zero with the 
spacetime held fixed.

%
%
%
%
\sectiono{A local mechanism to cancel the divergence}

\subsection{The restricted lambda model}

The lambda model is formulated to cancel the effects of local 
handles at short two dimensional distance.  The construction of the 
lambda model will be formal, at first, because of the artificial 
restriction to marginal scaling fields $\phi_{i}(z,\bar z)$ in the 
description of the divergent effects of handles.

To cancel the divergent bi-local insertion made by a local handle, 
the marginal coupling constants $\lambda^{i}$ are made into local 
sources $\lambda^{i}(z,\bar z)$ which are coupled to the marginal 
scaling fields $\phi_{i}(z,\bar z)$ by inserting
\eq
\label{eq:lambda-marginal-insertion}
\me^{-\int \dif^{2}z \, \mu^{2} \frac1{2\pi} \,
\lambda^{i}(z,\bar z) \, \phi_i(z,\bar z)}
\en
into the general nonlinear model of the worldsurface.  Then the 
sources $\lambda^{i}(z,\bar z)$ are set fluctuating with gaussian 
propagator
\eq
\expval{\,\lambda^i(z_1,\bar z_1) \; \lambda^j(z_2,\bar z_2)\,}
= - T\,g^{ij}
\,
\ln ({\Lambdazero^2} \abs{z_{1}-z_{2}}^{2})
\:.
\en
The fluctuating sources $\lambda^{i}(z,\bar z)$, 
coupled to the marginal scaling fields $\phi_{i}(z,\bar z)$, 
produce at leading order the insertion
\eq
\frac12 
\int \dif^{2}z_1 \mu^{2} \, \frac1{2\pi}
\;
\int \dif^{2}z_2 \mu^{2} \, \frac1{2\pi}
\;
\phi_{j}(z_2,\bar z_2)
\;
\expval{\,\lambda^i(z_1,\bar z_1) \; \lambda^j(z_2,\bar z_2)\,}
\;
\phi_{i}(z_1,\bar z_1)
\en
which cancels the effects of a single local handle at two 
dimensional distances near $\Lambdazero^{-1}$.  Exponentiated, the 
insertions of the lambda propagator cancel the effects of 
arbitrarily many independent local handles on the worldsurface.  But 
multiple handles are independent only when widely separated on the 
worldsurface.  The sub-leading effects of colliding local handles 
remain to be cancelled.

The gaussian fluctuations are generated by inserting into the 
general nonlinear model of the worldsurface a functional integral 
over the sources $\lambda^{i}(z,\bar z)$
\eq
\label{eq:lambda-first}
\int \Dlambda
\;
\me^{
-\int \dif^{2}z \, \frac1{2\pi} \,
T^{-1} g_{ij}
\,
\partial \lambda^{i}
\,
\bar \partial \lambda^{j}
}
\;
\me^{-\int \dif^{2}z \, \mu^{2} \, \frac1{2\pi} \,
\lambda^{i}(z,\bar z) \phi_i(z,\bar z)}
\en
The sources $\lambda^{i}(z,\bar z)$ have become dimensionless, 
massless, scalar quantum fields.  The propagator of a massless 
scalar field in two dimensions is logarithmic, so must be normalized 
at a characteristic two dimensional distance, which is 
$\Lambdazero^{-1}$.

The sub-leading effects of colliding local handles are cancelled by 
making non-gaussian corrections to the fluctuations.  The 
corrections are generated by adding interaction terms to the 
gaussian action
\eq
\label{eq:pert-action}
S(\lambda) =
\int \dif^{2}z \, \frac1{2\pi} \,
\bigg (
T^{-1} g_{ij}
\,
\partial \lambda^{i}
\,
\bar \partial \lambda^{j}
+  T^{-1} g_{ij,k} \lambda^{k}
\,
\partial \lambda^{i}
\,
\bar \partial \lambda^{j}
+  T^{-1} g_{ij,kl} \lambda^{k} \lambda^{l}
\,
\partial \lambda^{i}
\,
\bar \partial \lambda^{j} +
\cdots
\bigg )
\:.
\en
Only local interactions are needed to cancel the short distance 
effects of the local handles.  A rough argument is that collisions 
between handles produce the insertions of scaling fields that are to 
be cancelled by the interactions, only local interactions are 
needed.  A better argument is given later.  The interaction terms 
must be scale invariant because the effects of the handles are given 
by scale invariant integrals over worldsurface parameters.  The 
interaction terms must therefore all contain two derivatives of the 
dimensionless scalar fields $\lambda^{i}(z,\bar z)$, multiplied by 
any number of scalar fields.  Infinitely many such interaction terms 
are possible.  The infinite number of coefficients are calculated, 
in principle, from the worldsurface integrals for multiple handles.  
Fortunately there is a much simpler way.

The marginal coupling constants $\lambda^{i}$ are parameters for the 
scale invariant perturbations of the reference general nonlinear 
model that was chosen initially.  The $\lambda^{i}$ are local 
coordinates on the manifold $\M{\infty}$ of scale invariant general 
nonlinear models, which is the manifold of spacetimes.  The sources 
$\lambda^{i}(z,\bar z)$ are therefore the components of a map 
$\lambda(z,\bar z)$ from the worldsurface to the manifold 
$\M{\infty}$, written in coordinates.

The reference general nonlinear model is a point $\lambda_{1}$ in 
$\M{\infty}$, the origin of the coordinate system, the point with 
coordinates $\lambda_{1}^{i}=0$.  If the sources were nonzero 
constants $\lambda^{i}(z,\bar z)=\lambda_{2}^{i}$, their effect 
would be to change the general nonlinear model to a nearby scale 
invariant general nonlinear model $\lambda_{2}$ in $\M{\infty}$.  
The fluctuating sources $\lambda^{i}(z,\bar z)$ describe spacetime 
fluctuating locally on the worldsurface.

Once spacetime is set fluctuating in two dimensions, the mechanism 
that cancels the divergence must operate within any local 
fluctuation.  Within a local fluctuation, the worldsurface might be 
in a nearby spacetime $\lambda_{2}$ in $\M{\infty}$.  The 
fluctuating spacetime $\lambda(z,\bar z)$ can be considered to be 
nearly constant, locally in two dimensions, because the fields 
$\lambda^i(z,\bar z)$ are dimensionless, and their variations in two 
dimensions are suppressed in the functional integral.

Within a local fluctuation to a spacetime $\lambda_{2}$, the short 
distance effects of local handles are given by the handle gluing 
matrix $T\, g^{ij}(\lambda_{2})$ of the general nonlinear model 
$\lambda_{2}$.  Local fluctuations around $\lambda_{2}$ will be 
needed to cancel the effects of these local handles.  The gaussian 
fluctuations around $\lambda_{2}$ will be governed by the metric 
$T^{-1}g_{ij}(\lambda_{2})$ which is the inverse of the handle 
gluing matrix in the general nonlinear model $\lambda_{2}$.  The 
non-gaussian corrections at $\lambda_{2}$ are given by an infinite 
series of interaction terms, as in equation~\ref{eq:pert-action}, 
with coefficients $T^{-1}g_{ij,k}(\lambda_{2})$, 
$T^{-1}g_{ij,kl}(\lambda_{2})$, and so on.

To cancel the effects of local handles, once spacetime is set 
fluctuating locally in two dimensions, there must be a cancelling 
set of local fluctuations around each point $\lambda$ in 
$\M{\infty}$.  For each point $\lambda$ in $\M{\infty}$, the 
cancelling mechanism is a functional integral over maps 
$\lambda(z,\bar z)$ from the worldsurface to a coordinate 
neighborhood of $\lambda$ in $\M{\infty}$.

But the fluctuations around $\lambda_{2}$ are completely determined 
by the fluctuations around $\lambda_{1}$, and {\it vice versa,} 
since one set of fluctuations is obtained from the other simply by a 
translation of coordinates in $\M{\infty}$.  The cancelling 
mechanisms for two nearby points, $\lambda_{1}$ and $\lambda_{2}$, 
must be equivalent, under the dictionary that translates sources 
$\lambda^{i}(z,\bar z)$ in $\lambda_{1}$ to equivalent sources in 
$\lambda_{2}$.  The cancelling mechanisms must operate 
simultaneously, and coherently, in all the spacetimes $\lambda$ in 
the manifold $\M{\infty}$.

A coherent collection of such functional integrals over 
dimensionless scalar fields is a two dimensional nonlinear 
model~\citeF. The target manifold of the nonlinear model is the 
manifold $\M{\infty}$.  The field of the nonlinear model is a map 
$\lambda(z,\bar z)$ from the worldsurface to $\M{\infty}$.  The 
metric coupling of the nonlinear model is completely determined by 
the gaussian fluctuations at each point $\lambda$ in $\M{\infty}$.  
The handle gluing matrix $T\, g^{ij}(\lambda)$ in each spacetime 
$\lambda$ gives all the information needed to determine the higher 
order interactions of the fluctuations.  The action of the nonlinear 
model is globally defined as a function of the map $\lambda(z,\bar 
z)$ from the worldsurface to $\M{\infty}$,
\eq
\label{eq:lambda-action}
S(\lambda) = \int \dif^{2}z \, \frac1{2\pi}
\,
T^{-1} g_{ij}(\lambda)
\,
\partial \lambda^{i}
\,
\bar \partial \lambda^{j}
\:.
\en
The coherence condition on the local fluctuations avoids a laborious 
calculation of the effects of collisions of multiple handles in a 
fixed spacetime.

The local mechanism that is inserted to cancel the divergence is the 
functional integral
\eq
\label{eq:lambda-second}
\int \Dlambda
\; \me^{- S(\lambda)} \;
\me^{-\int \dif^{2}z \, \mu^{2} \, \frac1{2\pi} \,
\lambda^{i}(z,\bar z) \phi_i(z,\bar z)}
\:.
\en
This is a two dimensional nonlinear model whose target manifold is 
the manifold of spacetimes $\M{\infty}$.  The metric coupling is the 
natural metric $T^{-1} g_{ij}(\lambda)$ on the manifold of 
spacetimes, the inverse of the handle gluing matrix.  The field is 
the {\em lambda field,} $\lambda(z,\bar z)$.  The small fluctuations 
around a given reference spacetime are described in coordinates by 
the {\em lambda fields,} $\lambda^{i}(z,\bar z)$.

This nonlinear model completely accomplishes the cancelling of the 
short distance effects of the local handles.  The argument for 
complete cancelation is based on the coherence condition over the 
manifold of spacetimes $\M{\infty}$.  Suppose that some part of the 
divergence is left uncancelled.  For each spacetime $\lambda$ in 
$\M{\infty}$, the uncancelled divergence would be cancelled by 
additional interactions among the lambda fields $\lambda^{i}(z,\bar 
z)$.  The gaussian part of the divergence is already cancelled, by 
design, so the additional interactions must be at least tri-linear 
in the lambda fields $\lambda^{i}(z,\bar z)$.  There must be 
coherence of these remaining interactions as $\lambda$ varies in 
$\M{\infty}$, so the additional interactions must involve only 
derivatives of the lambda fields.  Otherwise, varying $\lambda$ 
would produce quadratic interaction terms.  The additional 
interactions must be scale invariant in two dimensions.  Finally, 
the additional interactions cannot increase at long two dimensional 
separations because the effects being cancelled are made by handles 
in collision.  There are no interactions compatible with all these 
conditions.

The lambda fields $\lambda^{i}(z,\bar z)$ are dimensionless scalar 
fields in two dimensions.  Once they are set fluctuating, large 
fluctuations are inevitable.  Locality in two dimensions requires 
that {\em all} configurations $\lambda(z,\bar z)$ participate in the 
functional integral, not merely the configurations that can be 
represented as perturbations $\lambda^{i}(z,\bar z)$ around a 
constant spacetime $\lambda$.  The functional integral must contain 
field configurations $\lambda(z,\bar z)$ that make large excursions 
in the manifold of spacetimes.  The most interesting configurations 
will be those that wrap around nontrivial topological features in 
the manifold of spacetimes, producing semi-classical nonperturbative 
effects.

On the other hand, because the lambda fields are dimensionless 
scalars, every fluctuation can be regarded locally as almost 
constant in two dimensions.  Every fluctuation, however large, can 
be regarded as pieced together out of locally almost constant 
fluctuations.

The action $S(\lambda)$ is defined by the coherent family of actions 
for small fluctuations around constant configurations of 
$\lambda(z,\bar z)$.  But $S(\lambda)$, as given by 
equation~\ref{eq:lambda-action}, is well-defined globally, for all 
maps $\lambda(z,\bar z)$ to the manifold of spacetimes.  It does not 
depend on any choice of reference point $\lambda_{1}$ in the target 
manifold $\M{\infty}$, nor on any choice of coordinates for the 
target manifold.  The global definition of the nonlinear model is 
equivalent to the local definition.

Also needed is a global construction of the worldsurface, as a 
functional of the map $\lambda(z,\bar z)$.  Each local two 
dimensional region can be constructed by inserting sources 
$\lambda^{i}(z,\bar z)$ into some general nonlinear model, as in 
equation~\ref{eq:lambda-marginal-insertion}.  The local regions can 
then be patched together, in principle, to construct the global 
worldsurface, as a function of the map $\lambda(z,\bar z)$.  But I 
do not give any effective method for doing such patching.  Instead, 
in section~\ref{sect:d=2+epsilon}, I point out a way to avoid 
worldsurface calculations entirely.

The cancelling mechanism described by 
equation~\ref{eq:lambda-second} is the {\em restricted} lambda 
model.  It is only a formal mechanism.  It works only 
perturbatively, and only in a generic submanifold of the target 
manifold $\M{\infty}$, and only at sufficiently short two 
dimensional distances $\Lambdazero^{-1}$.  The formal nature of this 
mechanism is due to the assumption that the spacetime $\lambda$ can 
be held fixed while the two dimensional cutoff distance 
$\Lambdazero^{-1}$ is taken to zero.  Once spacetime has been set 
fluctuating in two dimensions, this assumption becomes untenable.  
The two dimensional cutoff distance $\Lambdazero^{-1}$ must be held 
fixed while the spacetime fluctuates.  It is possible that 
fluctuations of $\lambda(z,\bar z)$ will reach places in 
$\M{\infty}$ where some scaling fields $\phi_{i}(z,\bar z)$ become 
only slightly irrelevant.  Some anomalous dimensions $\ad(i)$ will 
becomve very small.  The calculation of the divergence produced by a 
local handle must then be revised to include the insertions of 
slightly irrelevant scaling fields.  To cancel the divergence, it 
will be necessary to extend the target manifold of the lambda model 
beyond $\M{\infty}$.

\subsection{Macroscopic spacetimes}

As the spacetime $\lambda$ fluctuates, it may come upon places 
within $\M{\infty}$ where spacetime becomes large in some or all of 
its dimensions.  In such a {\em macroscopic} spacetime, there are 
many large distance wave modes which are not zero modes, but which 
have $\ad(i)$ very small, small enough that the corresponding 
scaling fields $\phi_{i}(z,\bar z)$ are not suppressed by the 
factors $(\mu \Lambdazero^{-1})^{\ad(i)}$ in the effects of a local 
handle.  These scaling fields $\phi_{i}(z,\bar z)$ are almost 
marginal.  It is no longer possible to separate distinctly the 
marginal scaling fields from the irrelevant scaling fields in the 
analysis of the degenerating handle.  It is no longer possible to 
ignore as cutoff independent the contribution of the irrelevant 
scaling fields, and describe the divergence entirely in terms of the 
marginal scaling fields.  The slightly irrelevant scaling fields 
must be included in the analysis of the divergence.

Write the spacetime metric on the macroscopic dimensions of 
spacetime as $\frac1{\alphaprime} h_{\mu\nu}(x)$, explicitly 
proportional to $1/\alphaprime$.  The linear size of the macroscopic 
spacetime spacetime goes as $(\alphaprime)^{-1/2}$.  The macroscopic 
spacetime becomes infinitely large as $\alphaprime \rightarrow 0$.  
The spacetime wave operators acting on the massless spacetime fields 
are proportional to $\alphaprime$, up to corrections that are higher 
order in $\alphaprime$.  The eigenvalues $\ad(i)$ go to zero as 
$\alphaprime$.  The characteristic spacetime distances $L(i)$ of the 
massless wave modes go as $(\alphaprime)^{-1/2}$.  More and more of 
the massless wave modes $\lambda^{i}$ become almost marginal 
coupling constants in the general nonlinear model.

As spacetime fluctuates in the lambda model, the parameter 
$\alphaprime$ might approach zero.  In the handle insertion, the 
coefficient of the almost marginal scaling fields would approach a 
logarithm of the two dimensional cutoff distance $\Lambdazero^{-1}$.  
Once the spacetime is set fluctuating at fixed two dimensional 
cutoff distance, it becomes impossible to separate the marginal 
coupling constants from the irrelevant coupling constants.

The manifold of spacetimes $\M{\infty}$ can be completed, and made 
locally compact, by adding a set of points $\M{\infty}_{d}$ which 
corresponds to the limits $\alphaprime \rightarrow 0$, subject to 
some identifications in the remaining parameters of the spacetime 
metric, which lose their significance in the limit.  The completed 
manifold of spacetimes is the union
\eq
\barMinfinity = \M{\infty} \cup \M{\infty}_{d}
\:.
\en
The submanifold $\M{\infty}_{d}$ might be called the {\em locus of 
decompactification}.  The macroscopic spacetimes are the points 
$\lambda$ in $\M{\infty}$ which lie near the locus of 
decompactification.

The prototype for this completion of $\M{\infty}$ is the manifold of 
toroidal two dimensional spacetimes.  A two dimensional spacetime 
torus has Kahler metric proportional to a complex number $\sigma$.  
The spacetime volume is $\Imag(\sigma)$.  When $\Imag(\sigma)$ is 
large, the manifold of spacetimes is parametrized by the complex 
parameter $q=\me^{2\pi i \sigma}$.  The locus of decompactification 
is the single point $q=0$.  The real part of $\sigma$ loses 
significance in the general nonlinear model in the limit 
$q\rightarrow 0$.

Besides the macroscopic spacetimes, there are also exceptional 
submanifolds within $\M{\infty}$ where some scaling fields that are 
generically irrelevant become marginal.  Such a scaling fields has 
anomalous dimension $\ad(i)=0$ on the exceptional submanifold, but 
is {\em not} a tangent vector to the manifold $\M{\infty}$.  The 
beta function $\beta(\lambda)$ vanishes to first order in the 
coupling constants, but becomes nonzero at some higher order.  Near 
the exceptional submanifold, the anomalous dimension $\ad(i)$ is 
slightly larger than $0$.

There are also combinations of these phenomena, places where 
spacetime becomes macroscopic in some dimensions and goes to an 
exceptional point in other dimensions.  These are the circumstances 
under which large distance spacetime wave modes get small nonzero 
masses $m(i)$.

Call the {\it singular locus} the entire submanifold 
$\barMinfinity$ where some generically nonzero anomalous dimensions 
$\ad(i)$ go to zero, where some generically irrelevant scaling 
fields $\phi_{i}(z,\bar z)$ become marginal.  The singular locus 
includes the locus of decompactification.  Almost marginal coupling 
constants $\lambda^{i}$ occur in the spacetimes which lie near the 
singular locus.

When the almost marginal coupling constants $\lambda^{i}$ are set 
fluctuating, the manifold of spacetimes $\M{\infty}$ is {\em 
thickened} near the singular locus.  Near the singular locus, the 
manifold $\M{\infty}$ is extended in the directions parametrized by 
the slightly irrelevant coupling constants $\lambda^{i}$.

\subsection{Gaussian fluctuations of quasi-marginal sources}

The short distance effects of a degenerating local handle must be 
re-calculated, to include the scaling fields that are slightly 
irrelevant.  The reference spacetime is still assumed to be in 
$\M{\infty}$.  The general nonlinear model is still assumed to be 
scale invariant.  Again, the integral over the handle thickness 
parameter $\abs{q}$ in equation~\ref{eq:degen-handle} is bounded 
below by the two dimensional cutoff $\abs{q}^{1/2} > 
\Lambdazero^{-1}$, and above by the separation between the endpoints 
of the handle, $\abs{q}^{1/2} < \abs{z_1-z_2}$.  The complete short 
distance contribution of the local handle, after integrating over 
the parameter $q$, is the bi-local insertion
\eqa
\label{eq:almost-divergence}
\lefteqn{\frac12 
\int \dif^{2}z_1\,\mu^{2} \frac1{2\pi}
\;
\int \dif^{2}z_2\,\mu^{2} \frac1{2\pi}
}
\nonumber \\[2ex]
& &
\phi_{i}(z_1,\bar z_1)
\;
T\,g^{ij}
\;
\left [
\frac{(\mu^{2}\abs{z_{1}-z_{2}}^{2})^{\ad(i)}- 
(\mu^{2}\Lambdazero^{-2})^{\ad(i)}}
{\ad(i)}
\right ]
\;
\phi_{j}(z_2,\bar z_2)
\:.
\ena
The indices $i,j$ now range over the entire collection of marginal 
and slightly irrelevant spin $0$ scaling fields.  The logarithmic 
divergence appears in the limit $\ad(i)\rightarrow 0$.

To cancel the effects of the degenerating local handle, again insert 
local sources $\lambda^{i}(z,\bar z)$ in the worldsurface,
\eq
\label{eq:source-insertion}
\me^{-\int \dif^{2}z \, \mu^{2} \frac1{2\pi} \,
\lambda^{i}(z,\bar z) \, \phi_i(z,\bar z)}
\en
and again set the sources fluctuating with a gaussian propagator,
which now is
\eq
\label{eq:lambda-propagator}
\expval{\,\lambda^i(z_1,\bar z_1)
\; \lambda^j(z_2,\bar z_2)\,}
=  T\,g^{ij}
\;
\left [
\frac{(\mu^{2} \Lambdazero^{-2})^{\ad(i)}
 - (\mu^{2} \abs{z_{1}-z_{2}}^{2})^{\ad(i)}}
{\ad(i)}
\right ]
\:.
\en
This gaussian propagator is scale invariant, given that 
$\lambda^{i}$ has scaling dimension $-\ad(i)$.  Only the additive 
normalization constant depends on $\Lambdazero^{-1}$.

Even though the lambda fields are not all dimensionless, their 
fluctuations are still described by a two dimensional nonlinear 
model.  But now the metric coupling of the nonlinear model depends 
on the two dimensional distance.
At distances $\abs{z_{1}-z_{2}}$ close to $\Lambdazero^{-1}$, the 
lambda propagator, equation~\ref{eq:lambda-propagator}, is 
approximately
\eqa
\expval{\,\lambda^i(z_1,\bar z_1)
\; \lambda^j(z_2,\bar z_2)\,}
\approx - T\,g^{ij}\, (\mu \Lambdazero^{-1})^{2\ad(i)} \;
\ln (\Lambdazero^{2} \abs{z_{1}-z_{2}}^{2})
\:.
\ena
This is the gaussian propagator of a nonlinear model with
a metric coupling
\eq
T^{-1}g_{ij}(\Lambdazero) =
(\Lambdazero \mu^{-1})^{2\ad(i)}
\, T^{-1}g_{ij} 
\en
that varies with the two dimensional distance $\Lambdazero^{-1}$.

Now consider the lambda propagator at a two dimensional distance 
$\Lambda^{-1}$ longer than $\Lambdazero^{-1}$ but still much shorter 
than $\mu^{-1}$.  For $\abs{z_{1}-z_{2}}$ near $\Lambda^{-1}$, the 
lambda propagator is
\eqa
\expval{\,\lambda^i(z_1,\bar z_1) \; \lambda^j(z_2,\bar z_2)\,}
&\approx&
- T\,g^{ij}\, (\mu \Lambda^{-1})^{2\ad(i)} \;
\ln (\Lambda^{2} \abs{z_{1}-z_{2}}^{2})
\nonumber \\[2ex]
& & \;\;{} + T\,g^{ij}\, 
\left [
\frac{(\mu^{2} \Lambdazero^{-2})^{\ad(i)}
 - (\mu^{2} \Lambda^{-2})^{\ad(i)}}
{\ad(i)}
\right ]
\:.
\ena
After the constant term is subtracted, this is the gaussian 
propagator of a nonlinear model with metric coupling
\eq
T^{-1}g_{ij}(\Lambda) =
(\Lambda \mu^{-1})^{2\ad(i)}
\, T^{-1}g_{ij} 
\:.
\en
The constant term
\eq
T\,g^{ij}\,
\left [
\frac{(\mu^{2} \Lambdazero^{-2})^{\ad(i)}
 - (\mu^{2} \Lambda^{-2})^{\ad(i)}}
{\ad(i)}
\right ]
=
T\,g^{ij}\,
\left [
\frac{\me^{-L_{0}^{2}\ad(i)}
- \me^{-L^{2}\ad(i)}}
{\ad(i)}
\right ]
\en
makes a contribution to the renormalization of the effective lambda 
model, and to the renormalization of the effective general nonlinear 
model.  For spacetime wave modes $\lambda^{i}$ at spacetime 
distances lying between $L_{0}$ and $L$, $L^{2}\ad(i)$ is small and 
$L_{0}^{2} \ad(i)$ is large, so the constant term is
\eq
- T\,g^{ij}\, \frac{1}{\ad(i)}
\:.
\en
Except for the minus sign, this is the tree-level spacetime 
propagator for the wave modes at distances between $L_{0}$ and $L$.  
The minus sign is there because, as the two dimensional distance 
increases from $\Lambdazero^{-1}$ to $\Lambda^{-1}$, the wave modes 
from spacetime distance $L_{0}$ down to $L$ are being {\em 
integrated in}.  String theory works in the opposite direction, from 
$L$ to $L_{0}$.  In string theory, spacetime wave modes are 
integrated {\em out} by making contractions using the spacetime 
propagator, with positive sign.  The difference in sign expresses 
the cancelling between lambda fluctuations and worldsurface handles.  
Integrating out the spacetime wave modes from $L$ up to $L_{0}$ {\em 
undoes} the integrating in that is done in the lambda model, goint 
from $L_{0}$ down to $L$.

When the characteristic two dimensional distance increases from 
$\Lambdazero^{-1}$ to $\Lambda^{-1}$, the lambda propagator becomes
\eq
 T\,g^{ij}\, (\mu \Lambda^{-1})^{2\ad(i)} \;
\;
\left [
\frac{1 - (\Lambda^{2} \abs{z_{1}-z_{2}}^{2})^{\ad(i)}}
{\ad(i)}
\right ]
\en
while, in the handle insertion, the coefficient of the scaling 
fields
\eq
 T\,g^{ij}\, (\mu \Lambda^{-1})^{2\ad(i)} \;
\;
\left [
\frac{(\Lambda^{2} \abs{z_{1}-z_{2}}^{2})^{\ad(i)} - 1}
{\ad(i)}
\right ]
\:.
\en
Comparing the two expressions shows that the string worldsurface is 
at two dimensional distances longer than $\Lambda^{-1}$, while the 
lambda model operates at two dimensional distances shorter than 
$\Lambda^{-1}$.  The handle insertion makes sense in the regime 
$\abs{z_{1}-z_{2}} > \Lambda^{-1}$, where handles contribute 
positively.  In the short distance regime, $\abs{z_{1}-z_{2}} < 
\Lambda^{-1}$, the handle insertion is defined only by analytic 
continuation.  On the other hand, the lambda propagator makes sense 
for $\abs{z_{1}-z_{2}} < \Lambda^{-1}$.  There are fluctuations at 
all two dimensional distances $y$ from $\Lambdazero^{-1}$ up to 
$\Lambda^{-1}$.  The fluctuations contribute positively to the 
lambda correlations when $y>\abs{z_{1}-z_{2}}$
\eq
\label{eq:propagator-integral}
\expval{\,\lambda^i(z_1,\bar z_1) \; \lambda^j(z_2,\bar z_2)\,}
=
\int_{\Lambdazero^{-1}}^{\Lambda^{-1}}
\dif y
\; \frac2{y}
\;
\theta(y - \abs{z_{1}-z_{2}})
\;
T \, g^{ij} \, (\mu y)^{2\ad(i)}
\:.
\en
This formula for the lambda propagator parallels the integral over 
handle thickness.  It exhibits the lambda fluctuations as a random 
process indexed by the two dimensional distance,
\eq
\lambda^i(z,\bar z)
= \int_{\Lambdazero^{-1}}^{\Lambda^{-1}}
\dif y \; \lambda^i(y, z,\bar z)
\en
\eqa
\expval{\,\lambda^i(y_{1},z_1,\bar z_1) \;
\lambda^j(y_{2},z_2,\bar z_2)\,}
&=&
\theta(y_{1} - \abs{z_{1}-z_{2}})\;\delta (y_{1} - y_{2})
\nonumber \\[1ex]
&& \;\;
\;
T \, g^{ij} \, (\mu y_{1})^{\ad(i)} \, (\mu y_{2})^{\ad(j)}
\:.
\ena
The lambda propagator is defined for $\abs{z_{1}-z_{2}} > 
\Lambda^{-1}$ only by continuation.

The gaussian fluctuations are generated by inserting into the 
general nonlinear model of the worldsurface a functional integral 
over the sources $\lambda^{i}(z,\bar z)$,
\eq
\label{eq:lambda-gaussian-quasi}
\int \Dlambda
\;
\me^{
-\int \dif^{2}z \, \frac1{2\pi} \,
T^{-1} g_{ij}(\Lambda)
\,
\partial \lambda^{i}
\,
\bar \partial \lambda^{j}
}
\me^{-\int \dif^{2}z \, \mu^{2} \, \frac1{2\pi} \,
\lambda^{i}(z,\bar z) \phi_i(z,\bar z)}
\:.
\en
This is the gaussian approximation to a nonlinear model whose metric 
coupling is explicitly scale dependent.

The gaussian fluctuations of the lambda field $\lambda^{i}(z,\bar 
z)$ at two dimensional distance $\Lambda^{-1}$ are suppressed by the 
factor
\eq
(\mu\Lambda^{-1})^{\ad(i)} = \me^{-L^{2}\ad(i)}
\en
which is just the suppression of the irrelevant coupling constants 
in the renormalized general nonlinear model of the worldsurface.  
Only the quasi-marginal coupling constants fluctuate significantly, 
the coupling constants $\lambda^{i}$ with $L^{2}\ad(i)$ not larger 
than, say, $400$.  These are the coupling constants that parametrize 
the extension of $\M{\infty}$ into $\M{L}$.  The spacetime wave 
modes are cut off in the ultraviolet.  The spacetime wave modes 
$\lambda^{i}$ fluctuate only at characteristic spacetime distances 
$L(i)$, given by $L(i)^{2} = 1/\ad(i)$, that are larger than the 
ultraviolet spacetime distance $L/20$.

The fluctuations of the quasi-marginal coupling constants 
$\lambda^{i}$ extend the target manifold of the lambda model from 
$\M{\infty}$ into $\M{L}$.  The target manifold becomes the manifold 
of spacetimes $\M{L}$, which is a {\em thickening} of the manifold 
of spacetimes $\M{\infty}$ near the locus of decompactification, and 
near the rest of the singular locus.  The thickening is controlled 
in the spacetime ultraviolet at spacetime distance $L$.  The control 
is in place before $\lambda$ is set fluctuating, having been 
provided by the renormalization of the general nonlinear model.  The 
thickening is suppressed away from the singular locus by the 
renormalization of the general nonlinear model.  Away from the 
singular locus, the target manifold of the lambda model is simply 
$\M{\infty}$.  The meaning of {\it away from} is set by the 
spacetime distance $L$, which derives from the ratio $\mu 
\Lambda^{-1}$ of the two dimensional distances in the 
renormalization of the general nonlinear model.  The meaning of {\it 
near the locus of decompactification} is set by the spacetime 
distance $L$.  A macroscopic spacetime is a spacetime of linear size 
much larger than $L$.

Once the fluctuations of the quasi-marginal coupling constants 
$\lambda^{i}$ extend away from $\M{\infty}$ into $\M{L}$, the 
cancelling mechanism must act in any spacetime $\lambda$ in $\M{L}$.  
The cancelling mechanism must act on worldsurfaces described by 
general nonlinear models that are not exactly scale invariant.  In 
retrospect, it was an oversimplification to start the analysis of 
the string theory divergence in a spacetime $\lambda_{1}$ that was 
an exact solution of $\beta=0$.  At the nonzero two dimensional 
cutoff distance $\Lambdazero^{-1}$, the possible general nonlinear 
models form the manifold $\M{L_{0}}$.  The initial spacetime 
$\lambda_{1}$ should have been chosen from the manifold of 
spacetimes $\M{L_{0}}$.

\subsection{The full lambda model}

Reconsider the gaussian fluctuations in a spacetime in $\M{\infty}$, 
described by equation~\ref{eq:lambda-gaussian-quasi}.  Define 
running coupling constants
\eq
\lambdar^{i} = (\mu^{-1} \Lambda)^{\ad(i)} \lambda^{i}
\en
which couple to scaling fields
\eq
\phi^{\Lambda}_i(z,\bar z) = (\mu \Lambda^{-1})^{2+\ad(i)}
\, \phi_i(z,\bar z)
\en
which are normalized at two dimensional distance $\Lambda^{-1}$.  
The gaussian fluctuations at distance $\Lambda^{-1}$ take the form
\eq
\int \Dlambdar
\;
\me^{
-\int \dif^{2}z \, \frac1{2\pi} \,
T^{-1} g_{ij}
\,
\partial \lambdar^{i}
\,
\bar \partial \lambdar^{j}
}
\;
\me^{-\int \dif^{2}z \, \Lambda^{2} \frac1{2\pi} \,
\lambdar^{i}(z,\bar z) \phi^{\Lambda}_i(z,\bar z)}
\en
which is the same at every two dimensional distance $\Lambda^{-1}$.  
This is the generalized scale invariance of the lambda model, in the 
gaussian approximation.

The effects of the local handle also take the same form at every two 
dimensional distance $\Lambda^{-1}$, when the states flowing through 
the ends of a handle are represented by the scaling fields 
$\phi^{\Lambda}_i$.  The handle gluing matrix takes the same form 
$T\, g^{ij}$ at every two dimensional distance $\Lambda^{-1}$.  The 
bi-local handle insertion for $\abs{z_1-z_2}$ near $\Lambda^{-1}$ is
\eq
\frac12 
\int \dif^{2}z_1\,\Lambda^{2} \frac1{2\pi}
\;
\int \dif^{2}z_2\,\Lambda^{2} \frac1{2\pi}
\;
\phi^{\Lambda}_{i}(z_1,\bar z_1)
\;
T\,g^{ij}
\;
\ln ({\Lambda^2} \abs{z_1-z_2}^2)
\;
\phi^{\Lambda}_{j}(z_2,\bar z_2)
\en
at every two dimensional distance $\Lambda^{-1}$.

Now consider a spacetime $\lambda_{1}$ in $\M{L}$.  The general 
nonlinear models in $\M{L}$ near $\lambda_{1}$ are parametrized by 
the quasi-marginal coupling constants $\lambda^{i}$.  The general 
nonlinear model near $\lambda_{1}$ is given by the insertion
\eq
\me^{-\int \dif^{2}z \, \mu^{2} \frac1{2\pi} \,
\lambda^{i} \phi_i(z,\bar z)}
\en
which is interpreted as a perturbation of the general nonlinear 
model $\lambda_{1}$ with coefficients $\lambda^{i}-\lambda_{1}^{i}$,
\eq
\me^{-\int \dif^{2}z \, \mu^{2} \frac1{2\pi} \,
\lambda_{1}^{i} \phi_i(z,\bar z)}
\;
\me^{-\int \dif^{2}z \, \mu^{2} \frac1{2\pi} \,
(\lambda^{i}-\lambda_{1}^{i}) \phi_i(z,\bar z)}
\:.
\en

The renormalized general nonlinear model at short two dimensional 
distance $\Lambda^{-1}$ depends on $\Lambda^{-1}$ only through the 
running coupling constants $\lambdar^{i}(\Lambda/\mu,\lambda)$, 
which satisfy the full renormalization group equation
\eq
\Lambda \partialby{\Lambda}_{\left / \mu,\lambda\right.} \lambdar^{i}
= \beta^{i}(\lambdar)
\:.
\en
The running coupling constants couple to the two dimensional quantum 
fields $\phi^{\Lambda}_i(z,\bar z)$ that are normalized at the short 
two dimensional distance $\Lambda^{-1}$.  The general nonlinear 
models are equally well described by insertion of the running 
coupling constants
\eq
\me^{-\int \dif^{2}z \, \mu^{2} \frac1{2\pi} \,
\lambda^{i} \phi_i(z,\bar z)}
=
\me^{-\int \dif^{2}z \, \Lambda^{2} \frac1{2\pi} \,
\lambdar^{i} \, \phi^{\Lambda}_i(z,\bar z)}
\:.
\en
The scale dependence of the renormalized general nonlinear model is 
expressed by the full renormalization group equation
\eq
\left (
\Lambda \partialby{\Lambda}_{\left / \lambdar\right.}
+ \beta^{i}(\lambdar) \partialby{\lambdar^{i}}
\right )
\;
\me^{-\int \dif^{2}z \, \Lambda^{2} \frac1{2\pi} \,
\lambdar^{i} \, \phi^{\Lambda}_i(z,\bar z)}
=0
\:.
\en

Once the quasi-marginal coupling constants $\lambda^{i}$ are set 
fluctuating locally in two dimensions, there will be local 
fluctuations into spacetimes where the general nonlinear model of 
the worldsurface is not scale invariant.  The effect of a local 
handle in that region of the worldsurface is a bi-local insertion of 
local fields in the scale non-invariant general nonlinear model.

Consider a general nonlinear model $\lambda_{1}$ in $\M{L}$.  At two 
dimensional distances close to $\Lambda^{-1}$, the departure from 
scale invariance is slight, because the quasi-marginal coupling 
constants are nearly marginal.  The departure from scale invariance 
becomes significant only over a range of two dimensional distances.  
The analysis of the effects of a local handle at $\abs{z_1-z_2}$ 
near $\Lambda^{-1}$ is just as in an exactly scale invariant 
worldsurface.  The general nonlinear model $\lambda_{1}$ is 
described at two dimensional distance $\Lambda^{-1}$ by the running 
coupling constants
\eq
\lambda_{1,r}^{i} = \lambdar^{i}(\mu \Lambda^{-1},\lambda_{1})
\:.
\en
The effects of the local handle are given by the bi-local insertion
\eq
\frac12 
\int \dif^{2}z_1\,\Lambda^{2} \frac1{2\pi}
\;
\int \dif^{2}z_2\,\Lambda^{2} \frac1{2\pi}
\;
\phi^{\Lambda}_{i}(z_1,\bar z_1)
\;
T\,g^{ij}(\lambda_{1,r})
\;
\ln ({\Lambda^2} \abs{z_1-z_2}^2)
\;
\phi^{\Lambda}_{j}(z_2,\bar z_2)
\en
where $T\,g^{ij}(\lambda_{1,r})$ is the handle gluing matrix at two 
dimensional distance $\Lambda^{-1}$ in the spacetime $\lambda_{1}$.

The gaussian mechanism that cancels the effects of the handle 
consists of sources $\lambdar^{i}(z,\bar z)$ fluctuating in two 
dimensions around the constant values $\lambda_{1,r}^{i}$,
\eq
\int \Dlambdar
\;
\me^{
-\int \dif^{2}z \, \frac1{2\pi} \,
T^{-1} g_{ij}(\lambda_{1,r})
\,
\partial \lambdar^{i}
\,
\bar \partial \lambdar^{j}
}
\;
\me^{-\int \dif^{2}z \, \Lambda^{2} \frac1{2\pi} \,
\lambdar^{i}(z,\bar z) \, \phi^{\Lambda}_i(z,\bar z)}
\en
where the metric coupling $T^{-1} g_{ij}(\lambda_{1,r})$ is the 
inverse of the handle gluing metric at two dimensional distance 
$\Lambda^{-1}$ in the spacetime $\lambda_{1}$.

It remains to patch together the collection of gaussian functional 
integrals, consistently, over the manifold of spacetimes $\M{L}$.  
Again, the interactions are completely determined by the collection 
of gaussian functional integrals around the points $\lambda_{1}$ in 
$\M{L}$, and by the condition that the interactions be coherent 
under shifting of the origin of coordinates in $\M{L}$.  Again, the 
model is a nonlinear model.  The target manifold is $\M{L}$.  The 
fields of the nonlinear model are the maps $\lambdar(z,\bar z)$ from 
the worldsurface to the manifold of spacetimes $\M{L}$.

The nonlinear model, the lambda model, is the functional integral 
over maps $\lambdar(z,\bar z)$
\eq
\label{eq:lambda-model}
\int \Dlambdar
\; \me^{- S(\lambdar)} \;
\me^{-\int \dif^{2}z \, \Lambda^{2} \frac1{2\pi} \,
\lambdar^{i}(z,\bar z) \, \phi^{\Lambda}_i(z,\bar z)}
\en
with action
\eq
S(\lambdar) = \int \dif^{2}z \, \frac1{2\pi}
\,
T^{-1} g_{ij}(\lambdar)
\,
\partial \lambdar^{i}
\,
\bar \partial \lambdar^{j}
\:.
\en
Again, the coherence condition on the local fluctuations avoids a 
laborious calculation of the effects of collisions of multiple 
handles in a fixed spacetime.

The lambda model is manifestly scale invariant in the generalized 
sense, as written in terms of the field $\lambdar(z,\bar z)$.  
Re-written in terms of the field $\lambda(z,\bar z)$, the lambda 
model is
\eq
\label{eq:lambda-model-2}
\int \Dlambda
\; \me^{- S(\Lambda,\lambda)} \;
\me^{-\int \dif^{2}z \, \mu^{2} \frac1{2\pi} \,
\lambda^{i}(z,\bar z) \, \phi_i(z,\bar z)}
\en
\eq
S(\Lambda,\lambda) = \int \dif^{2}z \, \frac1{2\pi}
\,
T^{-1} g_{ij}(\Lambda, \lambda)
\,
\partial \lambda^{i}
\,
\bar \partial \lambda^{j}
\:.
\en
The metric coupling depends on the two dimensional distance 
$\Lambda^{-1}$, but only through a transformation of the target 
manifold
\eq
T^{-1}g_{ij}(\Lambda, \lambda)
=
\frac{\partial \lambdar^{k}}{\partial \lambda^{i}}
\;
T^{-1}g_{kl}(\lambdar) 
\;
\frac{\partial \lambdar^{l}}{\partial \lambda^{j}}
\:.
\en
The metric coupling satisfies the natural renormalization group 
equation
\eq
\label{eq:gij}
\left ( \Lambda \partialby{\Lambda} + \betastar \right )
\; T^{-1} g_{ij}(\Lambda,\lambda) = 0
\en
where
\eq
\betastar (T^{-1} g_{ij})
=
\beta^{k} \partial_{k} (T^{-1} g_{ij})
+ (\partial_{i}\beta ^{k}) \, T^{-1} g_{kj}
+ T^{-1} g_{ik}\,  (\partial_{j}\beta ^{k})
\en
is the infinitesimal change of the metric coupling under the flow 
generated by the vector field $\beta^{i}(\lambda)$.

The generalized scale invariance of the full lambda model is of 
course consistent with the generalized scale invariance of the 
gaussian fluctuations around a scale invariant general nonlinear 
model.  There, the linearized beta function is
\eq
\beta^{i}(\lambda) = \ad(i) \lambda^{i} + \cdots
\en
and the infinitesimal equation for generalized scale invariance is
\eq
\left ( \Lambda \partialby{\Lambda} + \ad(i) + \ad(j) \right )
\; T^{-1} g_{ij}(\Lambda) = 0
\en
in the gaussian approximation.

The lambda model differs from the nonlinear models with generalized 
scale invariance as originally contemplated\citeF, in that the {\em 
classical} metric coupling of the lambda model depends nontrivially 
on the two dimensional distance.  The flow on the target manifold, 
generated by the vector field $\beta^{i}(\lambda)$, is present 
already in the classical lambda model, instead of arising from the 
quantum corrections.

Given $\beta^{i}(\lambda)$, the metric coupling 
$T^{-1}g_{ij}(\Lambda,\lambda)$ can be determined entirely from its 
value at one specific short two dimensional distance, for example 
its value $T^{-1}g_{ij}(\Lambdazero,\lambda)$ at two dimensional 
distance $\Lambdazero^{-1}$, by integrating the renormalization 
group equation~\ref{eq:gij}.  The data that gives the couplings of 
the lambda model can be determined entirely at small two dimensional 
distance in the renormalized general nonlinear model.  The long two 
dimensional distance $\mu^{-1}$ enters only in the determination of 
the target manifold $\M{L}$, by the decoupling of the irrelevant 
coupling constants in the renormalization of the general nonlinear 
model.

The lambda model is formulated at each two dimensional distance 
$\Lambda^{-1}$ as a nonlinear model whose metric coupling depends on 
$\Lambda^{-1}$.  The metric coupling at two dimensional distance 
$\Lambda^{-1}$ governs the fluctuations of the lambda field 
$\lambda(z,\bar z)$ at that distance.  The lambda model is built up 
incrementally in the two dimensional distance, from the cutoff 
$\Lambdazero^{-1}$ to longer two dimensional distances 
$\Lambda^{-1}$, using the nonlinear model at each distance to make 
the next incremental step.  The building of the lambda model 
expresses the fundamental principle of renormalization, that 
information propagates locally in the distance scale.

\subsection{Identification of the metric coupling}
\label{sect:metric-coupling}

The metric coupling $T^{-1}g_{ij}$ of the lambda model is defined as 
the inverse of the handle gluing matrix, $T\,g^{ij}$, because the 
lambda model is designed to cancel the effects of the local handles.  
But the handle gluing matrix is not a directly accessible object in 
the general nonlinear model of the worldsurface.  For calculation, 
it is useful to express the handle gluing matrix in terms of more 
usual field theory quantities.

First consider a scale invariant general nonlinear model.  Make a 
worldsurfurce by connecting a pair of 2-spheres to each other by a 
handle.  This worldsurface is equivalent to a single 2-sphere.  
Place a scaling field $\phi_{i}(z_{1},\bar z_{1})$ in one of the 
2-spheres, and a second scaling field $\phi_{j}(z_{2},\bar z_{2})$ 
in the other 2-sphere.  Calculate the partition function function of 
the worldsurface, summing over scaling fields at the ends of the 
handle, at points $z_{3}$ and $z_{4}$.  Schematically, the result is
\eq
\Z\expval{\,\phi_i(\mathrm{1}) \; \phi_k(\mathrm{3})\,}
\;T\,g^{kl}\;
\Z\expval{\,\phi_l(\mathrm{4}) \; \phi_j(\mathrm{2})\,}
\:.
\en
This can also be calculated as the partition function of the single 
2-sphere containing the two scaling fields,
\eq
\Z\expval{\,\phi_i(\mathrm{1}) \; \phi_j(\mathrm{2})\,}
\:.
\en
The equivalence of the two calculations implies that the metric 
coupling is identical to the un-normalized two point expectation 
value of the scaling fields at separation 
$\abs{z_{1}-z_{2}}=\mu^{-1}$,
\eq
T^{-1} g_{ij} =
\Z\expval{\,\phi_i(z_{1},\bar z_{1}) \; \phi_j(z_{2},\bar z_{2})\,}
\:.
\en
The two point expectation value at arbitrary separation is
\eq
\Z\expval{\,\phi_i(z_{1},\bar z_{1})
\; \phi_j(z_{2},\bar z_{2})\,}
= T^{-1} g_{ij}
\;
(\mu \abs{z_1-z_2})^{-2-\ad(i)-2-\ad(j)}
\:.
\en
The scale-dependent metric $T^{-1} g_{ij}(\Lambda)$ is given by the 
two point expectation value at separation 
$\abs{z_{1}-z_{2}}=\Lambda^{-1}$
\eq
T^{-1} g_{ij}(\Lambda) =
\Z\expval{\,\phi_i(z_{1},\bar z_{1})
\; \phi_j(z_{2},\bar z_{2})\,}
\;
(\mu^{-1}\Lambda)^{4}
\en
\eq
T^{-1} g_{ij} = \Z\expval{\,\phi^{\Lambda}_i(z_{1},\bar z_{1}) \; 
\phi^{\Lambda}_j(z_{2},\bar z_{2})\,}
\:.
\en

These un-normalized expectation values are normalized to geive 
ordinary expectation values.  The normalizing factor is the 
partition function without insertions
\eq
\Z\expval{1} = T^{-1}
\en
which is the factor $T^{-1}$ in the metric coupling $T^{-1}g_{ij}$.
The normalized metric $g_{ij}$ is identical to the 
normalized two point expectation value
\eq
g_{ij} =
\expval{\,\phi^{\Lambda}_i(z_{1},\bar z_{1})
\; \phi^{\Lambda}_j(z_{2},\bar z_{2})\,}
\en
at $\abs{z_{1}-z_{2}}=\Lambda^{-1}$.
Equivalently, the normalized metric
is the coefficient of the identity operator in the 
operator product expansion
\eq
\phi^{\Lambda}_i(z_{1},\bar z_{1})
\; \phi^{\Lambda}_j(z_{2},\bar z_{2})
=
(\Lambda \abs{z_1-z_2})^{-2-\ad(i)-2-\ad(j)}
g_{ij} \, 1
+ \cdots
\:.
\en

When the spacetime is macroscopic, it makes sense to calculate the 
volume $V$ of spacetime, which is a large number.  The spacetime 
coupling constant $\gst$ in the macroscopic spacetime is given by
\eq
T^{-1} = \gst^{-2} \,V
\:.
\en
The number $V$ is the factor in the partition function 
$\Z\expval{1}$ that comes from the integral over the zero mode of 
the worldsurface position $x^{\mu}(z,\bar z)$ in the macroscopic 
spacetime.  The metric coupling at the macroscopic spacetime is
written
\eq
T^{-1} g_{ij} = \gst^{-2} \,V \, g_{ij}
\:.
\en
The metric $V \, g_{ij}$ is properly normalized to be a local inner 
product on the wave modes of the spacetime fields, an integral over 
the macroscopic spacetime of a product of the two spacetime wave 
modes.

From the two dimensional point of view, the metric coupling should 
be written $T^{-1} g_{ij}$, because this form makes sense in the 
general spacetime, macroscopic or not.  The form $\gst^{-2} \, V\, 
g_{ij}$ only makes sense when there is a macroscopic spacetime, in 
which case it is the appropriate form for expressing effects that 
are local in the macroscopic spacetime.

Now consider a general spacetime $\lambda$ in $\M{L}$.  The general 
nonlinear model $\lambda$ is not scale invariant.  Again, the 
departure from scale invariance is not significant at two 
dimensional distances $\abs{z_{1}-z_{2}}$ close to $\Lambda^{-1}$.  
The argument identifying the gluing matrix can be repeated, since it 
depends only on the properties of the worldsurface at two 
dimensional distance $\Lambda^{-1}$, where the worldsurface appears 
scale invariant to a first approximation.  The metric coupling at 
two dimensional distance $\Lambda^{-1}$, the inverse of the handle 
gluing matrix, is again identified with $T^{-1}$ times the 
coefficient of the identity in the operator product
at $\abs{z_{1}-z_{2}}=\Lambda^{-1}$,
\eq
\phi^{\Lambda}_i(z_{1},\bar z_{1})
\;  \phi^{\Lambda}_j(z_{2},\bar z_{2})
=
g_{ij}(\lambdar) \, 1
+ \cdots
\:.
\en
The metric coupling is also again given by the un-normalized two 
point expectation value at $\abs{z_{1}-z_{2}}=\Lambda^{-1}$
\eq
T^{-1} g_{ij}(\lambdar) = \Z\expval{\, 
\phi^{\Lambda}_{i}(z_{1},\bar z_{1})
\; \phi^{\Lambda}_{j}(z_{2},\bar z_{2})\,} 
\en
but this formula obscures the crucial point that the metric coupling 
$T^{-1} g_{ij}(\lambdar)$ is a purely short distance property of the 
worldsurface, because its inverse, the handle gluing matrix, is a 
purely short distance property of the worldsurface.

Except for the factor $T^{-1}$, the metric coupling $T^{-1} 
g_{ij}(\lambdar)$, defined as the inverse of the handle gluing 
matrix, is identified with the intrinsic metric on the space of two 
dimensional quantum field theories used to prove the gradient 
property of 
$\beta^{i}(\lambda)$~\cite{Zamolodchikov-1,Zamolodchikov-2}.

$T$ is taken to be a fixed number.  It will have to be fixed at an 
extremely small numerical value, if the volume $V$ of macroscopic 
spacetime is to turn out proportional to $T^{-1}$.  I leave 
untouched the question of whether the value of $T$ is fixed by a 
dynamical mechanism within the lambda model.

%
%
%
%
\sectiono{$d = 2 + \epsilon$ dimensions}
\label{sect:d=2+epsilon}

The first step in calculating the quantum corrections to the scaling 
behavior of the lambda model is to calculate the scale variation of 
the renormalized general nonlinear model in the presence of sources 
$\lambda^{i}(z,\bar z)$ at a short two dimensional distance 
$\Lambda^{-1}$.  The method is to calculate to second order in the 
sources, in an arbitrary spacetime $\lambda_{1}$ in $\M{L}$,
then patch together the results to get the result.

First assume a spacetime in $\M{\infty}$.  The general nonlinear 
model is scale invariant.  Insert sources $\lambda^{i}(z,\bar z)$.  
If the sources were constant, there would be no dependence on 
$\Lambda^{-1}$.  Renormalization eliminates all dependence on the 
short two dimensional distance.  The variation with respect to 
$\Lambda^{-1}$ depends only on the derivatives of the sources 
$\lambda^{i}(z,\bar z)$.

The calculation is done to second order in the sources, keeping only 
terms containing derivatives of the sources, giving
\eqa
\label{eq:scale-variation-Minfinity}
\Lambda \partialby{\Lambda}_{\lambda} \; 
\me^{-\int \dif^{2}z \, \mu^{2} \, \frac1{2\pi} \, 
\lambda^{i}\phi_i} =
\int \dif^{2}z \, \frac1{2\pi}
\left (- \frac12 T \right )
T^{-1} g_{ij}(\Lambda)
\,
\partial \lambda^{i}
\,
\bar \partial \lambda^{j}
\:.
\ena
The computation is
\begin{displaymath}
\Lambda \partialby{\Lambda}_{\left / \lambda \right .} \;
\frac12 \int \dif^{2}z_{1} \, \mu^{2} \, \frac1{2\pi} \,
\int \dif^{2}z_{2} \, \mu^{2} \, \frac1{2\pi} \,
\theta(\abs{z_{1}-z_{2}}-\Lambda^{-1})
\; \lambda^{i}(\mathrm{1}) \lambda^{j}(\mathrm{2}) 
\phi_{i}(\mathrm{1}) \phi_{j}(\mathrm{2})
\end{displaymath}
\eqa
&=&  
\frac12 \int \dif^{2}z_{1} \, \mu^{2} \, \frac1{2\pi} \,
\int \dif^{2}z_{2} \, \mu^{2} \, \frac1{2\pi} \,
\Lambda^{-1} \delta(\abs{z_{1}-z_{2}}-\Lambda^{-1})
\nonumber \\[1ex]
&& \qquad \lambda^{i}(\mathrm{1}) \lambda^{j}(\mathrm{2}) 
g_{ij} (\mu \abs{z_{1}-z_{2}})^{-2-\ad(i)-2-\ad(j)}
\nonumber \\[2ex]
&=& 
\int \dif^{2}z \, \frac1{2\pi} \,
\left (- \frac12 \right )
(\Lambda \mu^{-1})^{\ad(i)+\ad(j)}
g_{ij}
\,
\partial \lambda^{i}
\,
\bar \partial \lambda^{j}
\nonumber \\[2ex]
&=& \int \dif^{2}z \, \frac1{2\pi} \, \left (- \frac12 T \right ) 
T^{-1} g_{ij}(\Lambda) \, \partial \lambda^{i} \, \bar \partial 
\lambda^{j} \:. 
\ena
The two scaling fields $\phi_{i}\phi_{j}$ are replaced by their 
expectation value because all other operators that contribute to the 
product are down by powers of $\Lambda^{-1}$.  The sources are 
assumed to vary only locally in two dimensions, which allows 
integration by parts.

Rewritten in terms of the running coupling constants $\lambdar^{i}$, 
and the sources $\lambdar^{i}(z,\bar z)$, the scale variation is
\eq
D = \Lambda \partialby{\Lambda}_{\left /  \lambda \right .}
= \Lambda \partialby{\Lambda}_{\left /  \lambdar\right .} + 
\beta^{i}(\lambdar) \partialby{\lambdar^{i}}
\:.
\en

\eq
D \; \me^{-\int \dif^{2}z \, \Lambda^{2} \, \frac1{2\pi} 
\, \lambdar^{i} \phi^{\Lambda}_i}
=
\int \dif^{2}z \, \frac1{2\pi} \,
\left (- \frac12 T \right )
T^{-1} g_{ij}
\,
\partial \lambdar^{i}
\,
\bar \partial \lambdar^{j}
\en

Next, the calculation is repeated for a general nonlinear model 
$\lambda_{1}$ in $\M{L}$, using the approximate scale invariance at 
two dimensional distances $\abs{z_{1}-z_{2}}$ near $\Lambda^{-1}$.  
Then the quadratic calculations are patched together coherently over 
$\M{L}$ to get the full scale variation formula
\eqa
D \; \me^{-\int \dif^{2}z \, \Lambda^{2} \, \frac1{2\pi} 
\, \lambdar^{i}(z,\bar z) \phi^{\Lambda}_i(z,\bar z)}
&=&
\me^{-\int \dif^{2}z \, \Lambda^{2} \, \frac1{2\pi} 
\, \lambdar^{i}(z,\bar z) \phi^{\Lambda}_i(z,\bar z)}
\nonumber \\[1ex]
\label{eq:global-scale-variation}
& & \;\; \int \dif^{2}z \, \frac1{2\pi} \,
\left (- \frac12 T \right )
T^{-1} g_{ij}(\lambdar)
\,
\partial \lambdar^{i}
\,
\bar \partial \lambdar^{j}
\:.
\ena
The result is a correction to the metric coupling 
$T^{-1}g_{ij}(\lambdar)$ of the lambda model, which is proportional 
to the metric coupling itself, with a coefficient $T/2$.  That is, 
the entire effect of the general nonlinear model on the scaling 
behavior of the lambda model is to give the metric coupling a 
scaling dimension of $T/2$.

If the lambda model were continued from two dimensions to dimension 
$d=2+T/2$, the same effect would be obtained.  The metric coupling 
of a nonlinear model in $d=2+\epsilon$ dimensions has scaling 
dimension $\epsilon$.

The general nonlinear model can now be dispensed with.  The lambda 
model interacting with the general nonlinear model is equivalent to 
the lambda model by itself in dimension $2+T/2$.  This technical 
device gives a way to avoid the technical difficulty of calculating 
properties of the general nonlinear model in the presence of sources 
when the lambda field $\lambda(z,\bar z)$ makes large excursions in 
the manifold of spacetimes.

%
%
%
%
%
\sectiono{A formula for $S(\lambda)$}
\label{sect:formula}

\subsection{The global scale variation formula}

The scale variation of the general nonlinear model in the presence 
of sources, equation~\ref{eq:global-scale-variation}, gives a 
formula for the action functional $S(\lambdar)$ of the lambda model,
\eq
\label{eq:action-scale-variation}
-2 \,T^{-1} \, D \;
\me^{-\int \dif^{2}z \, \Lambda^{2} \, \frac1{2\pi} \,
\lambdar^{i} \phi^{\Lambda}_i}
=
\me^{-\int \dif^{2}z \, \Lambda^{2} \, \frac1{2\pi} \,
\lambdar^{i} \phi^{\Lambda}_i}
\; S(\lambdar)
\:.
\en
Calculating $S(\lambdar)$ from the scale variation of the general 
nonlinear model is equivalent to calculating the short distance 
effects of the local handle, then designing the lambda propagator to 
cancel those effects, then finding the action $S(\lambdar)$ that 
produces the needed lambda propagator.  The equivalence between the 
two procedures for finding $S(\lambdar)$ rests on the identification 
of the handle gluing matrix $T\,g^{ij}$ with the inverse of the 
un-normalized 2-point expectation value of the fields $\phi_{i}$.

This is only a formula for $S(\lambdar)$.  It does not explain the 
lambda model as a mechanism.  The scale variation formula only 
happens to give the same result as the handle calculation.  But such 
a simple expression as equation~\ref{eq:action-scale-variation} for 
the action of the lambda model invites speculation that the formula 
has a deeper explanation.  Perhaps there is a way of writing the 
integral over string worldsurface parameters which makes it obvious 
that the effects of local handles are cancelled by the lambda model 
as defined by the scale variation formula.  Perhaps there is a more 
complete model of the string worldsurface in which the lambda model 
does not have to be inserted by hand, but arises automatically.

Such speculations are not immediately useful.  For now, it is enough 
to design the lambda model in order to cancel the local handles, and 
use the scale variation formula as an effective way to calculate the 
action functional $S(\lambdar)$ from the short distance properties 
of the general nonlinear model.

\subsection{The local scale variation and the gradient property}

Let the characteristic two dimensional distance $\Lambda^{-1}$ vary 
in two dimensions, defining a riemannnian metric
\eq
\dif s^{2} = \Lambda(z,\bar z)^{2} \, \abs{dz}^{2}
\en
with scalar curvature density
\eq
\Lambda^{2} \, R_{2}(\Lambda)(z,\bar z)
= {-}4 \partial\bar \partial\, \ln (\Lambda^{2})
\:.
\en
The general nonlinear model is renormalized locally in each two 
dimensional neighborhood.  The renormalization depends covariantly 
on the two dimensional riemannian metric.  The characteristic short 
two dimensional distance is $\Lambda(z,\bar z)^{-1}$ at the point 
$z$.

The local scale derivative is
\eq
D(z,\bar z)= 
\Lambda(z,\bar z)
\partialby{\Lambda(z,\bar z)}_{\left / \lambdar \right .}
+\beta^{i}(\lambdar(z,\bar z))
\,
\partialby{\lambdar^{i}(z,\bar z)}
\en
\eq
D = \int \dif^{2}z \, D(z,\bar z)
\:.
\en
The local scale variation of the general nonlinear model must be 
dimensionless, it must be covariant in the two dimensional metric, 
it must be a local functional of $\Lambda(z,\bar z)$ and 
$\lambdar(z,\bar z)$, and it must vanish if both $\Lambda(z,\bar z)$ 
and $\lambdar(z,\bar z)$ are locally constant.  It must take the 
form
\eqa
D(z,\bar z) \; \me^{-\int  \Lambda^{2} \, \frac1{2\pi} 
\, \lambdar^{i}\phi^{\Lambda}_i}
&=&
\me^{-\int  \Lambda^{2} \, \frac1{2\pi} 
\, \lambdar^{i}\phi^{\Lambda}_i}
\\[2ex]
\label{eq:local-scale-variation}
&& \frac1{2\pi} \,
\left (- \frac12 T \right )
\left [
T^{-1} g_{ij}(\lambdar)
\,
\partial \lambdar^{i}
\,
\bar \partial \lambdar^{j}
- \frac14 \Lambda^{2} R_{2}(\Lambda) T^{-1}a(\lambdar)
\right ]
\nonumber 
\ena
where $T^{-1}a(\lambdar)$ is some function on $\M{L}$.

The local scale derivatives commute,
\eq
{}[\,D[\epsilon_{1}], \, D[\epsilon_{2}] \,] = 0
\en
where
\eq
D[\epsilon] = \int \dif^{2}z \, \epsilon(z,\bar z)\,
D(z,\bar z)
\:.
\en
The commutator of two local scale derivatives acting on the
general nonlinear model is
\[
{}[\,D[\epsilon_{1}], \, D[\epsilon_{2}] \,]
\;
\me^{-\int  \Lambda^{2} \, \frac1{2\pi} 
\, \lambdar^{i}\phi^{\Lambda}_i}
=
\me^{-\int  \Lambda^{2} \, \frac1{2\pi} 
\, \lambdar^{i}\phi^{\Lambda}_i}
\;
\]
\[
\frac12 T
\,
\int \dif^{2}z \, \frac1{2\pi}
\;
\bigg\{
(\epsilon_{1} \partial \epsilon_{2}
-\epsilon_{2} \partial \epsilon_{1})
\left [
\beta^{i} \, T^{-1}g_{ij}(\lambdar) \, \bar \partial \lambdar^{j}
-  \bar \partial \, T^{-1}a(\lambdar)
\right ]
\]
\vskip0.5ex
\eq
\mbox{} +
\left [
\partial \lambdar^{i}\,
T^{-1}g_{ij}\, \beta^{j}(\lambdar)  - \partial \, T^{-1}a(\lambdar)
\right ]
(\epsilon_{1} \bar \partial \epsilon_{2}
-\epsilon_{2} \bar \partial \epsilon_{1})
\bigg \}
\en
This must vanish, because the local scale derivatives commute, so 
\eq
0 =
\beta^{i} \, T^{-1}g_{ij}(\lambdar) \, \bar \partial \lambdar^{j}
-  \bar \partial \, T^{-1}a(\lambdar)
\en
\eq
0 =
\partial \lambda_r^{i} \,
T^{-1}g_{ij}\, \beta^{j}(\lambdar)  - \partial \, T^{-1}a(\lambdar)
\:.
\en
The function $T^{-1}a(\lambdar)$ depends only on $\lambdar$, so
\eq
0= \beta^{i} \, T^{-1}g_{ij} - \partial_{j} (T^{-1}a)
\en
\eq
0= T^{-1}g_{ij}\, \beta^{j}  - \partial_{i} (T^{-1}a)
\en
proving that the vector field $\beta^{i}$ on $\M{L}$ is the gradient 
of the function $T^{-1}a$ with respect to the metric $T^{-1}g_{ij}$.
The function $T^{-1}a(\lambdar)$ is the {\em potential function}.

This proof of the gradient property, using the commutativity of the 
local scale derivatives, is equivalent to the original 
proof~\cite{Zamolodchikov-1,Zamolodchikov-2}.  The first attempts to 
show the gradient property of the beta function of the general 
nonlinear model failed because the dilaton couplings were missing 
from the general nonlinear model~\citeF. Given the dilaton couplings 
to the two dimensional scalar curvature~\cite{Fradkin-Tseytlin}, the 
gradient property of $\beta^{i}(\lambda)$ was shown by direct 
calculation, in the limit of large target spacetime~\cite{CFMP}.  
The beta function $\beta^{i}(\lambda)$ was shown to be the gradient 
of the classical spacetime field theory action that reproduced the 
large distance tree-level string scattering amplitudes.  That same 
spacetime field theory action had been identified with the 
coefficient of the two dimensional scalar curvature in the local 
scale variation of the general nonlinear 
model~\cite{Fradkin-Tseytlin}.  The proof of the gradient 
property~\cite{Zamolodchikov-1,Zamolodchikov-2} can be interpreted 
as an explanation of the coincidence between the 
calculations~\cite{Fradkin-Tseytlin,CFMP} which found the same 
spacetime field theory action both as the coefficient of $R_{2}$ in 
the scale variation and as the potential function whose gradient was 
$\beta^{i}(\lambda)$.  The 
proof~\cite{Zamolodchikov-1,Zamolodchikov-2} was based on the 
axiomatic properties of two dimensional quantum field theory.  It 
introduced an intrinsic metric to the space of two dimensional 
quantum field theories, $g_{ij}(\lambdar)$, rather than the ad hoc 
metric defined on the spacetime wave modes that was used 
previously\cite{Friedan-1,Friedan-2,Friedan-3,CFMP}.  The present 
version of the proof, using the commutativity of the local scale 
derivatives acting on the general nonlinear model, displays the 
historical genesis of the proof.  The present version of the proof 
is particularly suited to the general nonlinear model at nonzero two 
dimensional distance $\Lambda^{-1}$, where the general nonlinear 
model is parametrized by slightly irrelevant coupling constants.  
The axiomatics of two dimensional quantum field theory do not quite 
apply.

\subsection{The local scale variation formula}

The local scale variation gives a formula for the covariant action 
density of the lambda model
\eq
\label{eq:action-local-scale-variation}
-2 \,T^{-1} \, D(z,\bar z) \;
\me^{-\int \Lambda^{2} \, \frac1{2\pi} \,
\lambdar^{i} \phi^{\Lambda}_i}
=
\me^{-\int \Lambda^{2} \, \frac1{2\pi} \,
\lambdar^{i} \phi^{\Lambda}_i}
\; \L(\lambdar)(z,\bar z)
\en
\vskip1ex
\eqa
\L(\lambdar)(z,\bar z) &=&
\frac1{2\pi} \,
\left [
T^{-1} g_{ij}(\lambdar)
\,
\partial \lambdar^{i}
\,
\bar \partial \lambdar^{j}
- \frac14 \Lambda^{2} R_{2}(\Lambda) \, T^{-1}a(\lambdar)
\right ]
\\[2ex]
S(\lambdar) &=&
\int \dif^{2}z \;   \L(\lambdar)(z,\bar z)
\:.
\ena
The renormalization of the general nonlinear model guarantees that 
the metric coupling $T^{-1}g_{ij}(\lambdar)$ and the potential 
function $T^{-1}a(\lambdar)$ depend on the two dimensional distance 
$\Lambda^{-1}$ only through the running couplings $\lambdar^{i}$.

This is the locally scale invariant form of a nonlinear model with 
generalized scale invariance\cite{CFMP}.
The local scale variation of the action
\eq
D(z,\bar z) \; S(\lambdar) =
\frac1{2\pi} \,
\left [
\betastar T^{-1} g_{ij}(\lambdar)
\,
\partial \lambdar^{i}
\,
\bar \partial \lambdar^{j}
- \frac14 \Lambda^{2} R_{2}(\Lambda) \,  \beta^{k}\partial_{k}
T^{-1}a(\lambdar)
\right ] 
\:.
\en
is equivalent to the local change in the couplings $T^{-1}g_{ij}$ 
and $T^{-1}a$ that is produced by the flow along the vector field 
$\beta^{i}(\lambda)$ in the target manifold.  The potential function 
$T^{-1}a(\lambdar)$ plays the role that the dilaton potential plays 
in the general nonlinear model~\cite{Fradkin-Tseytlin,CFMP}.

The local scale variation formula, 
equation~\ref{eq:action-local-scale-variation}, gives an effective 
method to calculate the metric coupling $T^{-1}g_{ij}(\lambdar)$ and 
the potential function $T^{-1}a(\lambdar)$, the couplings of the 
locally covariant lambda model.  This data determines the couplings 
of the lambda model at every short two dimensional distance 
$\Lambda^{-1}$.

%
%
%
%
%
\sectiono{The \apm}

A nonlinear model such as the lambda model is specified by two 
pieces of data, the metric coupling, which is a riemannian metric on 
the target manifold, and the \apm\, which is a measure on the target 
manifold~\citeF. In the functional integral, 
equation~\ref{eq:lambda-model-2}, defining the lambda model
\eq
\int \Dlambda
\; \me^{- S(\Lambda,\lambda)} \;
\me^{-\int \dif^{2}z \, \mu^{2} \, \frac1{2\pi} \,
\lambda^{i}\phi_i}
\en
the functional measure $\Dlambda$ on the lambda fields is formally 
a product over the points $(z,\bar z)$ of the worldsurface, at 
characteristic two dimensional distance $\Lambda^{-1}$,
\eq
\Dlambda = \prod_{(z,\bar z)} \drho(\Lambda, \lambda(z,\bar z))
\en
where the measure at each point
\eq
\drho(\Lambda, \lambda(z,\bar z)) =
\dvol(\Lambda, \lambda(z,\bar z))
\,\rho (\Lambda,\lambda(z,\bar z))
\en
is the \apm, written as the metric volume element 
$\dvol(\Lambda,\lambda)$ associated to the metric coupling 
$T^{-1}g_{ij}(\Lambda,\lambda)$ multiplied by a function 
$\rho(\Lambda,\lambda)$.

The \apm\ of the lambda model is a normalized measure on the 
manifold of spacetimes $\M{L}$.  It describes the distribution of 
fluctuations at two dimensional distances shorter than 
$\Lambda^{-1}$.  It is dynamically determined, fixed by the two 
dimensional generalized scale invariance of the lambda model.

More concretely, the \apm\ governs the values of the lambda field at 
short distance on the worldsurface.  The \apm\ at two dimensional 
distance scale $\Lambda^{-1}$ is the distribution of the values of 
the field $\lambda(z,\bar z)$.
For any function $f(\lambda)$ on the manifold of spacetimes $\M{L}$,
\eq
\int \drho(\Lambda,\lambda) \, f(\lambda)
= \expval{\, f(\lambda(z,\bar z) ) \,}
\en
where the expectation value is in the functional integral over all 
lambda fluctuations at two dimensional distances up to 
$\Lambda^{-1}$.

The calculations of the \apm\ in this section will be tree-level 
calculations, done at leading order in $T$, ignoring quantum 
corrections.  The same calculations will be used, in the same form, 
in section~\ref{sect:effective-apm} below, to find the effective 
\apm\ of the effective lambda model, which includes the quantum 
corrections.

\subsection{Diffusion in $\lambda$}

The \apm\ of a scale invariant nonlinear model is completely 
determined by the renormalization group of the model.  As the two 
dimensional distance $\Lambda^{-1}$ increases, the \apm\ diffuses 
outward in the target manifold, because of the fluctuations.  The 
\apm\ is built outward from short two dimensional distance towards 
longer distance, as the fluctuations at longer distances are added 
in.  If the nonlinear model is scale invariant, then the \apm\ 
diffuses to the equilibrium measure of the diffusion process.  It 
does not matter what arbitrary measure is used for the \apm\ at the 
cutoff two dimensional distance $\Lambdazero^{-1}$.  The \apm\ will 
diffuse to the equilibrium measure at distances $\Lambda^{-1}$ much 
longer than the cutoff.  When the cutoff distance $\Lambdazero^{-1}$ 
is taken to zero, only the equilibrium \apm\ is visible.  In the 
lambda model, whose scale invariance is of the generalized kind, the 
action takes the same form at every two dimensional distance 
$\Lambda^{-1}$ when expressed in the running variables 
$\lambdar^{i}$, so the equilibrium \apm\ takes the same form also 
when expressed in terms of the running variables $\lambdar^{i}$.

The diffusion process is calculated in the usual way, expanding to 
second order about a reference point $\lambda_{1}$ in the target 
manifold,
\eq
f(\lambda)=
f(\lambda_{1}) + \lambda^{i} \partial_{i}f(\lambda_{1})
+ \frac12 \lambda^{i} \lambda^{j}
\del_{i} \partial_{j}f(\lambda_{1}) +\cdots 
\en
\eqa
- \Lambda \partialby{\Lambda} \,
\expval{\, f \,}
& = &
- \Lambda \partialby{\Lambda}
\;
\frac12
\expval{\,
\lambda^{i}(z,\bar z) \lambda^{j}(z,\bar z) \,}
\del_{i} \partial_{j}f(\lambda_{1}) \nonumber \\[2ex]
& = &
T\,g^{ij}(\Lambda,\lambda_{1})
\del_{i} \partial_{j} f(\lambda_{1})
\:.
\ena
Only the propagator contributes to the scale variation because only 
two derivatives of the lambda fields $\lambda^{i}(z,\bar z)$ occur 
in the interaction terms of the nonlinear model.  The same scale 
variation formula can be obtained by canonically quantizing the 
lambda model, in the radial quantization, where the target manifold 
laplacian operator occurs as the zero mode piece of the dilation 
generator.

Patching coherently over the target manifold gives the diffusion 
equation on expectation values
\eq
- \Lambda \partialby{\Lambda} \,
\expval{\, f \,}
=
\expval{\, T\,g^{ij}(\Lambda,\lambda) \,
\del_{i} \partial_{j} f  \,}
\en
or, equivalently, the diffusion equation directly on the measure
\eq
\label{eq:diffusion-lambda}
-\Lambda \partialby{\Lambda}_{\left / \lambda \right .} \rho
=
T\,g^{ij}(\Lambda,\lambda) \, \del_{i} \partial_{j} \rho
\:.
\en
Changing variables to the running coupling constants $\lambdar^{i}$,
the \apm\ is rewritten
\eq
\dvol(\lambdar) \, \rhor(\Lambda,\lambdar)
=
\dvol(\Lambda,\lambda) \, \rho(\Lambda,\lambda)
\:.
\en
The diffusion equation, rewritten in the variables 
$\lambdar^{i}$, is
\eq
-\Lambda \partialby{\Lambda} \, \rhor
=
\del_{i} 
\left [
T\,g^{ij}(\lambdar) \partial_{j} + \beta^{i}(\lambdar)
\right ]
\,
\rhor
\:.
\en
As $\Lambda^{-1}$ increases, the distribution of fluctuations 
diffuses outwards on the target manifold, while the running coupling 
constants are driven by $-\beta^{i}(\lambdar)$ towards the fixed 
point submanifold where $\beta=0$.

Using the gradient property, the diffusion equation is written
\eq
-\Lambda \partialby{\Lambda} \, \rhor
=
\del_{i}
\,
T\,g^{ij}
\,
\left (
\partial_{j} + \partial_{j}(T^{-1}a)
\right )
\,
\rhor
\:.
\en
The equilibrium \apm\ is simply
\eq
\dvol(\lambdar) \; \me^{-T^{-1}a(\lambdar)}
\:.
\en
It satisfies the first order equation
\eq
0 = (\partial_{i} +  T^{-1}g_{ij} \, \beta^{j})
\; \me^{-T^{-1}a(\lambdar)}
\en
which is the equation of motion $\beta=0$.

The lambda propagator, normalized at two dimensional distance 
$\Lambda^{-1}$, at a scale invariant general nonlinear model, is
\eq
\expval{\,\lambda^i(\mathrm{1}) \; \lambda^j(\mathrm{2})\,}
=
T\,g^{ij}\,
\left [
\frac{
(\mu^{2} \Lambda^{-2})^{\ad(i)}
- (\mu^{2} \abs{z_{1}-z_{2}}^{2})^{\ad(i)})
}
{\ad(i)}
\right ]
\:.
\en
Evaluating at $\abs{z_{1}-z_{2}}=\Lambdazero^{-1}$ gives
\eq
\expval{\,\lambda^i\, \lambda^j\,} = 
T\,g^{ij}\, 
\left [
\frac{
(\mu^{2} \Lambda^{-2})^{\ad(i)}
- (\mu^{2} \Lambdazero^{-2})^{\ad(i)})
}
{\ad(i)}
\right ]
\en
which is the solution of the diffusion 
equation~\ref{eq:diffusion-lambda} starting at the cutoff distance 
$\Lambdazero^{-1}$ with initial condition the delta function measure 
at $\lambda^{i}=0$.  The integral equation for the lambda 
propagator, equation~\ref{eq:propagator-integral}, is the solution 
of the diffusion equation.

\subsection{Gaussian approximation}

The potential function 
at a scale invariant general nonlinear model
has the gaussian approximation
\eq
T^{-1} a(\lambdar) = 
\frac12 \lambdar^{i} \, T^{-1}g_{ij} \, \ad(i) \,
\lambdar^{j} + \cdots
\en
which gives the tree-level two point correlation function
\eq
\expval{\, \lambdar^{i} \lambdar^{j} \, }
= T g^{ij} \frac1{\ad(i)}
\en
in the \apm.  This matches the tree-level one point expectation 
value of $f(\lambdar)=\lambdar^{i}\lambdar^{j}$ in the lambda model, 
which is the lambda propagator
\eq
\expval{\,\lambdar^i(\mathrm{1}) \; \lambdar^j(\mathrm{2})\,}
=
T\,g^{ij}\, 
\left [
\frac{
1
- (\Lambda^{2} \abs{z_{1}-z_{2}}^{2})^{\ad(i)})
}
{\ad(i)}
\right ]
\en
evaluated at $\abs{z_{1}-z_{2}}=0$.

The potential function written
in terms of the original renormalized
coupling constants $\lambda^{i}$ is
\eq
T^{-1} a(\Lambda,\lambda)
= 
\frac12 (\mu^{-1}\Lambda)^{\ad(i)}\, \lambda^{i} \, T^{-1}g_{ij}
\,\ad(i)\, 
(\mu^{-1}\Lambda)^{\ad(j)}\lambda^{j} + \cdots
\en
or
\eq
T^{-1} a(\Lambda,\lambda)
= 
\frac12 \me^{L^{2}\ad(i)}\lambda^{i} \, T^{-1}g_{ij}\, \ad(i) \,
\me^{L^{2}\ad(j)}\lambda^{j} + \cdots
\en
which exhibits the spacetime ultraviolet cutoff at $\ad(i) = 
L^{-2}$, at least in a naive way.  The actual implementation of the 
ultraviolet cutoff by the renormalization of the general nonlinear 
model depends on the decoupling of the $L$-irrelevant coupling 
constants in the interactions.  The \apm\ does manifestly accomplish 
the basic task of keeping the irrelevant running coupling constants 
$\lambdar^{i}$ from becoming large, which keeps the irrelevant 
renormalized coupling constants $\lambda^{i}$ close to zero, which 
is a necessary condition for the renormalization to be effective.

\subsection{Spacetime quantum field theory}

Near a macroscopic spacetime, the manifold of spacetimes $\M{L}$ is 
parametrized by the spacetime wave modes $\lambdar^{i}$ at spacetime 
distances larger than $L$.  The \apm\ takes the form of a functional 
integral over the spacetime wave modes.  The potential function can 
be written
\eq
T^{-1} a(\lambdar) = \gst^{-2}\, V\, a(\lambdar)
\en
where $\gst$ is the spacetime coupling constant, and $V\, 
a(\lambdar)$ is an integral over spacetime of a local functional of 
the spacetime fields.  The \apm
\eq
\dvol(\lambdar)\, \me^{-\gst^{-2}V a(\lambdar)}
\en
is the functional integral of a spacetime quantum field theory whose 
classical action is the potential function $\gst^{-2} \,V\, 
a(\lambdar)$, and whose classical equation of motion is 
$\beta^{i}(\lambdar)=0$.  It is the gradient property which that the 
classical action principle and implies that the classical spacetime 
equation of motion is $\beta=0$.

It was known by explicit calculation\cite{Fradkin-Tseytlin,CFMP} 
that the potential function $T^{-1}a(\lambdar)$ in a macroscopic 
spacetime has the form of a classical action functional of a 
spacetime field theory.  It was shown\cite{Lovelace-2} that the 
classical equation of motion $\beta=0$ in a macroscopic spacetime 
generates the tree-level string scattering amplitudes at large 
spacetime distance.  There should be a direct argument from the 
local scale variation, equation~\ref{eq:local-scale-variation}, 
showing that the potential function $T^{-1}a(\lambda)$ is the 
generating functional for the tree-level string scattering 
amplitudes at large spacetime distance.  It should be possible to 
show directly that the coefficient of the two dimensional curvature 
density $\Lambda^{2}R_{2}$ in the local scale variation gives 
precisely the generating functional for the one particle irreducible 
tree-level string scattering amplitudes.

The \apm\ governs the string worldsurface at short distance, which 
is to say, roughly, that the \apm\ governs strings when they look 
like points.  In particular, when a handle degenerates to a node, 
the two dimensional curvature density accumulates in a 
delta-function at the node
\eq
\frac1{4 \pi} \Lambda^{2}R_{2}(\Lambda) = - 2 
\delta^{2}(z,\bar z)
\en
containing the contribution $-2$ to the Euler number of the 
worldsurface.  The covariant action of the lambda model, 
equation~\ref{eq:action-local-scale-variation}, then contains a 
discrete contribution at the node which is exactly 
$T^{-1}a(\lambdar)$.  The lambda model therefore inserts
the \apm\ at the node 
\eq
\int \dvol(\lambdar) \; \me^{-T^{-1}a(\lambdar)}
\:.
\en
The \apm\ thus controls the propagation of strings at large 
spacetime distance.  There should be a similarly direct, general 
argument that the \apm\ governs the large distance string 
scattering.

%
%
%
%
%
%
%
\sectiono{The effective lambda model}

\subsection{$S_{\eff}(\lambdaeff)$}

Integrating out the fluctuations of the lambda fields at short two 
dimensional distances from the cutoff distance $\Lambdazero^{-1}$ up 
to $\Lambda^{-1}$ produces an effective general nonlinear model of 
the worldsurface at two dimensional distances longer than 
$\Lambda^{-1}$.  The effective model of the worldsurface is 
constructed out of two dimensional surface elements at distance 
$\Lambda^{-1}$, so it depends only on $\Lambda^{-1}$ and on 
effective running coupling constants $\lambdaeff^{i}$.  The 
effective general nonlinear model satisfies an effective 
renormalization group equation
\eq
\Deff
\;
\me^{-\int \dif^{2}z \, \Lambda^{2} \, \frac1{2\pi} \,
\lambdaeff^{i} \phieff_i(z,\bar z)}
=0
\en
\eq
\Deff = \Lambda \partialby{\Lambda}_{\left / \lambdaeff\right.} + 
\betaeff^{i}(\lambdaeff) \partialby{\lambdaeff^{i}}
\:.
\en
The effective model of the worldsurface is still approximately scale 
invariant at two dimensional distances near $\Lambda^{-1}$, so the 
calculations and arguments can be carried over from the general 
nonlinear model,
\eqa
\lefteqn{
\Deff \; \me^{-\int \dif^{2}z \Lambda^{2} \frac1{2\pi} 
\, \lambdaeff^{i}(z,\bar z) \phieff_i(z,\bar z)}
=
}
\nonumber \\[1ex]
&& \me^{-\int \dif^{2}z \Lambda^{2} \frac1{2\pi} 
\, \lambdaeff^{i}(z,\bar z) \phieff_i(z,\bar z)}
\int \dif^{2}z \, \frac1{2\pi} \,
\left (- \frac12 \Teff \right )
\Teff^{-1} \geff_{ij}(\lambdaeff)
\,
\partial \lambdaeff^{i}
\,
\bar \partial \lambdaeff^{j}
\:.
\ena
The effective metric $\Teff^{-1} \geff_{ij}(\lambdaeff)$ is the 
inverse of the handle gluing matrix for the effective model of the 
worldsurface.

The effective lambda model and the effective model of the 
worldsurface evolve in tandem under the two dimensional 
renormalization group, in the sense that the effective fluctuations 
of the lambda fields at each two dimensional distance $\Lambda^{-1}$ 
must automatically cancel the effects of local handles in the 
effective model of the worldsurface at two dimensional distance 
$\Lambda^{-1}$.  The tandem renormalization principle states that 
$\Teff^{-1} \geff_{ij}(\lambdaeff)$, obtained from the scale 
variation of the effective model of the worldsurface, is the 
effective metric coupling of the effective lambda model at two 
dimensional distance $\Lambda^{-1}$,
\eq
\Seff(\lambdaeff) = \int \dif^{2}z \, \frac1{2\pi}
\,
\Teff^{-1} \geff_{ij}(\lambdaeff)
\,
\partial \lambdaeff^{i}
\,
\bar \partial \lambdaeff^{j}
\:.
\en
Calculating the scale variation of the effective model of the 
worldsurface is equivalent to calculating the effective action 
$\Seff(\lambdaeff)$ of the lambda model,
\eq
\Deff \;
\me^{-\int \dif^{2}z \, \Lambda^{2} \, \frac1{2\pi} \,
\lambdaeff^{i}(z,\bar z) \phieff_i(z,\bar z)}
=
\me^{-\int \dif^{2}z \, \Lambda^{2} \, \frac1{2\pi} \,
\lambdaeff^{i}(z,\bar z) \phieff_i(z,\bar z)}
(-\frac12 \Teff )\, \Seff(\lambdaeff) \;
\:.
\en
Calculating the local scale variation of the effective model of the 
worldsurface is equivalent to calculating the covariant action 
density of the effective lambda model,
\eqa
\Deff(z,\bar z)
&= &
\Lambda(z,\bar z)\partialby{\Lambda(z,\bar z)}_{\left / 
\lambdaeff\right.}
+\betaeff^{i}(\lambdaeff(z,\bar z))
\partialby{\lambdaeff^{i}(z,\bar z)}
\\[2ex]
\label{eq:action-local-scale-variation-eff}
\Deff(z,\bar z) \;
\me^{-\int  \Lambda^{2} \, \frac1{2\pi} 
\, \lambdaeff^{i}\phieff_i}
&=&
\me^{-\int  \Lambda^{2} \, \frac1{2\pi} 
\, \lambdaeff^{i}\phieff_i}
\;
(-\frac12 \,\Teff )\, \; \Leff(\lambdaeff)(z,\bar z)
\\[2ex]
\Leff(\lambdaeff)(z,\bar z)
&=&
\frac1{2\pi} \,
\left [
\Teff^{-1} \geff_{ij}(\lambdaeff)
\,
\partial \lambdaeff^{i}
\,
\bar \partial \lambdaeff^{j}
- \frac14 \Lambda^{2} R_{2}(\Lambda)\, \Teff^{-1}a(\lambdaeff)
\right ]
\\[2ex]
\Seff(\lambdaeff)
&=&
 \int \dif^{2}z \;   \Leff(\lambdaeff)(z,\bar z)
\:.
\ena
The commutativity of local scale derivatives again implies the 
gradient property
\eq
0 = \betaeff^{i} \, \Teff^{-1}\geff_{ij}
 - \partial_{j} (\Teff^{-1}\aeff)
\en
\eq
0 = \Teff^{-1}\geff_{ij}\, \betaeff^{j}  
- \partial_{i} (\Teff^{-1}\aeff)
\:.
\en

The results of integrating out the lambda fluctuations at two 
dimensional distances up to $\Lambda^{-1}$ are described by the 
effective classical action $\Seff(\lambdaeff)$ of the effective 
lambda model.  As before, at an effective spacetime solving 
$\betaeff(\lambdaeff)=0$, there is a coordinate system of effective 
coupling constants $\lambdaeff^{i}$ in which the matrix of effective 
anomalous dimensions is diagonalized,
\eq
\Teff^{-1}\aeff(\lambdaeff)= \Teff^{-1}\aeff(0)
+ \frac12
\lambdaeff^{i} \, \Teff^{-1}\geff_{ij}\, \adeff(i)\, \lambdaeff^{j}
+ \cdots
\en
\eq
\betaeff^{i}(\lambdaeff)=\adeff(i) \, \lambdaeff^{i}+ \cdots
\:.
\en
The effective lambda propagator is, as before,
\eq
\label{eq:effective-propagator}
\expval{\,\lambdaeff^i(z_{1},\bar z_{1}) 
\; \lambdaeff^j(z_{2},\bar z_{2})\,}
=
\Teff\,\gsubeff^{ij}\, 
\left [ 
\frac{
1- (\Lambda^{2} \abs{z_{1}-z_{2}}^{2})^{\adeff(i)}
}
{\adeff(i)}
\right ]
\:.
\en

\subsection{Self-sufficiency of the lambda model}

For the classical lambda model, the scale variation of the general 
nonlinear model, equation~\ref{eq:action-local-scale-variation}, 
gives the couplings, the metric coupling $T^{-1}g_{ij}(\lambdar)$ 
and the potential function $T^{-1}a(\lambdar)$.
The scale variation formula gives the same couplings 
that are derived from the formulation of the lambda model as the 
local mechanism that cancels the short distance effects of local 
handles.

The effective metric coupling $\Teff^{-1}\geff_{ij}(\lambdaeff)$ and 
the effective potential function $\Teff^{-1}\aeff(\lambdaeff)$ can 
be calculated by constructing the effective model of the 
worldsurface, then taking the scale variation, 
equation~\ref{eq:action-local-scale-variation-eff}.  But the 
calculation can be done in the opposite direction.  The effective 
lambda model can be constructed, as an autonomous nonlinear model in 
$2+\epsilon$ dimensions.  Then the effective scale variation 
formula, equation~\ref{eq:action-local-scale-variation-eff}, can be 
used to find the short distance properties of the effective model of 
the worldsurface.  In principle, calculations in the effective model 
of the worldsurface can be entirely avoided, unless there is a 
reason to be interested in string scattering at small spacetime 
distance.

The manifold of renormalized general nonlinear models define the 
lambda model.  Once defined, the lambda model becomes 
self-sufficient.  All calculations of large distance physics can be 
done entirely within the lambda model.  No worldsurface calculations 
are needed.  On the other hand, it might well be useful to have a 
second means of calculating the effective couplings of the lambda 
model, from the effective model of the worldsurface.

\subsection{Tautological scale invariance}

The generalized scale invariance of the effective lambda model 
derives from the existence of the effective renormalizable model of 
worldsurface and from the tandem renormalization property.  
Enormously strong constraints are put on the renormalization of the 
lambda model.  The lambda model is not a {\em general} nonlinear 
model.  Its target manifold and its couplings are extremely special, 
mathematically natural objects.  They have remarkable properties, 
which are realized in the scale invariance of the effective lambda 
model.

The scale invariance of the effective lambda model derives from the 
parametrization of the effective model of the worldsurface by 
effective running coupling constants $\lambdaeff^{i}$, flowing under 
an effective renormalization group generated by an effective beta 
function $\betaeff^{i}(\lambdaeff)$.  The existence of such a 
parametrization of the effective general nonlinear model is a 
consequence of locality in the two dimensional distance scale, by 
the usual argument of effective field theory.  The processes which 
accomplish change of two dimensional distance in the effective model 
of the worldsurface do not themselves depend on the distance, but 
depend only on the effective couplings constants at each distance 
$\Lambda^{-1}$.  These processes now include integrating out the 
fluctuations of the lambda fields taking values in $\M{L}$, using 
the effective lambda action given by the scale variation of the 
effective model of the worldsurface.  All dependence on the two 
dimensional distance is absorbed into a flowing of the effective 
coupling constants $\lambdaeff^{i}$.  All local properties of the 
effective general nonlinear model at distance $\Lambda^{-1}$, such 
as the effective metric coupling $\Teff^{-1}\geff_{ij}(\lambdaeff)$ 
and the potential function $T^{-1}\aeff(\lambdaeff)$ depend only on 
the effective coupling constants, and are independent of the two 
dimensional distance $\Lambda^{-1}$.

The effective lambda model is {\em tautologically} scale invariant.  
Its scale invariance follows automatically from its tandem relation 
to the effective model of the worldsurface.

\subsection{The effective \apm}
\label{sect:effective-apm}

The overall distribution of fluctuations in the effective lambda 
model at distances shorter than $\Lambda^{-1}$ is described by an 
effective \apm\ on the manifold of spacetimes $\M{L}$,
\eq
\drhoeff(\Lambda,\lambdaeff) = \dvoleff(\lambdaeff) \,
\rhoeff(\Lambda,\lambdaeff)
\en
\vskip1ex
\eq
\int \dvoleff(\lambdaeff) \,
\rhoeff(\Lambda,\lambdaeff)
f(\lambdaeff)
= \expvaleff{\, f(\lambdaeff(z,\bar z) ) \,}
\en
where the expectation value is evaluated in the effective lambda 
model at two dimensional distance $\Lambda^{-1}$.

All the considerations that applied to the classical lambda model 
carry over to the effective lambda model.  The effective \apm\ 
satisfies an effective diffusion equation, which takes the same form 
as the tree-level diffusion equation.  The generalized scale 
invariance of the effective lambda model implies that the diffusion 
equation has stationary coefficients,
\eqa 
- \Lambda 
\partialby\Lambda_{\left / \lambdaeff\right.}
 \, \rhoeff(\Lambda,\lambdaeff) &=& \deleff_{i} 
\left ( \Teff\gsubeff^{ij}\, \partial_{j} 
+\betaeff^{i} \right ) \, \rhoeff \\
&=& \deleff_{i} \, \Teff\gsubeff^{ij} \, \left 
(  \partial_{j} + \partial_{j}(\Teff^{-1}\aeff)
\right ) \, \rhoeff
\:.
\ena
The effective \apm\ is the equilibrium measure
\eq
\dvoleff(\lambdaeff) \; \me^{-\Teff^{-1}\aeff(\lambdaeff)}
\en
which satisfies the equation of motion $\betaeff =0$,
\eq
0 = \left (\partial_{i}
+\Teff^{-1}\geff_{ij}\, \betaeff^{j}
\right )
\; \me^{-\Teff^{-1}\aeff(\lambdaeff)}
\:.
\en
The effective \apm\ is a measure on the manifold of spacetimes 
$\M{L}$.  If it concentrates near a macroscopic spacetime, then it 
will take the form
\eq
\dvoleff(\lambdaeff) \; \me^{-\gst^{-2} \, V \,\aeff(\lambdaeff)}
\en
which will be an effective quantum field theory of the spacetime 
physics at distances larger than $L$.  The gradient property of the 
effective beta function $\betaeff^{i}(\lambdaeff)$ implies the 
quantum action principle in the spacetime quantum field theory, and 
the quantum equation of motion $\betaeff=0$.

The effective potential function is the effective classical action 
of the spacetime quantum field theory.  Correlation functions in the 
\apm\ are classical calculations in the effective \apm.  For 
example, at an effective spacetime solving $\betaeff=0$, the 
two-point correlation function in the effective \apm\ is the 
effective lambda propagator, equation~\ref{eq:effective-propagator}, 
at $z_{1}=z_{2}$,
\eq
\expvaleff{\,\lambdaeff^{i}\,\lambdaeff^{j} \,} = 
T\,\gsubeff^{ij} \, \frac1{\adeff(i)}
\:.
\en

\subsection{Complementarity with effective string theory}

The effective \apm\ concentrates at the effective spacetimes 
$\lambdaeff$ in $\M{L}$ where $\betaeff(\lambdaeff)=0$.  The 
effective model of the worldsurface is scale invariant.  If the 
spacetime is macroscopic, then the effective model of the 
worldsurface can be used to calculate effective string scattering 
amplitudes at distances larger than $L$.  The same relation will 
exist between the effective string scattering amplitudes and the 
effective spacetime action as in the tree-level theory, by the same 
arguments.

The effective \apm\ of the lambda model is a spacetime quantum field 
theory.  It describes the spacetime physics at distances larger than 
$L$ by the effective field equation $\betaeff=0$, just as the 
uncorrected \apm\ is a classical spacetime field theory describing 
the spacetime physics at distances larger than $L$ by the classical 
field equation $\beta=0$.  The tandem renormalization principle 
guarantees that the effective spacetime action is the effective 
potential function derived from the effective model of the 
worldsurface, which is also the generating functional for the 
effective string scattering amplitudes at spacetime distances on the 
order of $L$.  The effective string scattering amplitudes therefore 
agree with the scattering amplitudes calculated from the effective 
spacetime quantum field theory.

Again, in principle there is no need to calculate the effective 
string scattering amplitudes except as a description of hypothetical 
physics at small distance in spacetime.  The effective \apm\ can be 
calculated entirely within the lambda model, and gives all the large 
distance physics.  In particular, it gives all the large distance 
string scattering amplitudes, via the effective spacetime quantum 
field theory.

Because the lambda model is designed to cancel the effects of local 
handles, string theory calculations proceeding from $L$ to larger 
spacetime distances would reverse the evolution of the \apm\ down 
from larger spacetime distances to $L$, if nonperturbative string 
theory calculations could be done.  In the absence of a 
nonperturbative formulation of string theory, all that can be said 
is that the perturbative evolution of the \apm\ is consistent with 
the string loop expansion, calculated using the effective model of 
the worldsurface.  This will ensure that the effective action of the 
spacetime quantum field theories produced by the lambda model 
depends on the characteristic spacetime distance $L$ in a way that 
is consistent with perturbative spacetime quantum field theory, 
whenever the perturbative theory is accurate.

At issue is the validity of the strange method by which the lambda 
model is to build spacetime quantum field theories.  A spacetime 
quantum field theory, as a measure on the wave modes of the 
spacetime fields, is to be built up starting from the wave modes at 
the largest spacetime distances.  As $\Lambda^{-1}$ increases, as 
$L$ becomes smaller, the spacetime wave modes at smaller and smaller 
spacetime distances are gradually included.  This is opposite to the 
method used by renormalizable spacetime quantum field theory, which 
builds from small spacetime distance to large.  The lambda model 
must be capable of reproducing the spectacular numerical successes 
that have been achieved in the real world by perturbative 
renormalizable spacetime quantum field theory.  The lambda model 
cannot possibly be useful unless the method by which it builds 
spacetime quantum field theory is consistent with that of 
perturbative renormalizable spacetime quantum field theory.

%
%
%
%
%
\sectiono{The fermionic spacetime wave modes}
\label{sect:fermionmodes}

The lambda model needs certain basic capabilities in the general 
nonlinear model of the worldsurface.  Compact riemannian background 
spacetimes must be accomodated.  There must be fermionic coupling 
constants $\lambda^{i}$ on an equal footing with the bosonic 
$\lambda^{i}$, so that the manifold of spacetimes $\M{L}$ will be a 
graded manifold.  The \apm\ will then be a measure on fermionic as 
well as bosonic spacetime wave modes, which can be the functional 
integral of a spacetime quantum field theory containing both bosonic 
and fermionic fields.

As far as I know, the only model of the worldsurface that has these 
capabilities is the model that is constructed starting with a 
superconformal worldsurface in which two dimensional 
super-reparametrization invariance is implemented using worldsurface 
superconformal ghost fields~\cite{Polyakov}.  The GSO projection 
then throws away the two dimensional spinor fields and the two 
dimensional supersymmetry, producing a two dimensional conformally 
invariant worldsurface.  The resulting ordinary, scale invariant 
model of the worldsurface is covariant in spacetime.  After the GSO 
projection, the spinor components of the worldsurface superconformal 
ghost fields are incorporated into the fermionic vertex operators, 
which are the two dimensional scaling fields $\phi_{i}(z,\bar z)$ 
that represent the on-shell fermionic spacetime wave 
modes~\cite{FMS}.  The GSO projection removes the tachyonic string 
modes, eliminating all scaling fields of scaling dimension less than 
$2$ from flowing through degenerating handles.  The drawback of the 
covariant worldsurface is the ambiguous characterization it gives of 
the two dimensional scaling fields.  The scaling fields occur in a 
multiplicity of equivalent linear spaces, called {\em pictures,} 
requiring picture independence to be verified in global worldsurface 
calculations.

A technical obstacle stands in the way of using the covariant 
worldsurface to construct a graded manifold of general nonlinear 
models that can serve as the target manifold of the lambda model.  
The fermionic scaling fields occur in different pictures from the 
bosonic scaling fields.  A unified description is needed of the 
bosonic and fermionic coupling constants $\lambda^{i}$, to serve as 
graded coordinates on the graded manifold of background spacetimes.

Only the on-shell fermionic vertex operators were needed for the 
covariant string perturbation theory.  The on-shell fermionic vertex 
operators are the scaling fields that represent the on-shell 
fermionic string states.  The lambda model needs all the fermionic 
scaling fields, on-shell and off-shell.  All the marginal and nearly 
marginal fermionic scaling fields have to be constructed, since all 
the fermionic string states flow through degenerating handles.  The 
fermionic scaling fields have to be constructed so that they appear 
in the worldsurface on an equal footing with the bosonic scaling 
fields, effectively in the same picture.  There must be a single 
metric $T^{-1} g_{ij}$ on all the scaling fields, symmetric in the 
bosonic directions and antisymmetric in the fermionic directions.  
In the operator product
\eq
\phi_{i}(\rm 1)\,\phi_{j}(\rm 2)
=T^{-1}g_{ij} \abs{z_{1}-z_{2}}^{-2-\ad(i)-2-\ad(j)}
\mathrm{1} + \cdots
\en
the metric $T^{-1}g_{ij}$ is antisymmetric if and only if $\phi_{i}$ 
and $\phi_{j}$ are fermionic.

The fermionic scaling fields $\phi_{i}(z,\bar z)$ have to be 
constructed in the same picture as the bosonic scaling fields, with 
an antisymmetric metric arising from the operator product.  Then the 
bosonic and fermionic scaling fields can be coupled to bosonic and 
fermionic lambda fields $\lambda^{i}(z,\bar z)$, which can be 
interpreted as the even and odd components of a map $\lambda(z,\bar 
z)$ from the worldsurface to the graded manifold of spacetimes.

Two notable technical consequences will follow from the construction 
of the antisymmetric metric on the Ramond sector scaling fields.  

First, it seems that only on the heterotic 
worldsurfaces~\cite{heterotic} can the lambda model make sense.  A 
non-heterotic worldsurface would contain a Ramond-Ramond sector of 
bosonic scaling fields.  The metric $T^{-1}g_{ij}$ on the 
Ramond-Ramond sector would be the tensor product of two 
antisymmetric metrics, which cannot be positive definite.  The 
metric coupling of the lambda model would not then be positive 
definite on the bosonic part of the manifold of spacetimes.  For 
this purely technical reason, it seems that the lambda model can 
only work on the heterotic worldsurface.

Second, the spacetime equation of motion for the fermionic wave 
modes is a second order wave equation, just as it is for the bosonic 
modes.  The spacetime equation of motion takes the same form
\eq
0=\beta^{i}(\lambda) = \ad(i) \lambda^{i} + O(\lambda^{2})
\en
for all the wave modes $\lambda^{i}$, fermionic and bosonic.  The 
anomalous dimension is quadratic in the spacetime wave number, 
$\ad(i)= p(i)^{2}+m(i)^{2}$, for the fermionic wave modes, as well 
as the bosonic ones.  The unphysical states in the solutions of the 
second order wave equation are eliminated by a gauge symmetry, 
leaving the usual physical solutions of the first order Dirac 
equation.

The construction of the fermionic scaling fields is guided by the 
requirement that the linear space of scaling fields should match the 
space of states flowing through the handle, the need for an 
antisymmetric metric on the ferminonic scaling fields, and also the 
need to realize perturbative spacetime supersymmetry as a direct 
cancellation between the bosonic and fermionic lambda fields, whose 
simplest expression is the vanishing of the graded trace
\eq
\delta^{i}_{i} = T\, g^{ij} \; T^{-1}g_{ji}= 0
\en
which is the dimension of the graded manifold of spacetimes.  In the 
space of string states, the vanishing of the graded trace 
$\delta^{i}_{i} = 0$ follows from perturbative spacetime 
supersymmetry applied to the one-loop vacuum string diagram.  The 
same equation must hold in the corresponding space of scaling fields 
$\phi_{i}(z,\bar z)$ or in the corresponding space of coupling 
constants $\lambda^{i}$.

The rest of this section is purely technical.  The notation 
of~\cite{FMS}\ is used.

\subsection{The antisymmetric metric}

The covariant string worldsurface~\cite{FMS} is an ordinary bosonic 
worldsurface.  The spinor components $\beta(z), \gamma(z)$ of the 
superconformal worldsurface ghost fields\cite{Polyakov} are combined 
with the spacetime degrees of freedom to form the scaling fields 
$\phi_{i}(z,\bar z)$.  For simplicity, I only treat here the 
worldsurface in flat ten dimensional spacetime, and discuss only the 
$z$ dependent parts of the worldsurface scaling fields.  All the 
essential technical issues are resolved in this simplified context.  
The novel part of the construction of the fermionic scaling fields 
involves only the structure of the $\beta, \gamma$ ghost fields.  It 
is easily taken over to the general nonlinear model with a general 
target spacetime.

The space of $z$ dependent scaling fields splits into two subspaces, 
the Ramond sector and the Neveu-Schwarz sector.  The string states 
and the corresponding two dimensional scaling fields are described 
redundantly in an infinite set of pictures, labelled by the picture 
charge.  The Neveu-Schwarz sector is represented by the pictures of 
integer picture charge, the Ramond sector by the pictures of charge 
integer plus half.

In analyzing the effects of degenerating handles in the 
worldsurface, there is an obvious benefit to choosing those special 
pictures in which the scaling dimensions are bounded below.  In 
those special pictures, the scaling dimensions of the $z$ dependent 
fields are bounded below by 1.  The $z$ dependent fields of 
dimension 1 are combined with $\bar z$ dependent fields of dimension 
1 to form the marginal scaling fields, of scaling dimension 2.

For the Neveu-Schwarz sector, there is only one picture with bounded 
scaling dimensions, the picture of charge $-1$.  The $z$ dependent 
Neveu-Schwarz sector fields with picture charge $-1$ and scaling 
dimension 1 are, after GSO projection, the ten bosonic fields
\eq
\psi_\mu e^{-\phi} \quad \mu = 1,\cdots, 10
\en
plus two fermionic fields made entirely from the worldsurface ghost 
fields
\eq
\beta_{-1/2} e^{-\phi}, \qquad
\gamma_{-1/2} e^{-\phi}
\:.
\en
The field $\phi(z)$ is the bosonization of the $\beta \gamma$ 
current, $\beta \gamma = - \partial \phi$.  The exponentials of 
$\phi(z)$ correspond to the highest weight states of the $\beta, 
\gamma$ algebra, \eqa \beta_{n} e^{q\phi} &=& 0 \qquad n \ge -q-1/2 
\\
\gamma_{n} e^{q\phi} &=& 0 \qquad n\ge q+3/2
\:.
\ena
The operators $\beta_{n}$ and $\gamma_{n}$ lower the scaling 
dimension by $n$.  The only pictures with scaling dimension bounded 
below are $q=-1/2$, $q=-1$, and $q=-3/2$

The graded trace $\delta_{i}^{i}$ in flat spacetime is the product 
of two factors,
\eq
\delta_{i}^{i} = (\delta_{i}^{i})_{z}
(\delta_{i}^{i})_{\bar z}
\:.
\en
One factor is the graded trace over the $z$ dependent fields, the 
other factor comes from the $\bar z$ dependent fields.
The object will be to have
\eq
(\delta_{i}^{i})_{z} = 0  \:.
\en
The Neveu-Schwarz sector fields make a net contribution of $10-2=8$.  
The Ramond sector fields must make a contribution of $-8$.

The metric on the bosonic $z$ dependent fields of the \NS\ sector is 
symmetric and positive
\eq
\expval{\,
\psi_\mu e^{\phi}(z_1)
\;
\psi_\nu e^{\phi}(z_2)
\,}
=
\delta_{\mu\nu} \, (z_1-z_2)^{-2}
\:.
\en
while the metric on the pair of fermionic \NS\ fields is antisymmetric 
\eqa
\expval{\beta_{-1/2} e^{\phi}(z_1)\;\gamma_{-1/2} e^{\phi}(z_2)} 
&=&   (z_1-z_2)^{-2} \\[1ex]
\expval{\gamma_{-1/2} e^{\phi}(z_1)\;\beta_{-1/2} e^{\phi}(z_2)}
&=& - (z_1-z_2)^{-2}
\:.
\ena

The Ramond sector has {\em two} pictures in which the scaling 
dimensions are bounded below, the pictures of charges $-1/2$ and 
$-3/2$.  These two pictures are conjugate to each other in the 
metric on scaling fields, because a product of scaling fields can 
have nonzero expectation value on the 2-sphere, where the metric is 
calculated, only if the sum of the picture charges is $-2$.

Before GSO projection, the dimension 1 fields of picture charge 
$-1/2$ are
\eq
F_1(\gamma_0) e^{-\phi/2} S_\alpha  (z)
\en
and those of picture charge ${-}3/2$ are
\eq
F_2(\beta_0) e^{-3\phi/2} S_\beta (z)
\en
where $S_\alpha(z)$ is the spin field of the spacetime degrees of 
freedom, a 32 component spacetime spinor in $10$ dimensional 
spacetime; $\beta_0$ and $\gamma_0$ are the zero mode operators of 
the spinor ghost fields, satisfying the canonical commutation 
relations $[\gamma_0, \beta_0]=1$; and $F_{1,2}$ are arbitrary 
functions.

The spacetime spinor fields $S_\alpha (z)$ have $32$ components.  
The GSO projection cuts that number in half, to $16$.  Somehow, the 
infinite multiplicity of the $\beta_{0},\gamma_{0}$ zero mode 
representation must give another factor of $1/2$, to obtain the 
contribution of $-8$ to $(\delta_{i}^{i})_{z}$, in order to cancel 
the contribution of ${+}8$ from the \NS\ sector.

Several questions need to be answered.  Which of the two pictures 
should go at each end of a degenerating handle?  How can the Ramond 
sector fields, appearing in two different pictures, $-1/2$ and 
$-3/2$, play the same role as the \NS\ sector fields, appearing in 
the single charge $-1$ picture?  How can a single {\em 
antisymmetric} metric on a space of graded dimension $-8$ be made 
from the symmetric metric $h_{\alpha\beta}$ on the spacetime 
spinors?

All of these questions are answered by finding a formalism in which 
the two conjugate $q=-1/2$ and $q=-3/2$ pictures appear effectively 
in a single picture of charge $-1$.  The key is to represent the 
states of the quantized $\beta,\gamma$ ghost fields in terms of 
distributions~\cite{Verlinde-Verlinde}.  One crucial technical 
subtlety in the nature of these distributions has to be remarked.

Define the Ramond sector field
\eq
S_{\alpha}(\t,z) = \delta(\t-\gamma_{0}) e^{-\phi/2} 
S_{\alpha}(z)
\en
which depends on a spacetime spinor index $\alpha$ and a complex 
number $\t$.  $S_{\alpha}(\t,z)$ is a distribution in the 
complex number $\t$, satisfying
\eq
\int \dif \t \, \t^{m} \, S_{\alpha}(\t,z) =
\gamma_0^m e^{-\phi/2}S_\alpha (z)
\en
\eq
\int \dif \t \, \delta^{(n)}(\t) \, S_{\alpha}(\t,z) =
\beta_0^n e^{-3\phi/2}S_\alpha (z)
\en
where the latter follows from the identity~\cite{Verlinde-Verlinde}
\eq
e^{-3\phi/2} = \delta(\gamma_{0}) e^{-\phi/2}
\:.
\en
The metric is
\eq
\label{eq:fermionmetric}
\expval{S_{\alpha_{1}} (t_{1},z_{1})
\,
S_{\alpha_{2}} (t_{2},z_{2})
}
=
\K(\t_{1},\t_{2})\, h_{\alpha_{1},\alpha_{2}} \,
(z_{1}-z_{2})^{-2}
\en
where $h_{\alpha_{1},\alpha_{2}}$ is the symmetric metric on 
the spacetime spinors, and
\eq
\K(\t_{1},\t_{2}) = \delta(\t_{1}-\t_{2})
\:.
\en
The crucial technical subtlety is that this delta function 
distribution is an {\em odd} function of its argument,
\eq
\delta(\t_{1}-\t_{2})
= -\delta(\t_{2}-\t_{1})
\:.
\en
This is not the real delta function distribution which is a measure 
on the real line, to be integrated against functions of a real 
variable.  Rather, it is the formal delta function of a complex 
variable.  It is to be integrated against analytic functions of the 
complex variable according to the rule
\eq
\int \dif \t \, \delta (\t) f(\t) = f(0)
\:.
\en
This formal delta function can be written as an equivalence class of 
ordinary distributional 1-forms on the complex plane,
\eq
\delta (\t) = \dif\bar{\t} \, 
\delta^{2}(\t,\bar{\t})
\en
modulo $\partial/\partial\bar{\t}$ of an arbitrary distribution 
with compact support on the complex plane.  The formal delta 
function $\delta (\t)$ is a 1-form, therefore an odd object.  
It satisfies, for any nonzero complex number $a$,
\eq
\delta (a \t) = \dif(\bar a \bar{\t}) \,
 \delta^{2}(a \t,\bar a \bar{\t})
= a^{-1}\delta (\t)
\:.
\en
In particular
\eq
\delta (- \t) = -\delta (\t)
\:.
\en

To see that the formal delta function is needed for the 
distributional quantization of the $\beta(z)$, $\gamma(z)$ ghost 
fields, consider the identity
\eq
e^{-\phi}(z) = \delta(\gamma(z))
\en
which is justified by comparing the analytic operator product 
expansions
\eq
\gamma(z) \, e^{-\phi}(0) = z\, \gamma_{-1/2}e^{-\phi}(0) +\cdots
\en
\eq
\gamma(z) \,\delta(\gamma(0)) = z\, \partial\gamma(0)
\delta(\gamma(0)) + \cdots
\:.
\en
Then consider the operator product
\eq
e^{-\phi}(z) \, e^{-\phi}(0) = z^{-1} \, e^{-2\phi}(0) + \cdots
\en
which is translated, for all complex numbers $z$,
\eqa
\delta(\gamma(z)) \, \delta(\gamma(0)) &= &
\delta(z \partial\gamma(0)) \, \delta(\gamma(0)) + \cdots \nonumber \\[1ex]
&= & z^{-1} \, \delta(\partial\gamma(0)) \, \delta(\gamma(0)) 
 + \cdots
\ena
only if the formal delta function is used.  This calculation 
illustrates how the usual quantization of the $\beta$, $\gamma$ 
ghost fields\cite{FMS} is systematically translated into the 
language of formal delta functions\cite{Verlinde-Verlinde}.  More 
details of the translation are given in 
section~\ref{sect:delta-quant} below.

The metric $\K$ is antisymmetric,
\eq
\K(\t_{1},\t_{2}) = - \K(\t_{2},\t_{1})
\en
and it is an odd object, because it is a formal delta function.

The metric $h_{\alpha,\beta}$ on the spacetime spinors can also be 
interpreted as a distribution.  The spacetime spinors $s^{\alpha}$ 
are the functions of $5$ anticommuting variables 
$\that$.
There are $2^{5}=32$ linearly independent functions 
$s^{\alpha}(\that)$.
The symmetric metric $h_{\alpha\beta}$ is represented by
the distribution
\eqa 
\Khat(\that_{1},\that_{2}) &=& s^{\alpha_{1}}(\that_{1})\, 
h_{\alpha_{1}\alpha_{2}} s^{\alpha_{2}}(\that_{2}) \\[1ex]
&=&
\delta^{5}(\that_{1}+\that_{2})
\:.
\ena
The metric on the spacetime spinors $\Khat(\that_{1},\that_{2})$ is 
also an odd object, but symmetric.

The spinor fields $S_{\alpha}(z)$ are rewritten as functions of 
the 5 anticommuting variables $\that$,
\eq
S(\that,z) = S_{\alpha}(z) \, s^{\alpha}(\that)
\en
The Ramond sector scaling field is a function of $\t$ and $\that$,
\eq
S(\t,\that,z) = \delta(\t-\gamma_{0}) \, e^{-\phi/2} \, S(\that,z)
\:.
\en
The Ramond sector metric, equation~\ref{eq:fermionmetric}, is the 
product
\eq
\K (\t_{1},\t_{2}) \, \Khat (\that_{1},\that_{2})
= \delta(\t_{1}-\t_{2}) \, \delta^{5}(\that_{1}+\that_{2})
\:.
\en
It is even, as the product of two odd objects.  It is antisymmetric 
as the product of an antisymmetric metric and a symmetric metric.  
The Ramond sector field $S(\t,\that,z)$ is therefore fermionic.

The GSO transformation sends $\that\rightarrow -\that$ and 
$\t\rightarrow - \t$, so the GSO projected fields are
\eq
\Sgso(\t,\that,z) =
\frac12 S(\t,\that,z) + \frac12 S(-\t,-\that,z)
\:.
\en
The metric on the GSO projected fields is
\eq
\label{eq:fermion-metric}
\frac12
\delta(\t_{1}-\t_{2}) \, 
\delta^{5}(\that_{1}+\that_{2})
+
\frac12
\delta(\t_{1}+\t_{2}) \,
\delta^{5}(\that_{1}-\that_{2})
\:.
\en
When the scaling field $\Sgso(\t,\that,z)$ is smeared with a 
polynomial function of $\t$, it is in picture $-1/2$.  When it is 
smeared with $\delta(\t)$, or derivatives of $\delta(\t)$, it is in 
picture $-3/2$.  Effectively, in any worldsurface calculation, 
$\Sgso(\t,\that,z)$ is midway between the two pictures, which puts 
it in picture $-1$ along with the fields of the \NS\ sector.

The fermionic marginal scaling fields $\phi_{i}(z,\bar z)$ are 
formed by combining the Ramond sector fields $\Sgso(\t,\that,z)$ 
with bosonic scaling fields depending on $\bar z$, for example
\eq
\Sgso(\t,\that,z) \, \bar \partial x^{\mu}(\bar z)
\:.
\en
Fermionic coupling constants $\lambda^{i}$ couple to these fermionic 
scaling fields.

The merging of the two pictures $-1/2$ and $-3/2$ to form a {\em 
virtual} picture $-1$ removes the ambiguity in the assignment of a 
picture at each of the two ends of a local handle.  The handle can 
be represented as a sum of pairs of bosonic fields plus a sum of 
pairs of fermionic fields, each contracted with the handle gluing 
metric.  Whatever picture changing operators are needed near the 
local handle will serve simultaneously to define the insertions of 
both the fermionic and the bosonic scaling fields.

The antisymmetric metric is presented as a kernel in 
equation~\ref{eq:fermion-metric}.  It is a well-defined generalized 
function of the variables $\t_{1},\that_{1},\t_{2},\that_{2}$.  But 
no concrete vector space is defined, on which the antisymmetric 
metric acts as a bilinear inner product.  Formally, the Ramond 
sector fields $S(\t,\that,z)$ lie midway between picture $-1/2$ and 
picture $-3/2$.  This formal description serves all practical 
purposes, since calculations in the lambda model require only 
contractions of products of the metric and its inverse.  But the 
technical foundations of the theory would be more secure if the 
Ramond sector fields could be indexed by a concrete vector space.  
This should be a vector space of functions of $\t$, lying midway 
between the analytic functions and the formal delta functions, 
perhaps some space of half-forms.

\subsection{Lack of positivity in a Ramond-Ramond sector}

If the string worldsurface has a Ramond-Ramond sector, as in any of 
the non-heterotic string theories, there is a serious technical 
difficulty for the lambda model, because the metric on a 
Ramond-Ramond scaling fields is not positive definite.  For example, 
the Ramond-Ramond scaling fields
\eq
\Sgso(\t_{1},\that_{1},z) \,
\bar {\Sgso}(\t_{2},\that_{2},\bar z)
\en
have a metric that is the tensor product of two antisymmetric 
metrics.  Such a tensor product always has directions with negative 
metric.  Bosonic lambda fields would couple as sources to these 
Ramond-Ramond scaling fields.  The metric coupling on those bosonic 
lambda fields would be negative.  The action of the lambda model on 
these bosonic lambda fields would be unbounded below.  I do not see 
how the lambda model could be made to work then.  I do not see how 
there could be control over the short distance worldsurface 
fluctuations of the negative metric bosonic lambda fields, even if 
they are unphysical gauge artifacts.

This pathology seems to disqualify the non-heterotic string theories 
from being associated with a sensible large distance physics, at 
least as provided by the lambda model.  But the pathology is purely 
technical.  It should have a physical interpretation.  There should 
be an explanation in physical terms of what goes wrong with the 
large distance physics in non-heterotic string theories.

\subsection{$\delta_{i}^{i}=0$}

The goal now is to show that the Ramond sector fields contribute 
$-8$ to the graded trace.  The inverse of the kernel
\eq
\delta(\t_{1}-\t_{2}) \, 
\delta^{5}(\that_{1}+\that_{2})
\en
is the kernel
\eq
\dif^{5}\that_{1}\dif\t_{1}
\: \delta(\t_{1}-\t_{2}) \, 
\delta^{5}(\that_{1}+\that_{2})
\: \dif^{5}\that_{2} \dif\t_{2}
\en
because
\begin{displaymath}
\int_{\t_{2},\that_{2}}
\delta(\t_{1}-\t_{2}) \, 
\delta^{5}(\that_{1}+\that_{2})
\;
\dif^{5}\that_{2} \dif\t_{2}
\;
\delta(\t_{2}-\t_{3}) \, 
\delta^{5}(\that_{2}+\that_{3})
\;
 \dif^{5}\that_{3} \dif\t_{3}
\;=
\end{displaymath}
\eq
\delta(\t_{1}-\t_{3}) \, 
\delta^{5}(\that_{1}-\that_{3})
\dif^{5}\that_{3} \dif\t_{3}
\en
which is the kernel of the identity operator.  The inverse metric on 
the Ramond sector fields is then the GSO projection
\eq
\dif^{5}\that_{1} \dif\t_{1} 
\left [
\frac12
\delta(\t_{1}-\t_{2}) \, 
\delta^{5}(\that_{1}+\that_{2})
+
\frac12
\delta(\t_{1}+\t_{2}) \, 
\delta^{5}(\that_{1}-\that_{2})
\right ]
\dif^{5}\that_{2} \dif\t_{2}
\:.
\en
The contribution of the Ramond sector fields to the graded trace is
\eq
- \int \dif^{5}\that_{1} \dif\that_{1}
\left [
\frac12 \delta(\t_{1}-\t_{2}) \, 
\delta^{5}(\that_{1}-\that_{2})
+\frac12  \delta(\t_{1}+\t_{2}) \, 
\delta^{5}(\that_{1}+\that_{2})
\right ]_{\left / \that_{2}=\that_{1},\that_{2}=\that_{1}\right .}
\en
where the overall minus sign comes from the antisymmetry of the 
metric.
The first term inside the integral needs to be regularized
\eqa 
\left [ \delta(\t_{1}-\t_{2}) \, 
\delta^{5}(\that_{1}-\that_{2}) \right 
]_{\left / \that_{2}=\that_{1},\that_{2}=\that_{1}\right .}
&=& 
\lim_{y\rightarrow 1} \; \left [ \delta(\t_{1}-y \t_{1}) \, 
\delta^{5}(\that_{1}-y \that_{2}) \right ] \nonumber\\[1ex]
&=&
\lim_{y\rightarrow 1} \; 
\left [ 
(1-y)^{-1} \delta(\t_{1})
\,
(1-y)^{5} \delta^{5}(\that_{1}) \right ]
\nonumber\\[1ex]
&=& 0
\:.
\ena
The second term contributes
\eqa
- \int \dif^{5}\that_{1} \dif\that_{1}
\; \frac12  \delta(2\t_{1}) \, 
\delta^{5}(2\that_{1})
&=&
- \int \dif^{5}\that_{1} \dif\that_{1}
\; \frac12 \; 2^{-1} \delta(\t_{1}) \;
2^{5} \delta^{5}(\that_{1})
\nonumber \\[1ex]
&=&
-8
\ena
to the graded trace, as was to be shown.  The first factor $1/2$ is 
from the GSO projection.  The factor $2^{5}$ is from the trace over 
spacetime spinors.  The extra factor of $1/2$ comes from the trace 
over the states of the bosonic ghost zero modes $\beta_{0}$, 
$\gamma_{0}$.

\subsection{Second order wave equation}

In flat spacetime, the almost marginal fermionic scaling fields take 
the form
\eq
\phi_{i}(z,\bar z)=
\Sgso(\t,\that,z) \, \bar \partial x^{\mu}(\bar z)
\,\me^{ipx}
\en
indexed by $i=(\t,\that,\mu, p_{\mu})$.  The anomalous scaling 
dimension is $\ad(i)=p^{2}$.  The spacetime equation of motion 
$\beta^{i}(\lambda)=0$ linearizes to $\ad(i) \lambda^{i} =0$ which 
is the second order differential equation in spacetime $p^{2} 
\lambda^{i}=0$.  In a curved spacetime, the linearized equation of 
motion on the fermionic lambda modes becomes a covariant second 
order differential operator.

On the on-shell states, which satisfy $p^{2}=0$, the worldsurface 
BRS operator is $\t\!\pslash$, where $\pslash$ is the spacetime 
Dirac operator.  The physical states, in either of the two conjugate 
Ramond sector pictures, are the BRS cohomology classes.  In either 
picture, the BRS cohomology classes are the solutions of the first 
order spacetime Dirac equation.  The infinite multiplicity of the 
ghost zero modes is eliminated.

The \apm\ of the lambda model, interpreted as a spacetime quantum 
field theory, uses a second order differential wave equation on the 
fermionic fields, not the traditional Dirac equation.  But the 
physical content is the same.

\subsection{Quantizing the $\beta$, $\gamma$ ghost fields using the 
formal delta function}
\label{sect:delta-quant}

The $\beta, \gamma$ ghost fields are expanded in modes
\eq
\beta(z) = \sum_{n} z^{-n-3/2}\beta_{n}
\qquad
\gamma(z) = \sum_{n} z^{-n+1/2}\gamma_{n}
\en
where the index $n$ is integer in the Ramond sector, integer plus 
half in the NS sector.  The modes satisfy canonical commutation 
relations
\eq
{[} \gamma_{m},\,\beta_{n} ] = \delta_{m+n}
\:.
\en
The ground state of picture charge $q$ is the state $\ket{q}$, 
satisfying
\eqa
\gamma_{n} \ket{q} = 0
			&\qquad& n-q = 3/2, \, 5/2,\, \cdots \\[1ex]
\beta_{n} \ket{q} = 0
			&\qquad& n+q = -1/2, \, 1/2,\,\cdots
\:.
\ena
The ground state $\ket{0}$ is the $SL_{2}$ invariant state.

In the bosonization formalism for the $\beta, \gamma$
ghosts,
\eq
\ket{q} = e^{q\phi}(0) \, \ket{0} 
\en

The states are represented as distributional wave 
functions\cite{Verlinde-Verlinde}, say of the $\gamma_{n}$.  The 
$\beta_{n}$ act as derivative operators
\eq
\beta_{n}= -\partialby{\gamma_{-n}}
\:.
\en
The ground state of picture charge $q$ is
\eq
\ket{q} = 
\delta(\gamma_{3/2+q})
\,\delta(\gamma_{5/2+q})
\,\delta(\gamma_{7/2+q}) \, \cdots.
\en
The dual states are
\eq
\bra{q} = \cdots
\,\delta(\gamma_{-7/2-q})
\,\delta(\gamma_{-5/2-q})
\,\delta(\gamma_{-3/2-q})
\en
so
\eq
\inprod{-q-2\,}{\,q} = 1
\en
The states and dual states can be thought of as two classes of 
analytic subvarieties in the infinite dimensional analytic manifold 
whose coordinates are the $\gamma_{n}$.  The inner product is the 
intersection number of the subvarieties.

The field $\delta(\gamma(z))$ acts on the ground state $\ket{q}$
by
\eq
\delta(\gamma(z)) \,\ket{q} =
\delta(z^{-q}\gamma_{1/2+q}+\cdots)\,\ket{q}
= z^{q} \, \ket{q-1} + \cdots
\en
where the formal delta function must be used in order that the 
operator product expansion will be analytic.

The inner product on the zero mode wave functions is obtained by 
noting that
\eq
\bra{-1/2} F(\gamma_{0}) \ket{-1/2} = \int \dif\gamma_{0} 
\; F(\gamma_{0})
\en
then calculating
\eqa
\bra{-1/2} \,
\delta(\t_{1}-\gamma_{0})\,\delta(\t_{2}-\gamma_{0})
\, \ket{-1/2} &=& \int \dif\gamma_{0}
\; \delta(\t_{1}-\gamma_{0})\,\delta(\t_{2}-\gamma_{0})\nonumber\\[1ex]
&=& \delta(\t_{1}-\t_{2})
\:.
\ena
A formal integral representation of the formal delta function,
\eq
\delta(\gamma) = \int \dif t \, e^{t\gamma}
\en
allows such calculations as
\eqa
\delta(\beta(z)) \, \ket{q}
		&=& \delta(z^{q}\beta_{-q-3/2}+\cdots) \, \ket{q}\nonumber\\[1ex]
		&=& z^{-q}\,\delta(\beta_{-q-3/2})
					\,\delta(\gamma_{3/2+q}) \, \ket{q+1}+\cdots
					\nonumber\\[1ex]
		&=& z^{-q}\int \dif t \,
		\exp\left (-t\partialby{\gamma_{3/2+q}}\right )
					\, \delta(\gamma_{3/2+q}) \, \ket{q+1}+\cdots
					\nonumber\\[1ex]
		&=& z^{-q}\int \dif t 
					\, \delta(-t+\gamma_{3/2+q})\,  \ket{q+1}+\cdots
					\nonumber \\[1ex]
		&=& - z^{-q} \, \ket{q+1}+\cdots
\ena
which is the operator product needed to make the identification
\eq
e^{-\phi}(z) = - \delta(\beta(z))
\:.
\en
The identities
\eq
1 = \int \dif\gamma \, \delta(\gamma) 
= \int \dif\gamma \int \dif t \, e^{t\gamma}
= - \int \dif t \int \dif\gamma \,  e^{t\gamma}
= \int \dif t \, \delta(t)
\en
are justified by the fact that the formal expression $e^{t\gamma}$ 
is odd under exchange of $t$ and $\gamma$, because it implicitly 
contains the factor $\dif \bar t \dif\bar \gamma$.

%
%
%
%
%
\sectiono{Geometric identities on the manifold of spacetimes}

Scale invariance in $2+\epsilon$ dimensions in a nonlinear model 
such as the lambda model is expressed by the vanishing of the beta 
function, which is a geometric identity on the target manifold of 
the model~\citeF. To one loop, the geometric identity expressing 
ordinary scale invariance is
\eq
0 = - \epsilon\, T^{-1}g_{ij} + 2 R_{ij}
\:.
\en
The numerical coefficient 2 multiplying the Ricci tensor is due to 
the normalization of the action $S(\lambda)$, which is the same as 
the normalization of the general nonlinear model, which is designed 
to give anomalous dimensions of the form $p(i)^{2}$, with numerical 
coefficient 1.

Scale invariance of the generalized kind is expressed by a somewhat 
more elaborate geometric identity involving the potential function 
$T^{-1}a(\lambda)$ and whatever other couplings occur in the 
nonlinear model~\cite{Friedan-1,Friedan-2,Friedan-3,CFMP}.

The effective metric coupling $T^{-1}\geff_{ij}(\lambdaeff)$ and the 
effective potential function $T^{-1}\aeff(\lambdaeff)$ and whatever 
other effective couplings might arise will satisfy the geometric 
identities expressing generalized scale invariance.

I will not write these {\it meta} Einstein equations here.  The 
quantum corrections to the metric coupling and the other couplings 
will enter at each order, so the full import is in the exact 
equations, not in their truncation to one loop or to any finite 
number of loops.  The geometric identities will involve the 
effective potential function $T^{-1}\aeff(\lambdaeff)$, which is the 
effective action in the spacetime quantum field theory.  It will 
eventually be interesting to ask what significance the meta Einstein 
equations might have in the special spacetime quantum field theories 
produced by the lambda model.

\subsection{Geometric identites from perturbative spacetime 
supersymmetry}

Perturbative spacetime supersymmetry suppresses string loop 
corrections.  The lambda model is formulated to cancel the effects 
of the string loop corrections at large distance in spacetime.  
Perturbative spacetime supersymmetry of the string theory in a given 
spacetime $\lambda$ will be mirrored as a perturbative symmetry of 
the lambda fluctuations around the point $\lambda$ in the target 
manifold.  Perturbative spacetime supersymmetry will supress 
perturbative quantum corrections to the couplings of the lambda 
model at the point $\lambda$ in its target manifold.

In particular, perturbative spacetime supersymmetry preserves the 
degeneracy of the manifold of spacetimes $\M{\infty}$ against 
perturbative quantum corrections.  $\M{\infty}$ is the manifold of 
solutions of $\beta(\lambda)=0$.  The restricted lambda model is the 
formal, perturbative nonlinear model whose target manifold is 
$\M{\infty}$.  In the restricted lambda model, 
$\betaeff(\lambdaeff)=0$ perturbatively on $\M{\infty}$.  So the 
restricted lambda model will be scale invariant order by order in 
the loop expansion, in the ordinary sense of scale invariance.  The 
vanishing of the perturbative beta function for the metric coupling 
of the restricted lambda model means that the metric 
$T^{-1}g_{ij}(\lambda)$ on $\M{\infty}$ will satisfy a series of 
geometric identities, indicative of a very special geometry.  The 
fermionic directions in the manifold of spacetimes are essential for 
these identities, since the identities arise from cancellations 
between the bosonic and fermionic directions in the manifold of 
spacetimes.

The first cancellation is the vanishing of the (graded) trace
\eq
\delta^{i}_{i} = T^{-1}g_{ij} \; T \, g^{ji} = 0
\en
which states that the graded dimension of the manifold of spacetimes 
is zero.  The equation $\delta^{i}_{i}=0$ is easily recognized from 
the string loop expansion.  It is the condition that the one loop 
correction to the vacuum string amplitude is finite.  The one loop 
correction to the vacuum amplitude is
\eq
\label{eq:one-loop-vacuum}
\int \Z_{1}(q,\bar q)
\en
where $\Z_{1}(q,\bar q)$ is the partition function of the genus $1$ 
worldsurface, the complex one-torus parametrized by $q=e^{2\pi 
i\tau}$.  The integral is over the modular domain of the upper half 
complex $\tau$ plane.  The partition function is nonsingular 
everywhere except possibly at $q=0$, where the torus degenerates.  
The only place where the integral might diverge is at $q=0$.  The 
complex one-torus near $q=0$ is an almost degenerate handle 
connected to a 2-sphere.  The integral is cut off at $\abs{q}^{1/2}> 
\mu\Lambdazero^{-1}$.  The divergent part is
\eq
\Lambdazero\partialby{\Lambdazero}
\; \int \dif^{2}q \, \frac{1}{2\pi}
\;
\abs{q}^{-4}
\;
\abs{q}^{2+\ad(i)}
\;
T\,g^{ij}
\;
T^{-1} g_{ij}
= 
2
(\mu\Lambdazero^{-1})^{2\ad(i)}
\;
T\,g^{ij}
\;
T^{-1} g_{ij}
\:.
\en
Finiteness follows from the existence of a conserved, holomorphic 
supersymmetry current $Q_{S}(z)$ on the worldsurface, with 
charge operator $Q_{S}$, and the existence of a conjugate operator 
$Q^{\prime}_{S}$ such that~\cite{FMS}
\eq
{}[Q_{S}, \; Q^{\prime}_{S} ] = 1
\:.
\en
Finiteness in the limit $\Lambdazero^{-1}\rightarrow 0$ is precisely 
the condition $\delta_{i}^{i}=0$, where the graded trace is taken 
over the marginal coupling constants, those having $\ad(i)=0$.

A more subtle version of this argument should work locally in a 
spacetime $\lambda$ that lies in $\M{L}$, at nonzero short two 
dimensional distance $\Lambda^{-1}$.  The argument should give a 
version of the vanishing of the graded trace, $\delta_{i}^{i}=0$, 
that applies locally in spacetime.

\subsection{The meta Einstein equation on $\M{\infty}$}

The second geometric identity is a meta Einstein equation on 
$\M{\infty}$, expressing one loop scale invariance of the metric 
coupling of the restricted lambda model in $d=2+\epsilon$ 
dimensions, with $\epsilon=T/2$,
\eq
\label{eq:meta-einstein}
0 = - \frac{1}{2} g_{ij} + 2 \, R_{ij}
\:.
\en
The term $2\,R_{ij}$ is the usual one loop beta function of the 
nonlinear model.  The term $-\frac{1}{2} g_{ij}$ is the contribution 
from the scale variation of the general nonlinear model, 
equation~\ref{eq:scale-variation-Minfinity}.

It should be possible to derive the meta Einstein 
equation~\ref{eq:meta-einstein} directly from one loop finiteness of 
the string loop corrections.  Differentiating the finite one string 
loop vacuum correction, equation~\ref{eq:one-loop-vacuum}, with 
respect to the marginal coupling constants $\lambda^{i}$, gives the 
finiteness of the one string loop correction to the one point 
function.  This is the vanishing of the one loop correction to 
$\beta^{i}(\lambda)=0$.  Differentiating the finite one loop vacuum 
correction, equation~\ref{eq:one-loop-vacuum}, twice with respect to 
the marginal coupling constants $\lambda^{i}$, gives the finiteness 
of the one string loop correction to the two point function.  As 
before, the scale variation must then vanish.  The scale variation 
extracts the contribution of a degenerating handle attached to a 
2-sphere in which there are two scaling fields $\phi_{i}$, 
$\phi_{j}$.  This contribution is an integral of the four point 
expectation value on the 2-sphere, contracted with a handle gluing 
matrix, of the form
\eq
\int \;g^{kl}\,
\expval{\,\phi_{i}(1) ,\phi_{j}(2),\phi_{k}(3),\phi_{l}(4)\,}
\:.
\en
The Ricci tensor of the metric $g_{ij}$ can be calculated from the 
scale variation of the general nonlinear model with sources, 
equation~\ref{eq:scale-variation-Minfinity}.  The metric is a two 
point expectation value of scaling fields.  The curvature tensor is 
made from two derivatives of the metric, so the curvature is given 
by an integral of an expectation value of four scaling fields.  The 
Ricci tensor is then obtained by contracting the curvature tensor 
with the inverse metric, $g^{ij}$.

These results of these two calculations have the same form, so it is 
plausible that the meta Einstein equation can be derived explicitly 
from one loop string finiteness.  Heuristically, the one loop 
finiteness of the string loop corrections gives rise to an identity 
on the metric which involves two derivatives of the metric.  By 
covariance in $\M{\infty}$, this identity should be of the form of 
the meta Einstein equation.  Only the relative numerical coefficient 
$1/4$ between the two terms needs to be verified.

It should also be possible to verify the meta Einstein 
equation~\ref{eq:meta-einstein} by explicit calculation of the Ricci 
tensor of the metric $T^{-1}g_{ij}(\lambda)$ on $\M{\infty}$, at 
least in simple cases such as the manifold of toroidal spacetimes.

The restricted lambda model is perturbatively finite because of its 
generalized scale invariance, which is a basic property of the 
lambda model.  Perturbative spacetime supersymmetry is only an 
accidental property of individual spacetimes.  Perturbative 
spacetime supersymmetry simplifies the realization of generalized 
scale invariance in the lambda model, by maintaining the degeneracy 
of the manifold of spacetimes $\M{\infty}$ against perturbative 
corrections.  As a consequence, there are strong identities on the 
geometry of the manifold of perturbatively supersymmetric 
spacetimes.  For physics, perturbative spacetime supersymmetry is 
useful because, by maintaining the degeneracy against perturbative 
corrections, it guarantees that any effects that lift the degeneracy 
will be nonperturbatively small.

%
%
%
%
%
\sectiono{Lambda instantons}
\label{sect:lambda-instantons}

The dominant nonperturbative effects in the lambda model will be 
produced by harmonic surfaces in the space of string backgrounds, 
the {\em lambda instantons}.  A lambda instanton is a classical 
field configuration $\lambda_{H}(z,\bar z)$ which is a local minimum 
of the lambda model action, $S(\lambda)$.  These are the harmonic 
surfaces in the manifold of spacetimes.

There are at least two kinds of lambda instanton.  The {\em global} 
lambda instantons, are the harmonic surfaces in the manifold 
$\M{\infty}$.  The action $S(\lambda_{H})$ of a global lambda 
instanton is on the order of $T^{-1}$, so only collective effects of 
global lambda instantons will be significant.  I will describe here 
one elementary example of a global lambda instanton, in the manifold 
of toroidal spacetimes, and speculate on possible collective effects 
that might single out a macroscopic spacetime.

At a macroscopic spacetime $\lambda$ in $\M{L}$, there are {\em 
localized} lambda instantons, which are harmonic surfaces in the 
manifold $\M{L}$, localized in the macroscopic spacetime at 
spacetime distances on the order of $L$.  These are harmonic 
surfaces in the manifold of spacetime fields.  The action 
$S(\lambda_{H})$ of a localized lambda instanton is on the order of 
$\gst^{-2}=(V\,T)^{-1}$, so the localized lambda instantons could 
possibly produce interesting characteristic nonperturbative 
spacetime distances.  I only point out here that local lambda 
instantons exist.

\subsection{Example of a global lambda instanton}

Consider a family of spacetimes in $\M{\infty}$ with two toroidal 
dimensions.  Each spacetime is the product of a two dimensional real 
torus with a fixed eight dimensional manifold.  The two dimensional 
real torus is a complex 1-torus with a Kahler form proportional to 
a complex number $\sigma$ in the upper half complex plane.  The 
volume of the torus is the imaginary part, $\Imag(\sigma)$.  All of 
the other parameters describing the spacetime are held fixed, 
including the parameter describing the complex structure of the 
complex 1-torus.

The modular group is the group of fractional linear transformations 
of the upper half plane with integer coefficients, 
$\sigma\rightarrow (a\sigma+b)/(c\sigma + d)$.  The modular group is 
generated by $\sigma\rightarrow \sigma+1$ and $\sigma\rightarrow 
-1/\sigma$.  The two dimensional quantum field theories of the 
worldsurface parametrized by $\sigma$, $\sigma+1$, and $-1/\sigma$ 
are all equivalent.  So the family of spacetimes in $\M{\infty}$ is 
parametrized by the modular domain, which is the quotient of the 
upper half complex $\sigma$ plane by the action of the modular 
group.  The modular domain can be parametrized by the classical 
modular function $j(\sigma)$ whose values range over the entire 
complex plane when $\sigma$ ranges over the modular domain.  The 
family of toroidal spacetimes is parametrized by the complex $j$ 
plane.

The torus becomes macroscopic in the limit $\Imag(\sigma) 
\rightarrow \infty$.  In this limit, $j \approx e^{-2\pi i \sigma}$.  
The family of spacetimes can be compactified to a 2-sphere by 
appending the point $j=\infty$.  The compactified family of 
spacetimes is a complex curve of genus $0$, parametrized by the 
complex projective $j$ plane.

The {\it $j$-instanton} is the three parameter family of maps from 
the worldsurface to $\M{\infty}$
\eq
j(z,\bar z) = \frac{a z + b}{c z + d}
\en
parametrized by complex numbers $a,b,c,d$ satisfying $ad-bc=1$.  The 
$\bar j$-instanton is the complex conjugate map.  The three complex 
parameters are just the paramters of the group $SL_{2}(C)$, the 
conformal group of the instanton.  The $j$-instanton and the $\bar 
j$-instanton are each three parameter families of global lambda 
instantons.  They depend implicitly on all the other parameters of 
the spacetime, the parameters describing the fixed eight dimensional 
manifold and the complex structure of the torus.

Compactifying the family of 2-tori with the point $j=\infty$ adds a 
submanifold to $\M{\infty}$, described by all the other parameters 
of the spacetime besides $j$.  Near $j=\infty$, the two spacetime 
dimensions of the 2-torus become macroscopic.  The volume of the 
macroscopic spacetime is $V= \Imag(\sigma) \approx 
(2\pi)^{-1}\ln\abs{j}$.  The $j=\infty$ submanifold is part of the 
decompactification locus.

There are three distinguished points in the modular domain, the 
decompactification point $j=\infty$, and two orbifold points at 
$j=0$ and $j=1728$.  The point $j=0$ corresponds to $\sigma = 
e^{i\pi/3}$, which is the fixed point of the $\IZ_{3}$ subgroup of 
modular transformations generated by $\sigma \rightarrow 
1-\sigma^{-1}$.  The point $j=1728$ corresponds to $\sigma = i$, the 
fixed point of the $\IZ_{2}$ subgroup of modular transformations 
generated by $\sigma \rightarrow -\sigma^{-1}$.  The 
decompactification point $j=\infty$ can also be regarded as an 
orbifold point, left fixed by the full integer subgroup $\IZ$ 
generated by $\sigma\rightarrow\sigma+1$.

The three complex parameters of the $j$-instanton can be taken to be 
the three points on the worldsurface, $z_{3}$, $z_{2}$, 
$z_{\infty}$, where the $j$-instanton passes through the three 
orbifold points, $j(z_{3})=0$, $j(z_{2})=1728$ and 
$j(z_{\infty})=\infty$.  The $j$-instanton can be described as a 
configuration of three lambda defect operators, $\tau_{3}$, 
$\tau_{2}$, $\tau_{\infty}$,
\eq
\tau_{3}(z_{3},\bar z_{3}) \;
\tau_{2}(z_{2},\bar z_{2}) \;
\tau_{\infty}(z_{\infty},\bar z_{\infty})
\:.
\en
Similarly, the $\bar j$-instanton is described as a configuration of 
three complex conjugate defect operators $\bar \tau_{3}$, $\bar 
\tau_{2}$, $\bar \tau_{\infty}$.

At each of the orbifold points $j=0$, $j=0$ or $j=1728$, the 
orbifold group, or defect group, $\IZ_{3}$, $\IZ_{2}$ or $\IZ$, acts 
as a group of internal symmetries of the worldsurface.  The 
worldsurface in the spacetime $j=0$ has a $Z_{3}$ symmetry; the 
worldsurface in the spacetime $j=1728$ has an internal $Z_{2}$ 
symmetry.  There is no actual spacetime at the decompactification 
point $j=\infty$, so the action of the orbifold group, the integers 
$\IZ$, has to be defined in the limit $j\rightarrow \infty$ as an 
internal symmetry group of the worldsurface.

Each defect operator $\tau$ or $\bar \tau$ pins the worldsurface to 
an orbifold point in the manifold of spacetimes.  The defect 
$\tau_{3}(z,\bar z)$ pins the point $z$ to the torus $j=0$.  The 
defect $\tau_{3}(z,\bar z)$ pins $z$ to the torus $j=0$.  The 
decompactifying defect, $\tau_{\infty}$ pins the point $z$ to the 
decompactification locus at $j=\infty$.

Each lambda defect operator is associated to an element in the 
corresponding orbifold or defect group.  The group element is the 
monodromy of the coupling constants $\lambda^{i}$ circling the 
defect operator on the worldsurface.  Away from the lambda defects, 
the scaling fields $\phi_{i}(z,\bar z)$ vary adiabatically over the 
lambda instanton $\lambda_{H}(z,\bar z)$, in a path independent 
fashion, because nearby general nonlinear models have the same 
degrees of freedom.  But when a path on the worldsurface circles 
around one of the lambda defect operators, the scaling fields 
$\phi_{i}$ are transformed among themselves by the element of the 
orbifold group carried by the defect operator.  The lambda defect 
acts on the worldsurface as the twist operator of the orbifolded 
spacetime.  The lambda defect operator twists locally by its 
orbifold group element, projecting on the invariant degrees of 
freedom, removing the non-invariant degrees of freedom, and adding 
twist fields as new effective degrees of freedom on the 
worldsurface.

The nonperturbative lambda model is a two dimensional gas of lambda 
defect operators.  At issue is the detailed dynamics of the defect 
gas.  Is it a plasma?  Or a neutral gas, with the defects all bound 
together?  Or a combination, a plasma of some defects and some bound 
systems of defects?

\subsection{Existence of localized lambda instantons}

Near a macroscopic spacetime, there will exist lambda instantons in 
the target manifold $\M{L}$ which are localized in bounded regions 
of the macroscopic spacetime.  These localized lambda instantons are 
the harmonic surfaces in the manifold of spacetime fields.

There is a standard topological argument for the existence of 
instantons~\cite{Singer}.  The localized lambda instantons are 
indexed by the second homotopy group, $\pi_{2}$, of the manifold of 
the target manifold $\M{L}$.  Every homotopy class of 2-spheres in 
$\M{L}$ should contain local minima of $S(\lambda)$.

The target manifold $\M{L}$ is the manifold of fields of the 
effective spacetime field theory.  The spacetime fields are 
localized, which means that they go to zero outside a bounded region 
of the macroscopic spacetime, or more generally become trivial 
there.  So the spacetime fields can be regarded as defined on a ball 
in $n$-dimensional euclidean space, where $n$ is the dimension of 
the macroscopic spacetime, and the boundary of the ball can be 
identified to a point.  Topologically, the spacetime fields can be 
regarded as defined on the $n$-sphere.

The manifold of spacetime fields is actually the manifold of gauge 
equivalence classes of spacetime tensor fields, including the metric 
tensor and the gauge fields.  The manifold of localized spacetime 
fields is the quotient manifold $\F_{n}/\G_{n}$, where $\F_{n}$ is 
the space of tensor fields on the $n$-sphere and $\G_{n}$ is the 
group of local gauge transformations on the $n$-sphere.

The second homotopy group $\pi_{2}(\F_{n}/\G_{n})$ is calculated 
using the long exact sequence:
\eq
\cdots \rightarrow \pi_{k}(\F_{n}) \rightarrow \pi_{k}(\F_{n}/\G_{n})
\rightarrow \pi_{k-1}(\G_{n})
\rightarrow \pi_{k-1}(\F_{n}) \rightarrow \cdots
\en
the relevant part of which is
\eq
\cdots \rightarrow \pi_{2}(\F_{n}) \rightarrow \pi_{2}(\F_{n}/\G_{n})
\rightarrow \pi_{1}(\G_{n})
\rightarrow \cdots
\:.
\en

Nontrivial topology in the manifold $\F_{n}$ of tensor spacetime 
fields comes only from the spacetime scalar fields.  The space of 
metrics and gauge fields is topologically trivial before gauge 
equivalence is taken into account.  The spacetime scalar fields take 
their values the parameters that describe the non-macroscopic 
dimensions of the spacetime.  These are the coupling constants that 
parametrize the decompactification locus $\M{L}_d$.  The 
scalar fields form a map from the $n$-sphere to the 
decompactification locus.  So
\eq
\pi_{2}(\F_{n}) = \pi_{n+2}(\M{L}_d)
\:.
\en
When $n=0$, this is the homotopy group that classifies the global 
lambda instantons.

Localized lambda instantons formed from the spacetime scalar fields 
might have interesting physical effects.  Locally in spacetime, they 
might pin to submanifolds of the decompactification locus where 
additional spacetime dimensions become macroscopic.

The localized lambda instantons formed from the spacetime metric and 
the spacetime gauge fields are indexed by the first homotopy group 
of the local gauge group, $\pi_{1}(\G_{n})$.  If the global internal 
gauge group is $G$, then the local gauge transformations are maps 
from the $n$-sphere to $G$.  They contribution $\pi_{n+1}(G)$ to 
$\pi_{1}(\G_{n})$.  The local gauge transformations of the spacetime 
metric are the maps from the $n$-sphere to itself, so they 
contribute $\pi_{n+1}(S^{n})$.  For $n=4$, these homotopy groups are 
typically nontrivial, so localized lambda instantons do exist.

It is not clear to me that this is a complete classification of the 
localized lambda instantons.  In order to find the example of a 
global lambda instanton described above, the $j$-instanton, it was 
necessary to complete the manifold of spacetimes by adding the 
decompactification locus at $j=\infty$.  Is there an analogous 
process of completion for the space of localized spacetime metrics 
and gauge fields modulo gauge equivalence, which would give rise to 
additional localized lambda instantons?

It seems quite possible that localized lambda defects will exist.  A 
localized lambda defect would occur at a point $z$ on the 
worldsurface where a localized lambda instanton $\lambda_{H}(z,\bar 
z)$ passes through a spacetime field configuration with symmetry.  
The symmetry subgroup of the local spacetime gauge group would be 
the defect group of the localized lambda defect.  The nontrivial 
closed path in the local gauge group $\G_{n}$ associated with the 
localized lambda instanton would then be composed of a sequence of 
path segments, each path segment implementing a defect twist.  The 
homotopy argument shows the existence of local lambda instantons.  
They still need to be constructed explicitly.  Then it can be 
determined whether they are smooth objects or composed of local 
lambda defects.

The localized lambda instantons in $\M{L}$ are made from the 
spacetime wave modes at spacetime distances greater than $L$.  The 
spacetime physics at distance $L$ will be affected by those 
localized lambda instantons that are made from the spacetime wave 
modes $\lambda^{i}$ at distances of the order of $L$.  Calculations 
of their effects will be done locally in spacetime, in local 
spacetime regions at distances of the order of $L$.

Taking $L\rightarrow\infty$ contracts $\M{L}$ to $\M{\infty}$, 
formally.  The localized lambda instantons in $\M{L}$ are pushed 
closer and closer to the decompactification locus $\M{\infty}_d$.  
There should be an interpretation of the limit defining a completion 
of $\M{\infty}$ that can stand for the target manifold of the lambda 
model at $L=\infty$.  The limit $L\rightarrow\infty$ will be a 
practical issue in calculations of the properties of decompactifying 
lambda defects, such as $\tau_{\infty}$.  Spacetime is macroscopic 
in the core of a decompactifying defect.  The core of the defect is 
dressed with localized lambda instantons in the macroscopic 
spacetime.  The limit $\Lambda^{-1}\rightarrow 0$ will see the 
center of the decompactifying defect, where the difficulties of the 
$L\rightarrow \infty$ limit will have to be resolved.

\subsection{Lambda instanton calculations}

To calculate the quantum corrections to lambda instanton 
configurations, some way is needed to calculate the contribution of 
the general nonlinear model in the presence of a nontrivial lambda 
field $\lambda_{H}(z,\bar z)$.  The general nonlinear model 
contributes at order $T^{0}$, the same order as the one loop 
corrections in the lambda model.  Each contributes a pre-factor 
multiplying the classical instanton contribution 
$e^{-S(\lambda_{H})}$.  Neither pre-factor is scale invariant 
separately, but only the combination.

In principle, the general nonlinear model in the presence of a 
lambda instanton can be made out of local two dimensional patches, 
the sources $\lambda^{i}(z,\bar z)$ being almost constant within 
each patch.  But I have no practical method of putting together the 
patches that could be used for calculation.  A possibly effective 
method of calculation might be to treat the general nonlinear model 
in the presence of a lambda instanton as a correlation function of 
lambda defects, then calculate using the monodromy properties of the 
defects.

It is this difficulty of calculation that motivates the proposal of 
section~\ref{sect:d=2+epsilon} to account for the general nonlinear 
model contribution to the lambda model by continuing the dimension 
from $d=2$ to $d=2+\epsilon$, dropping the general nonlinear model 
entirely, and determining the quantum corrections by finding the 
scale invariant fixed point of the effective lambda model in 
$d=2+\epsilon$ dimensions.

\subsection{Speculation about the nonperturbative structure}

I cannot resist indulging in some premature idle speculation about 
the nonperturbative lambda model.  Lambda instantons will make 
nonperturbative corrections to the beta function $\beta^i(\lambda)$ 
of the general nonlinear model.  It seems possible that these 
corrections will disturb the degeneracy of the manifold of 
spacetimes.  I see two ways this might happen.

In the first type of scenario, nonperturbative corrections to 
$\beta^{i}(\lambda)$ simply single out some particular spacetimes 
from the manifold of spacetimes.  These become the local minima of 
the effective potential function $T^{-1}\aeff$.  The \apm\ 
concentrates at these particular spacetimes, breaking the 
degeneracy.  Global lambda instantons might concentrate the \apm\ at 
a particular macroscopic spacetime, at a particular point near the 
locus of decompactification.  In that macroscopic spacetime, local 
lambda instantons might contribute terms to the local spacetime 
action $\gst^{-2}\,V\,\aeff$, violating perturbative spacetime 
supersymmetry and giving the perturbatively massless spacetime 
fields definite vacuum expectation values and small masses.  The 
original perturbative degeneracy of the manifold of spacetimes would 
come to be seen as merely accidental.

In the second type of scenario, the lambda instantons disorder the 
system.  A plasma of lambda defects would accomplish this.  The 
lambda defects would act as twist operators, projecting on the 
singlets of the defect group, removing the non-singlet degrees of 
freedom, and adding the twist degrees of freedom.  The degrees of 
freedom $\lambda^{i}$ would take entirely different effective forms.  
An effective target manifold $\M{L}_{\eff}$ would replace the 
original target manifold $\M{L}$.  The effective \apm\ might 
concentrate at particular places in the effective target manifold 
$\M{L}_{\eff}$.  Or something more complicated might happen, perhaps 
a hierarchy of disordered systems.

A lambda instanton makes logarithmically divergent corrections to 
the general nonlinear model when it is configured as a 2-sphere 
connected to the worldsurface by an almost degenerate handle.  The 
lambda instanton is a complex analytic curve of genus 0, so three 
complex parameters describe its configuration in the worldsurface.  
In the $j$-instanton, for example, the three complex parameters are 
the locations of the three lambda defects $\tau_{2}$, $\tau_{3}$, 
$\tau_{\infty}$.  In the logarithmically divergent configuration, 
one parameter becomes the point on the worldsurface where the 
instanton is attached.  One parameter is the thickness of the handle 
by which the instanton is attached.  The third parameter is the 
point on the lambda instanton where it is attached to the 
worldsurface.  The effective measure on the third parameter 
determines what states flow through the handle to appear on the 
worldsurface as corrections to $\beta^{i}(\lambda)$.  If the measure 
on the third parameter is concentrated at a smooth point on the 
lambda instanton, the first scenario will apply.  That point in the 
manifold of spacetimes will be singled out.  The lambda defects will 
appear entirely bound.  If the measure on the third parameter is 
concentrated at one or more of the lambda defects, the worldsurface 
will see a plasma of lambda defects.  Intermediate possibilities 
might better be described as an interacting gas of lambda defects on 
the worldsurface.  The effective measure on the configuration of the 
lambda instanton might depend on the spacetime distance $L$.  The 
interactions among the lambda defects might depend on $L$, so the 
effective form of the degrees of freedom might change with $L$.

For a local lambda instanton, in the first kind of scenario, where a 
particular spacetime is singled out on the lambda instanton, the 
lambda instanton will insert local fields with logarithmically 
divergent coefficients into the general nonlinear model.  If no 
spacetime supersymmetry generator can be globally defined over the 
lambda instanton, then the divergent insertions can violate 
spacetime supersymmetry.  Likewise, any other spacetime symmetry can 
be removed, if the symmetry generator cannot be defined as a 
single-valued object over the lambda instanton.  Perhaps even local 
spacetime gauge symmetry might be removed in this fashion.

In the disordered scenario, a plasma of lambda defects could 
distribute the \apm\ over the lambda instanton.  Alternatively, 
degeneracy could be broken by pinning to the orbifold spacetimes.  
For example, a plasma of global decompactifying lambda defects, like 
the defect $\tau_{\infty}$ of the $j$-instanton, would pin the 
system to the locus of decompactification.  The $\tau_{2}$ and 
$\tau_{3}$ defects would appear bound.  This would be a novel form 
of decompactification, described by the orbifolded general nonlinear 
model at the decompactification locus.  Such virtual orbifold models 
still need to be analyzed.  The simplest case to examine is the 
$\IZ$ orbifold of the 2-torus at $j=\infty$.  There would presumably 
be no definite global spacetime geometry.  Twisting by the defect 
$\tau_{\infty}$ would remove the angular parameter of the global 
spacetime geometry, the real part of $\sigma$, as a degree of 
freedom.  In general, lambda defects will disorder angular 
parameters in the neighborhood of the orbifold point in the manifold 
of spacetimes.  This is a tantalizing possibility.  Mechanisms that 
might remove angular degrees of freedom are especially interesting 
because of the problem of the $\theta$ parameter in QCD.

Symmetries such as spacetime supersymmetry might also be removed as 
a result of twisting in a plasma of local lambda defects.  If a 
spacetime supersymmetry generator winds nontrivially around a lambda 
defect, then the generator would be removed by the plasma of defects.

A form of spacetime gauge confinement could conceivably be produced 
by a plasma of local lambda defects twisting by elements of the 
local gauge group.  The plasma would disorder the local spacetime 
gauge group, projecting on gauge singlets.  The most interesting 
case to investigate is of course the $SU(3)$ local gauge group in 
four spacetime dimensions.  Perhaps this could provide a viable 
alternative to the hypothetical quantum field theoretic confinement 
of QCD{}.  It might even be possible to find effective methods of 
calculation, so that a confinement mechanism in the lambda model 
could be checked against the experimental data.  It would be 
essential that the dynamics of the lambda defects depend on $L$.  
When $L$ drops below a characteristic confinement distance, the 
lambda defects would have to bind, so that the perturbative 
spacetime gauge theory would become visible.

I would even guess at a general principle, that the lambda model 
always disorders at large enough values of $L$.  In the limit 
$L\rightarrow \infty$, I would expect the lambda model to explore 
the entire manifold of spacetimes.  The effective degrees of freedom 
at $L=\infty$ will not be those associated with any particular 
spacetime, but will be constructed from the entire manifold of 
spacetimes by the nonperturbative fluctuations in the lambda model.  
Physics in any individual spacetime will give only a partial view of 
the large distance physics.

Undoubtedly, these speculations are far too naive, and far too much 
influenced by the simple-looking example of the $j$-instanton.  The 
nonperturbative lambda model is likely to be a hugely complicated 
gas of interacting lambda defects and smooth lambda instantons.  The 
hope is that there are relatively simple regimes at spacetime 
distances $L$ which correspond to the distances in nature where 
relatively simple theoretical descriptions of physics have been 
found to apply.  My speculations are offered only as suggestions of 
a possible complexity and richness in the nonperturbative lambda 
model that will be a challenge to calculation, but might yield 
interesting physics.

%
%
%
%
\sectiono{Spacetime gauge invariance}

The lambda model needs a practical implementation of spacetime gauge 
invariance, including spacetime general covariance.  The manifold of 
spacetimes is the manifold of spacetime tensor fields modulo 
equivalence under spacetime gauge transformations.  In principle, 
the target manifold of lambda model is the manifold of gauge 
equivalence classes.  But the fields $\phi_{i}(z,\bar z)$ of the 
general nonlinear model couple to the wave modes of the spacetime 
tensor fields, not to the gauge equivalence classes.  The coupling 
constants $\lambda^{i}$ are the wave modes of the spacetime tensor 
fields.

Some of the fields $\phi_{i}(z,\bar z)$ make no difference when they 
perturb the action of the general nonlinear model.  These are the 
{\em redundant} fields.  The redundant fields are the derivatives of 
spin 1 fields in the general nonlinear model.  For every spin 1 
field $(\chi_{a}^{z}(z,\bar z),\chi_{a}^{\bar z}(z,\bar z))$ there 
is a redundant spin 0 field
\eq
\phi^{\mathit red}_{a} = \partial \chi_{a}^{z}
+ \bar \partial \chi_{a}^{\bar z}
\:.
\en
The redundant fields in a general nonlinear model $\lambda$ are 
certain linear combinations
\eq
\phi^{\mathit red}_{a} = G^{i}_{a}(\lambda) \phi_{i}
\en
of the fields $\phi_{i}$.  The redundant coupling constants are the 
coupling constants $\lambda^{i}$ that couple to the redundant 
fields.

In a macroscopic spacetime, the redundant coupling constants of the 
general nonlinear model are the gauge variations of the spacetime 
tensor fields.  For example, in a macroscopic spacetime with 
spacetime metric $h_{\mu\nu}(x)$, each vector field $v^{\mu}(x)$ on 
the macroscopic spacetime gives a spin 1 field
\eqa
v^{a}\chi_{a}^{z}(z,\bar z) 
&=& v^{\sigma}(x) \, h_{\sigma\nu}(x) \, \bar \partial 
x^{\nu} \nonumber \\[1ex]
v^{a}\chi_{a}^{\bar z}(z,\bar z) 
&=& v^{\sigma}(x) \, h_{\mu\sigma}(x) \, \partial 
x^{\mu}
\ena
whose derivatives give a redundant field spin 0 field
\eq
v^{a} G^{i}_{a} \phi_{i} = v\substar h_{\mu\nu} (x)
\partial x^{\mu} \bar \partial x^{\nu} 
\en
which represents the infinitesimal gauge transformation of the 
spacetime metric produced by the vector field $v^{\mu}(x)$.

The $G^{i}_{a}(\lambda)$ form a Lie algebra of vector fields on the 
manifold of spacetimes
\eq
G^{j}_{a}\partial_{j}G^{i}_{b} (\lambda) 
-G^{j}_{b}\partial_{j}G^{i}_{a} (\lambda) = F^{c}_{ab}G^{i}_{c} 
(\lambda) \:.
\en
The $F^{c}_{ab}$ are the structure constants of the Lie algebra of 
redundancy transformations in the general nonlinear model, which is 
the Lie algebra of local gauge transformations in spacetime.

The manifold of spacetimes is parametrized by the coupling constants 
$\lambda^{i}$ modulo the redundant coupling constants.  The lambda 
model must respect the equivalence relations given by redundancy in 
the general nonlinear model.  If the lambda model respects 
redundancy, then gauge invariance in any macroscopic spacetime will 
follow automatically.  In particular, the \apm\ will respect 
equivalence under redundancy.  The effective spacetime quantum field 
produced by the lambda model in a macroscopic spacetime will be 
gauge invariant.

The lambda field $\lambda(z,\bar z)$ can be pictured as a map to the 
manifold of redundancy equivalence classes, but the component lambda 
fields $\lambda^{i}(z,\bar z)$ would then couple ambiguously to the 
fields $\phi_{i}(z,\bar z)$, up to arbitrary admixtures of redundant 
fields.

Instead, let there be a lambda field $\lambda^{i}(z,\bar z)$ for 
each spin 0 field $\phi_{i}(z,\bar z)$ in the general nonlinear 
model, including the redundant fields.  Then introduce auxiliary 
spin 1 sources $(\xi^{a}_{z},\xi^{a}_{\bar z})$ to implement 
spacetime gauge invariance.  Couple each spin 1 field 
$(\chi_{a}^{z},\chi_{a}^{\bar z})$ in the general nonlinear model to 
a spin 1 source $(\xi^{a}_{z},\xi^{a}_{\bar z})$, adding
\eq
\int \dif^{2}z \, \mu^{2} \frac1{2\pi} \,
\left [ \xi^{a}_{z}(z,\bar z)\,\chi_{a}^{z}(z,\bar z)
+ \xi^{a}_{\bar z}(z,\bar z)\, \chi_{a}^{\bar z}(z,\bar z)
\right ]
\en
to the action of the general nonlinear model

The general nonlinear model is now locally invariant under 
infinitesimal gauge transformations
\eq
\lambda^{i} \rightarrow \lambda^{i} 
								+ \epsilon^{a}(z,\bar z)
								\, G^{i}_{a}(\lambda)
\en
if at the same time the auxiliary lambda fields are transformed
by
\eqa
\xi^{a}_{z}       &\rightarrow& \xi^{a}_{z} + \partial \epsilon^{a}
\nonumber \\[1ex]
\xi^{a}_{\bar z} &\rightarrow& \xi^{a}_{\bar z} + \bar \partial 
\epsilon^{a}
\:.
\ena
The action density of the general nonlinear model changes by the 
total derivative
\eq
\partial (\epsilon^{a} \chi_{a}^{z})
		+ \bar \partial (\epsilon^{a} \chi_{a}^{\bar z}) )
\:.
\en

A locally gauge invariant action $S(\lambda,\xi)$ for the lambda 
model will be determined by the scale variation of the general 
nonlinear model, as in section~\ref{sect:formula}.  But the form it 
will take is obvious.  The two dimensional derivatives 
$\partial\lambda^{i}$ and $\bar \partial \lambda^{i}$ are simply 
replaced in $S(\lambda)$ by the covariant derivatives
\eqa 
{\mathrm D} \lambda^{i} &=& \partial \lambda^{i} - 
G^{i}_{a}(\lambda) \xi^{a}_{z} \nonumber \\[1ex]
\bar {\mathrm D} \lambda^{j} &=&
\bar \partial \lambda^{j}- G^{j}_{a}(\lambda) \xi^{a}_{\bar z}
\:.
\ena

A localized lambda instanton is now a local minimum of 
$S(\lambda,\xi)$.  The covariant derivatives ${\mathrm D} 
\lambda^{i}$ and $\bar {\mathrm D} \lambda^{i}$ must go to zero as 
$z\rightarrow\infty$.  The auxiliary lambda field 
$(\xi^{a}_{z},\xi^{a}_{\bar z})$ is a 1-form on the complex plane 
with values in the Lie algebra of infinitesimal spacetime gauge 
transformations.  Its path-ordered integrals are group elements in 
the spacetime gauge group.  Along a closed contour around 
$z=\infty$, the indefinite path ordered integral from a fixed 
starting point gives a closed loop in the group $\G_{n}$ of 
spacetime gauge transformations, representing the element in 
$\pi_{1}(\G_{n})$ that indexes the localized lambda instanton.

The renormalization of the general nonlinear model respected general 
covariance in the target manifold~\cite{Friedan-3}.  No particular 
symmetry of the target manifold was assumed.  Renormalization of 
target manifold symmetry was subsumed in renormalization of target 
manifold general covariance.  The renormalization of general 
covariance in the target manifold was subject to possible 
obstructions which were cohomology classes on the target manifold, 
the nonlinear model anomalies.
	
The target manifold of the lambda model is the manifold of 
spacetimes.  The spacetime gauge symmetries are internal symmetries 
of the lambda model, analogous to spacetime symmetries in the 
general nonlinear model.  The renormalization of spacetime gauge 
symmetry in the lambda model is subsumed into the renormalization of 
reparametrization invariance in the manifold of spacetimes.  
Potential nonlinear model anomalies in the lambda model would 
obstruct renormalization of reparametrization invariance in the 
manifold of spacetimes, and might show themselves in spacetime 
quantum field theory as gauge anomalies.  It will have to be shown 
that the lambda model is free from anomalies.

%
%
%
%
%
\sectiono{What needs to be done}

The most urgent task now is to find all the local lambda instantons 
in explicit form, and develop concrete methods for calculating their 
contributions to the effective beta function of the general 
nonlinear model.  Temporarily assume a particular macroscopic 
spacetime and assume a fixed small value for the spacetime coupling 
constant $\gst$, in order to find out if the lambda model actually 
does remove spacetime supersymmetry, produce small nonzero masses, 
and lift the degeneracies that are local in the macroscopic 
spacetime.

Developing effective methods of calculation will require filling in 
details of my arguments for the structure of the theory, or finding 
better arguments.  The most essential elements include the principle 
of tandem renormalization and the effective renormalization group 
invariance of the effective general nonlinear model, which together 
imply the tautological scale invariance of the effective lambda 
model.  Also crucial is the identification the action $S(\lambda)$ 
with the scale variation of the general nonlinear model, which is 
used to establish the gradient property and the spacetime action 
principle in macroscopic spacetimes.

Details of the action $S(\lambda)$ also need to be filled in.  This 
should be straightforward, since $S(\lambda)$ is completely 
determined by the scale variation formula, 
equation~\ref{eq:action-local-scale-variation}.  Lambda fields 
$\lambda^{i}(z,\bar z)$ are introduced as sources for all the 
scaling fields $\phi_{i}(z,\bar z)$ that occur in the general 
nonlinear model of the worldsurface.  The action $S(\lambda)$ is 
read off from the scale variation of the general nonlinear model in 
the presence of those sources.

In particular, several special scaling fields $\phi_{i}(z,\bar z)$ 
occur in the string worldsurface, made entirely from worldsurface 
ghost fields.  The coupling constants $\lambda^{i}$ that couple to 
these special scaling fields play distinguished 
roles~\cite{Polchinski,La-Nelson}{}.  One special bosonic coupling 
constant $\lambda_{D}$ has the effect of shifting the value of the 
number $\ln(T)$.  It is conjugate in the metric 
$T^{-1}g_{ij}(\lambda)$ to a second special bosonic coupling 
constant $\lambda_{D}^{\prime}$, which is redundant, at least in a 
scale invariant worldsurface.  When these special coupling constants 
are made into lambda fields $\lambda_{D}(z,\bar z)$ and 
$\lambda_{D}^{\prime}(z,\bar z)$, it appears that 
$\lambda_{D}^{\prime}(z,\bar z)$ can be interpreted as the logarithm 
of the local two dimensional scale factor $\Lambda(z,\bar z)$ and 
acts as a Lagrange multiplier, enforcing local two dimensional scale 
invariance.  The combined coefficient of the two dimensional 
curvature density $\Lambda^{2}R_{2}(\Lambda)$ from the combined 
local lagrangians of the general nonlinear model and the lambda 
model, is the sum of the special coupling constant $\lambda_{D}$, 
the number $\ln(T)$, and the potential function $T^{-1}a(\lambda)$.  
This seems worth pursuing.  The details of the system of special 
lambda fields need to be worked out.  It might be that they play 
only a formal role in the lambda model.  But it is also possible 
that there will be some indication of how the number $T$ might be 
determined.

A second basic detail that needs filling in is the possible 
antisymmetric tensor coupling in the lambda action.  The heterotic 
worldsurface is chirally asymmetric.  The scale variation of the 
general nonlinear model of the heterotic worldsurface can contain a 
graded antisymmetric tensor coupling $\Theta b_{ij}(\lambda)$ in 
addition to the graded symmetric metric coupling $T^{-1}g_{ij}$.  It 
would be surprising if an antisymmetric coupling did not appear.

If calculation shows that the lambda model can in fact produce the 
needed local effects in a macroscopic spacetime, then there will be 
two obvious directions to take.  One will be a renewed search among 
the possible macroscopic spacetimes for a match to the standard 
model.  The lambda model will produce a local quantum field theory 
in each macroscopic spacetime.  Methods will be needed to winnow the 
macroscopic spacetimes for promising candidates to compare in detail 
with the standard model.

It will also become promising to investigate basic issues, including 
decompactification mechanisms, mechanisms that could determine the 
spacetime coupling constant $\gst^{2}=V\,T$, cosmological 
interpretation, the construction of real time, a mechanism that 
could fix the number $T$ or the dimension $d=2+\epsilon$, topology 
change, and the issue of security in the limit 
$\Lambda^{-1}\rightarrow 0$, $L\rightarrow\infty$.

The limit $L\rightarrow\infty$ raises two questions.  First is 
simply the existence of a scale invariant limit, without which the 
lambda model would have no foundation on which to build the large 
distance physics.  If the lambda model does have a scale invariant 
limit at $\Lambda^{-1}=0$, the question becomes, does degeneracy 
remain in the limit?  Whatever form the effective degrees of freedom 
$\lambdaeff^{i}$ take, do any of them have vanishing effective 
anomalous dimension $\adeff(i)=0$?  Are there marginal effective 
coupling constants $\lambdaeff^{i}$ in the short distance limit?  If 
not, if all degeneracy is lifted at $L=\infty$, then the logarithmic 
divergence will be removed.  The original purpose of the lambda 
model will be realized.  This last question evokes the historical 
roots of the two dimensional nonlinear model and the lambda model in 
the ideas of Bloch, Hohenburg, Mermin, Wagner and Coleman about the 
logarithmic divergences of spin waves in $d=2$ dimensions, and the 
physical consequences of their impossibility for two dimensional 
physics.

%
%
%
%
%
\sectiono{Discussion}

The lambda model is a theory of physics which has a fundamental unit 
of spacetime distance and works entirely at large distance compared 
to that fundamental unit.  The lambda mode appears capable of 
explaining some of the most basic principles of physics.  It appears 
capable of constructing quantum mechanics in spacetime, and 
determining the hamiltonian.  The lambda model appears capable of 
doing this without assumptions about physics at experimentally 
inaccessible spacetime distances near the Planck length.

It seems futile to speculate about small distance physics without 
having in hand a coherent and testable theory of large distance 
physics, given the enormous gulf between the Planck length and the 
length scales of practical experiments.  Without a means of reliably 
predicting observable large distance physics, how can a speculative 
theory of small distance physics be checked against the real world?  
There is considerable room for surprise in the roughly $14$ or $15$ 
orders of magnitude between the Planck length and the smallest 
distances where theories can be checked.  It might be worth 
remembering that past explorations over $14$ or $15$ orders of 
magnitude in distance discovered such surprises as quantum mechanics 
and the elementary particles.  What could possibly justify 
theoretical assumptions about physics across such an enormous gulf 
of spacetime distances, if those theoretical assumptions cannot lead 
to definite statements that can be checked in the real world?

The lambda model presents the possibility of exceptions to the 
well-supported principle that the physics of the large is completely 
explained by the physics of the small.  Under conditions of 
degeneracy, the lambda model may produce nonperturbative effects in 
spacetime which are not explicable on atomistic principles.  If such 
effects can be derived from the lambda model, and confirmed by 
experiment, it will be a salutary reminder that knowledge in physics 
is always incomplete, no matter how striking the success of existing 
theory.  There is a temptation to extrapolate successful theories 
far beyond the extent of their demonstrated reliability, especially 
after the past successes of atomistic physics.  A theory which 
succeeds at describing all available experimental results in a 
certain regime of distances, such as the standard model of particle 
physics does now, is assumed to explain {\em in principle} all the 
complicated phenomena observed at larger distances, if only the 
necessary difficult calculations could be carried out.  Even when 
many such complicated phenomena are successfully explained, there is 
no guarantee that all large distance phenomena will be explained.  
There still remains a remote possibility that subtle unexpected 
effects are yet to be observed.  To search at random for such 
effects is unlikely to be useful.  Guidance is needed from a highly 
credible theory.  The lambda model is proposed as a theory that 
might be capable of acquiring such credibility and also predicting 
unexpected phenomena.

The crucial advantage that the lambda model might have over a 
fundamentally atomistic model of physics is the security that the 
lambda model could give at large distance in spacetime by building 
physics from the limit $L = \infty$ downwards in $L$.  Infrared 
security in the lambda model would eliminate the need to guess at 
the nature of microscopic physics at unobservably small distances in 
spacetime.  The need for some such infrared security is suggested by 
the miniscule value of the observed cosmological constant, which 
seems inexplicable in any atomistic version of spacetime physics.

The lambda model is an attempt to make a weakly coupled theory of 
physics.  Weak coupling means that the spacetime coupling constant 
$\gst$ should be a reasonably small number, say on the order of 
$1/10$.  The value of the number $T$ is a separate matter.  The 
lambda model undoubtedly needs $T$ to be an extremely small number.  
The dimension $d=2+\epsilon$ must be very near 2, otherwise the 
entire analysis and physical interpretation of the lambda model 
would break down.  The spacetime coupling constant $\gst$ emerges 
only in a macroscopic spacetime of volume $V$, by the relation 
$\gst^{2}= V\, T$.  The lambda model does not seem to require that 
$\gst$ be small.  The lambda model might well be a strongly coupled 
two dimensional quantum field theory in some spacetime regimes.  The 
lambda model might still be useful there, if it happens to be an 
integrable two dimensional field theory.  The most obvious prospects 
of the theory, however, seem to call for weak coupling.  For 
example, it is difficult to imagine how a spectrum of exponentially 
large spacetime distances could arise without a small spacetime 
coupling constant.

I retain a naive hope that a weakly coupled theory of large distance 
physics can succeed in explaining the standard model of the 
elementary particles.  It is remarkable that all observed couplings 
of the standard model are in fact weak at the smallest distances 
accessible to experiment.  The weakness of all the observed 
couplings is one of the most striking results from high energy 
experimental physics.  It seems to me misguided to turn away from 
the possibility of a weakly coupled theory before having in hand a 
coherent method to determine large distance physics.  A systematic 
weakly coupled theory of large distance physics would be so useful 
that nothing but a definitive demonstration of infeasibility should 
forestall the attempt.  In the end, of course, the assumption of 
weak coupling must be justified dynamically, since the spacetime 
coupling constant is a parameter of the manifold of spacetimes.

The lambda model is mathematically universal.  The target manifold, 
the metric coupling, the potential function are all mathematically 
natural objects.  The couplings of the lambda model satisfy 
mathematically natural differential equations on the manifold of 
spacetimes, expressing generalized two dimensional scale invariance.  
No arbitrary choices are made.

The lambda model is {\em not} universal in the in sense of quantum 
field theory.  As a nonlinear model, it is scale invariant in the 
generalized sense.  Its couplings are at a fixed point of the 
renormalization group of the general nonlinear model whose target 
manifold is the manifold of spacetimes.  The fixed point is not 
stable under the renormalization group.  If a small perturbation 
were made, the renormalization group would drive the nonlinear model 
far away from the fixed point.  There are infinitely many unstable 
directions.  Perturbing the action density by any function 
$f(\lambda)$ on the target manifold gives a dimension 2 
perturbation, which would grow quadratically in the two dimensional 
distance, freezing the lambda field to the minimum of the function 
$f(\lambda)$.  A dimension 2 perturbation would freeze the system 
into a fixed spacetime, suppressing the fluctuations of the lambda 
field that are needed to cancel the divergence due to local handles 
in that spacetime.  The logarithmic divergence would return.

The lambda model must be held at the fixed point.  All relevant 
perturbations of the lambda model must be tuned to zero.  There 
might be a formal apparatus in the lambda model, perhaps involving 
the special lambda field $\lambda_{D}^{\prime}$, that enforces this 
tuning.  Or there might be a deeper mechanical explanation.  If some 
mechanical model of the string worldsurface automatically gives rise 
to the lambda model, it would presumably hold the lambda model 
precisely at the fixed point.

However the tuning is done, by hand if necessary, it is possible to 
carry out the task because the couplings of the lambda model do not 
fluctuate.  The couplings of the lambda model are classical 
geometric quantities on the manifold of spacetimes.

The lambda model produces a probabilistic description of spacetime.  
It may single out a number of possible macroscopic spacetimes.  In 
each, the spacetime fields describing geometry and matter fluctuate 
according to a quantum field theory produced by the lambda model.  
But the geometry that defines the lambda model does {\em not} 
fluctuate.  The couplings of the lambda model take definite values 
satisfying classical differential equations on the manifold of 
spacetimes.  Strict causality, which was renounced in spacetime when 
quantum mechanics was discovered, might be regained at another level 
of abstraction.

If the theory works, part of the \apm\ on the manifold of spacetimes 
will be found to concentrate at a spacetime that matches our 
spacetime in its dimension, its cosmology, and in the phenomenology 
of its elementary particles.  Our spacetime might turn out to be 
only one of many where the \apm\ concentrates, and might carry only 
a small share of the total measure.  It would become a challenge to 
devise experiments that could detect the other possibilities.

%
%
%
\newpage
\def\acknowledgments{}
\section*{Acknowledgments}
\addcontentsline{toc}{section}{Acknowledgments}
%
%
\acknowledgments

Very early in my thinking about the lambda model, circa 1988, I was 
stimulated by a comment from V.~Periwal and a comment from 
S.~Shenker.  I am grateful for their help.

This work has been supported since 1989 by Rutgers, The State 
University of New Jersey through its New High Energy Theory Center.  
Since 1986, much of the work has been done in Reykjav\'{\i}k, during 
visits to Raunv\'{\i}sindastofnum H\'ask\'olans \'Islands, The 
Natural Science Institute of the University of Iceland.

The United States Department of Energy provided support from 1978 
until 2001, first through the Lawrence Berkeley Laboratory of the 
University of California, Berkeley during 1978-1980, then through 
the Fermi Institute of the University of Chicago during 1981-1989, 
then through the Rutgers New High Energy Theory Center during 
1989-2001.  The Centre d\'{e}s \'{E}tudes Nucleaire de Saclay, 
France provided support during 1980-1981.  A fellowship from the 
Sloan Foundation provided support during 1981-1983, and a fellowship 
from the MacArthur Foundation during 1987-1993.

I am grateful to all the people and institutions involved.  I am 
especially grateful to I.~Singer, Y.~Nambu and L.~Kadanoff for 
arranging support during the first years.

%
%
%

%
\end{document}